\documentclass[11pt]{article}
\usepackage[left=1in,top=1in,right=1in,bottom=1in,head=.1in,nofoot]{geometry}

\usepackage{setspace,url,bm,amsmath} 
\usepackage{makecell}
\usepackage{titlesec} 
\titlelabel{\thetitle.\quad} 

\usepackage[margin=20pt]{subcaption}

\RequirePackage{amsthm,amsmath,amsfonts,amssymb}
\RequirePackage[authoryear]{natbib}
\RequirePackage[colorlinks,citecolor=blue,urlcolor=blue]{hyperref}
\RequirePackage{graphicx}
\RequirePackage{mathrsfs}  
\RequirePackage{bm}
\RequirePackage{graphicx}
\RequirePackage{graphics}
\RequirePackage{multicol}
\RequirePackage{multirow}
\RequirePackage{enumitem}
\RequirePackage{booktabs}
\RequirePackage[table,xcdraw]{xcolor}
\usepackage{indentfirst}
\usepackage{authblk}

\usepackage{appendix}
\RequirePackage[plain,noend]{algorithm2e}

\theoremstyle{plain}
\newtheorem{theorem}{Theorem}
\newtheorem{corollary}{Corollary}
\newtheorem{lemma}{Lemma}

\newtheorem{condition}{Condition}

\theoremstyle{remark}

\newcommand{\bbE}{{\mathbb{E}}}
\newcommand{\bbV}{{\text{var}}}
\newcommand*{\QEDA}{\hfill\ensuremath{\blacksquare}}
\newcommand{\argmin}{{\operatorname{arg\,min}}}

\newcommand{\bZero}{{\boldsymbol 0}}

\newcommand{\bM}{{\boldsymbol M}}
\newcommand{\bB}{{\boldsymbol B}}
\newcommand{\bD}{{\boldsymbol D}}

\newcommand{\bI}{{\boldsymbol I}}

\newcommand{\bS}{{\boldsymbol S}}
\newcommand{\bX}{{\boldsymbol X}}
\newcommand{\bY}{{\boldsymbol Y}}
\newcommand{\bV}{{\boldsymbol V}}
\newcommand{\bW}{{\boldsymbol W}}

\newcommand{\bZ}{{\boldsymbol Z}}

\newcommand{\bbeta}{{\boldsymbol \beta}}
\newcommand{\bmu}{{\boldsymbol \mu}}
\newcommand{\bepsilon}{{\boldsymbol \epsilon}}
\newcommand{\blambda}{{\boldsymbol \lambda}}
\newcommand{\bSigma}{{\boldsymbol \Sigma}}
\newcommand{\bLambda}{{\boldsymbol \Lambda}}
\newcommand{\bGamma}{{\boldsymbol \Gamma}}

\newcommand{\cD}{{\mathcal D}}
\newcommand{\cP}{{\mathcal P}}

\newcommand{\cS}{{\mathcal S}}

\newcommand{\mW}{{\mathscr W}}

\newcommand{\bbP}{{\mathbb P}}

\begin{document}

\title{\bf Bayesian criterion for Re-randomization}
\author[1]{Zhaoyang Liu$^{*}$}
\author[1]{Tingxuan Han\footnote{These authors have contributed equally to this work.}}
\author[2]{Donald B. Rubin}
\author[1]{Ke Deng\footnote{Corresponding author: kdeng@tsinghua.edu.cn}}
\affil[1]{Center for Statistical Science \& Department of Industrial Engineering, Tsinghua University}
\affil[2]{Yau Mathematical Sciences Center, Tsinghua University}
\date{}
\renewcommand\Authands{ and }

\maketitle

\begin{abstract}
{
Re-randomization has gained popularity as a tool for experiment-based causal inference due to its superior covariate balance and statistical efficiency compared to classic randomized experiments. 
However, the basic re-randomization method, known as ReM, and many of its extensions have been deemed sub-optimal as they fail to prioritize covariates that are more strongly associated with potential outcomes.
To address this limitation and design more efficient re-randomization procedures, a more precise quantification of covariate heterogeneity and its impact on the causal effect estimator is in a great appeal.
This work fills in this gap with a Bayesian criterion for re-randomization and a series of novel re-randomization procedures derived under such a criterion.
Both theoretical analyses and numerical studies show that the proposed re-randomization procedures under the Bayesian criterion outperform existing ReM-based procedures significantly in effectively balancing covariates and precisely estimating the unknown causal effect.
}
\end{abstract}

\section{Introduction}
\label{sec:intro}

Covariate imbalance is to be avoided in randomized experiments, even though complete randomization balances all potential confounding factors on average and therefore becomes the ``gold standard'' in causal inference \citep{Holschuh1980,Wu1981,Urbach1985,Imai2008,Cox2009,Morgan2012}.
A natural way to avoid treatment assignment allocations with unbalanced covariates is to reject a ``bad'' allocation, and redo the randomization until a ``good'' allocation with properly balanced covariates is obtained, before the experiment is conducted. 
This is called \textit{re-randomization}, which was first suggested by \cite{fisher1992arrangement}. 
The general framework including the sufficient condition to guarantee unbiased estimation and related benefits from doing re-randomization when using Mahalanobis distance as the metric to measure imbalance (referred to as ReM) is provided in \cite{Morgan2012}.

Subsequent to the initial work, relevant literatures have shown extended applications and theoretical properties of ReM. 
Unequal weights on covariates are considered using \emph{ReM in tiers of covariates} (ReM$_T$) in \cite{Morgan2015}, where covariates are partitioned into tiers according to their relative importance, with a tighter threshold of ReM for covariates in a  more important tier. After re-randomization in one tier, residuals orthogonal to all previous tiers of covariates can be used to calculate Mahalanobis distance in the next tier in ReM$_T$, leading to reduced computational complexity and refined statistical efficiency of ReM when a large number of covariates are involved. Applications of ReM in factorial and sequential designs are analyzed in \cite{Branson2016} and \cite{Zhou2018}. 
Asymptotic theory for the standard difference-in-means estimator of treatment effect under ReM is derived by an orthogonal decomposition of the final sampling distribution as a linear combination of a Gaussian random variable and a truncated Gaussian random variable \citep{Li2018,Li2020}. 
The asymptotic performance of ReM in stratified randomization and survey experiments are studied in \cite{wang2021rerandomization} and \cite{yang2021rejective}.
\cite{zhang2021pca} recommended applying ReM for the top-$k$ principal components of covariates only, leading to a variate of ReM named PCA-ReM, which reduces computational complexity of ReM significantly while retaining much of its benefits. 
Ridge-ReM proposed by \cite{Branson2021} utilizes modified Mahalanobis distance to deal with collinearities among covariates in situations with high-dimensional or highly-correlated covariates. 
Considering that a small acceptance probability will lead to heavy computational burden, \cite{zhu2022pair} proposed a pair-switching strategy to reduce the computational cost of ReM.

Beyond the ReM-based approaches, alternative strategies for re-randomization based on other criteria have been proposed as well.
\cite{Kallus2018} proposed an optimal allocation  based on a minimax criterion, which remains controversial, as discussed in \cite{Johansson2020} and \cite{Kallus2020}. 
To tackle with this problem, \cite{wang2022rerandomization} proposed a procedure allowing the covariate imbalance to diminish as sample size increases.
On the other hand, \cite{Li_2021kde} suggested to estimate the density function of covariates in the treatment group and the control group via kernel density estimation, and measure the degree of covariate imbalance by comparing the estimated density functions.

However, all these existing re-randomization methods are sub-optimal in sense that they fail to prioritize covariates that are more strongly associated with potential outcomes, while there is a clear insight that such covariates should enjoy higher priority to be balanced.
Although ReM$_T$ considers covariate heterogeneity by partitioning the covariates into tiers and assigning higher priority to balancing covariates in the leading tiers, such a solution is incomplete because it treats covariates in the same tier equally. 
A more precise quantification of covariate heterogeneity and its impact on the causal effect estimator would lead to a better solution.

This study fills in this gap by establishing a Bayesian criterion for re-randomization, which formulates our knowledge on the relative importance of different covariates to the potential outcomes via a prior distribution, and derives a novel re-randomization procedure referred to as ReB.
We found that ReB 
takes many existing ReM-based re-randomization procedures, e.g., ReM, PCA-ReM and Ridge-ReM, as special cases, forming a unified framework for studying various re-randomization approaches.
Theoretical analyses show that the proposed ReB outperforms the ReM-based re-randomization procedures in terms of getting a more accurate causal effect estimator, as long as the prior distribution is informative to highlight the relative importance of different covariates.

In case that no such a nice informative prior distribution is available, we propose a two-stage strategy to implement ReB, where a small pilot experiment is conducted in the first stage to establish the informative prior distribution we need, and ReB is conducted for the remaining samples in the second stage.
The final causal effect estimator can be obtained by properly integrating the two estimators obtained in both stages.
Theoretical analyses and simulation studies show that the two-stage ReB procedure also achieves superior covariate balance and thereby more precise estimation of causal effects than existing methods.
In the literature, the idea of using pre-experimental data in re-randomaization has been studied. 
For example, \cite{johansson2020rerandomization} proposed a rank-based balance measure for re-randomization and weight for each covariates was estimated using pre-experimental data; 
\cite{zhang2021response} introduced the response-adaptive design for re-randomization and took ethics into consideration. 
Compared to the proposed two-stage ReB procedure, however, these two methods do not make full use of the information obtained from the pre-experiment and suffer from the lack of theoretical guarantees.

The following part of this paper is organized as follows. 
Section \ref{sec:Preliminary} defines the basic notations and briefly reviews the repeated sampling properties of re-randomization criterion based on the Mahalanobis distance in the literature.
Section \ref{sec:ReB} proposes the Bayesian criterion for re-randomization, establishes the corresponding theoretical properties and discusses its connections to other re-randomization criteria in the literature. 
Section \ref{sec:Two-stage} introduces the framework of two-stage Bayesian re-randomization procedures and gives their asymptotic properties. 
Section \ref{sec:Simulation} and \ref{sec:RealDataAnalysis} support the proposed methodology with simulation studies and a real data application.
Finally, we conclude with a discussion in Section \ref{sec:Discussion}.
Detailed proofs and additional simulation results
are provided in the Supplementary Material.

\section{Notation and preliminaries}\label{sec:Preliminary}
Following the classic setting for causal inference in \cite{Rubin1974} and in \cite{Imbens2015}, we consider statistical inference for the \emph{average causal effect} (ACE) of an \emph{active treatment} $t$ with respect to a \emph{control treatment} $c$ over a finite population of units $\cP=\{1,\cdots,N\}$, i.e.,
\begin{equation*}
    \tau=\frac{1}{N}\sum_{i=1}^N\tau_i=\frac{1}{N}\sum_{i=1}^N\big(Y_i(1)-Y_i(0)\big),   
\end{equation*}
where $Y_i(1)$ and $Y_i(0)$ are the potential outcomes of unit $i\in\cP$ under treatment $t$ and control $c$ respectively, and $\tau_i=Y_i(1)-Y_i(0)$ is the unit-level causal effect.

Let $W_i$ be the \emph{treatment assignment indicator} of unit $i$ ($W_i=1$ if unit $i$ is assigned to receive the active treatment $t$, and $W_i=0$ if the control treatment $c$ is assigned),
and $\bW=(W_1,\cdots,W_N)^T$ be the \emph{treatment assignment vector} for the experiment. 
In a randomized experiment with $N_t$ and $N_c$ being the group size of the treatment group and control group respectively (apparently $N_t+N_c=N$), the treatment assignment vector is a random sample from the space:
\[\begin{split}
\mW=\left\{\bW\in\{0,1\}^N: \sum_{i=1}^N W_i=N_t\right\},
\end{split}\]
leading to the following unbiased estimator of  the causal estimand $\tau$ under the Stable Unit Treatment Value Assumption (SUTVA) introduced by \cite{Rubin1980}:
\begin{equation*}
    \hat\tau(\bW)=\frac{1}{N_t}\sum_{i=1}^NY_i(1)\cdot W_i-\frac{1}{N_c}\sum_{i=1}^NY_i(0)\cdot(1-W_i).
\end{equation*}

\subsection{Framework of re-randomization}
Further assume that unit $i$ is associated with a group of covariates $\bX_{i}=(X_{i1},\cdots,X_{ip})^T$, which may have prediction power for potential outcomes $(Y_i(1),Y_i(0))$, and thus the causal effect $\tau_i$. 
Let $\bX=(X_{ij})_{1\leq i\leq N,1\leq j\leq p}$ be the covariate matrix of the $N$ units in the population of interest.
A re-randomization procedure, which can be traced back to R.A. Fisher as pointed out by \cite{Morgan2012}, defines a deterministic \emph{acceptance rule}
\begin{equation*}
\phi(\boldsymbol{X}, \boldsymbol{W})=\begin{cases}
1, & \mbox{if assignment $\bW$ is acceptable,}\; \\
0,& \mbox{otherwise,} \;
\end{cases}
\end{equation*}
to reject ``bad'' assignments (i.e., ones making $\phi(\bX,\bW)=0$),
until a plausible assignment $\bW$ satisfying $\phi(\bX,\bW)=1$ is obtained. 
Acceptance rule $\phi(\bX,\bW)$ defines a subset of acceptable assignments $\mW_{\phi}\subseteq \mW$, which is referred to as the \emph{acceptance region} of rule $\phi$.

\cite{Morgan2012} suggested specifying the re-randomization procedure $\phi$ by controlling the Mahalanobis distance between
$$\bar{\bX}_t=\frac{1}{N_t}\sum_{i=1}^N\bX_{i}W_i\quad\mbox{and}\quad\bar{\bX}_c=\frac{1}{N_c}\sum_{i=1}^N\bX_{i}(1-W_i),$$
which is defined as
\begin{equation*}
    d_M \triangleq \bD^T\bSigma_\bD^{-1}\bD
\end{equation*}
with $\bD=\bar{\bX}_t-\bar{\bX}_c$ and $\bSigma_\bD=\operatorname{cov}\left(\bD\right)$ being the covariance matrix of $\bD$,
leading to the following \emph{re-randomization using the Mahalanobis distance} (ReM):
\begin{equation*}
\phi_M(\bX,\bZ)=I\big(\bD^T\bSigma_\bD^{-1}\bD\leq a\big),
\end{equation*}
where $I(\cdot)$ is the 0-1 indicator function, and the threshold $a$ can be determined based on the equation below for each specific acceptance rate $\alpha\in(0,1)$:
\begin{equation*}
    \bbP(\phi_M=1)=\bbP\big(\bD^T\bSigma_\bD^{-1}\bD\leq a\big)=\alpha.
\end{equation*}
Denote $(\hat\tau\mid\phi_M=1)$ as the difference-in-mean estimator of $\tau$ under the re-randomization mechanism $\phi_M$.
\cite{Morgan2012} showed that $(\hat\tau\mid\phi_M=1)$ is an unbiased estimator of $\tau$ satisfying
$\bbE(\hat\tau\mid\phi_M=1)=\tau$
as long as $N_t=N_c$;
and, 
$(\hat\tau\mid\phi_M=1)$ is a more efficient estimator of $\tau$ that the classic $\hat\tau$ under a completely randomized experiment with the following \emph{percent reduction in variance} (PRIV):
\begin{equation*}
    PRIV=1-\frac{\bbV(\hat\tau\mid\phi_M=1)}{\bbV(\hat\tau)}=100\times(1-v_{\alpha,p})R^2,
\end{equation*}
where $R^2$ represents the squared multiple correlation between the observed potential outcomes and the covariates within each treatment group, and
\begin{equation*}
v_{\alpha,p}\triangleq \frac{\bbE(\chi^2_p\mid\chi^2_p\leq\xi_{\alpha,p})}{p} = \frac{\bbP(\chi^2_{p+2}\leq \xi_{\alpha,p})}{\bbP(\chi^2_p\leq \xi_{\alpha,p})}=\frac{\bbP(\chi^2_{p+2}\leq \xi_{\alpha,p})}{\alpha},
\end{equation*}
with $\chi^2_p$ being a chi-square random variable with degree of freedom $p$ and $\xi_{\alpha,p}$ being the $\alpha$-quantile of $\chi^2_p$.

\subsection{Repeated sampling inference under re-randomization}\label{subsec:asymptotic}
Hereinafter, we use notation ``$\bX_n\overset{\cdot}{\sim}\bY_n$" for the fact that two sequences of random vectors $\bX_n$ and $\bY_n$ converge weakly to the same distribution with sample size $n$ going to infinite, and notation ``$\bX\sim\bY$" for the fact that two random vectors $\bX$ and $\bY$ follow the same distribution.

Under complete randomization, Theorem 3 in \cite{Li2017} implies that  $\sqrt{N}\big(\hat{\tau}-\tau,\bm{D}^T\big)$ has the following covariance matrix: 
\begin{equation}\label{eq:V}
    \bm V_N = \left(\begin{array}{cc}
        V_{\tau\tau,N} & \bm V_{\tau x,N}  \\
        \bm V_{x\tau,N} & \bm V_{xx,N} 
    \end{array}\right)
= \left(\begin{array}{cc}
r_1^{-1}S^2_{Y(1)}+r_0^{-1}S^2_{Y(0)}-S^2_{\tau}\quad & r_1^{-1}\bm S^2_{Y(1),\bm X}+r_0^{-1}\bm S^2_{Y(0),\bm X}  \\
r_1^{-1}\bm S^2_{\bm X,Y(1)}+r_0^{-1}\bm S^2_{\bm X,Y(0)}\quad & \bm (r_0r_1)^{-1}\bm S^2_{\bm X} 
\end{array}\right),
\end{equation}
where $r_0=N_c/N$ and $r_1=N_t/N$, finite-population variance and covariance matrices  $S^2_{Y(z)} = (N-1)^{-1} \sum_{i=1}^N \big(Y_i(z)-\bar{Y}(z)\big)^2,\quad S^2_{\tau} = (N-1)^{-1} \sum_{i=1}^N (\tau_i-\tau)^2, \quad \bm S^2_{\bm X} = (N-1)^{-1} \sum_{i=1}^N \big(\bm X_i-\bar{\bm X}\big)\big(\bm X_i-\bar{\bm X}\big)^T,\quad \bm S^2_{Y(z),\bm X}=\bm S^2_{\bm X,Y(z)} =
    (N-1)^{-1}\sum_{i=1}^N \big(Y_i(z)-\bar{Y}(z)\big)\big(\bm X_i-\bar{\bm X}\big)^T$, $z=0$ or $1$. 
Moreover, according to finite population central limit theorem, i.e., Theorem 5 in \cite{Li2017}, when Condition \ref{cond1} below is satisfied, $\sqrt{N}\big(\hat{\tau}-\tau,\bm{D}^T\big)$ will converge to a Gaussian distribution weakly as the sample size $N$ approaches infinity, i.e.,
\begin{equation}\label{eq:AssymptoticDistribution4HatTau_bD}
\sqrt{N}\big(\hat{\tau}-\tau,\bm{D}^T\big) \overset{\cdot}{\sim} (A,\bB^T), \text{ where }(A,\bB^T)  \sim
    \mathcal{N}(\bm 0,\bm V_N).
\end{equation}
For simplicity, we omit $N$ in finite-population covariance matrices $\bm V_{N}, V_{\tau\tau,N}, \bm V_{\tau x,N}, \bm V_{x\tau ,N}$ and $\bm V_{xx,N}$, and denote their limits as $\bm V_{\infty}, V_{\tau\tau,\infty}, \bm V_{\tau x,\infty}, \bm V_{x\tau ,\infty}$ and $\bm V_{xx,\infty}$ hereinafter.
\begin{condition}\label{cond1}
As $N \rightarrow \infty$, for $z=0,1$,
(1) $r_z$ has positive limits; (2) the finite population variances and covariances $S^2_{Y(z)}$, $S^2_{\tau}$, $\bm S^2_{\bm X}$ and $\bm S^2_{\bm X,Y(z)}$ have finite limiting values, and the limit of $\bm S^2_{\bm X}$ is nonsingular; and (3) $\max_{1\leq i \leq N} |Y_i(z)-\bar{Y}(z)|^2/N \rightarrow 0$ and $\max_{1\leq i \leq N} ||\bm X_i-\bar{\bm X}||^2/N \rightarrow 0$.
\end{condition}

Moreover, \cite{Li2018} has established more detailed asymptotic results for a special family of re-randomization mechanisms whose acceptance rule depends on $\sqrt{N}\bm D$ and $\bm V_{xx}$ only.
Let $\Phi$ be the family of all re-randomization mechanisms that is of the form $\phi\big(\sqrt{N}\bm D,\bm V_{xx}\big)$ and satisfies the Condition \ref{cond2} below.
Let $\mathcal{B}_{\phi} = \{\bm\mu:\phi(\bm\mu,\bm V_{xx})=1\}$ be the acceptance region for $\sqrt{N}\bm D$ induced by $\phi$, and $\mathcal{B}_{\phi,\infty} = \{\bm\mu:\phi(\bm\mu,\bm V_{xx,\infty})=1\}$ be the limit of $\mathcal{B}_{\phi}$.
\cite{Li2018} proved that \begin{equation}\label{eq:Asymptotics4Phi}
    \left.\left(\begin{array}{c}
        \sqrt{N}(\hat{\tau}-\tau)  \\
        \sqrt{N}\bm D 
    \end{array}\right)\right|\sqrt{N}\bm D \in \mathcal{B}_{\phi} \overset{\cdot}{\sim} 
    \left.\left(\begin{array}{c}
        A  \\
        \bm B 
    \end{array}\right)\right|\bm B \in \mathcal{B}_{\phi},\ \forall \phi \in \Phi, 
\end{equation}
where $(A,\bm B^T) \sim \mathcal{N}(\bm 0, \bm V)$ with its limit $(A_{\infty},\bm B_{\infty}^T) \sim \mathcal{N}(\bZero,\bV_{\infty})$.
\begin{condition}\label{cond2}
The covariate balance criterion $\phi(\cdot,\cdot)$ satisfies:
(1) $\phi(\cdot,\cdot)$ is almost surely continuous; (2) for $\bm B\sim N(0,\bm U)$, $\bbP(\phi(\bm B, \bm U)=1)>0$ for any $\bm U>0$, and $\bbV(\bm B|\phi(\bm B, \bm U)=1)$ is a continuous function of $\bm U$; (3) $\phi(\bm\mu,\bm U)=\phi(-\bm\mu,\bm U)$ for all $\bm \mu$ and $\bm U>0$.
\end{condition}

The symmetry condition for $\phi$ in Condition \ref{cond2} guarantees that $\hat\tau$ is an unbiased estimator of $\tau$ under any re-randomization mechanism $\phi\in\Phi$.
Moreover, based on \eqref{eq:Asymptotics4Phi}, 
we have
\[
\begin{split}
\big.\sqrt{N}(\hat{\tau}-\tau)\big|\sqrt{N}\bm D \in \mathcal{B}_{\phi} &\overset{\cdot}{\sim} \left. (A\right|\bm B \in \mathcal{B}_{\phi})\sim \epsilon + (\bm V_{\tau x}\bm V_{xx}^{-1}\bm B|\bm B \in \mathcal{B}_{\phi}),
\end{split}
\]
where $\epsilon \sim \mathcal{N}(0,V_{\tau\tau}-\bm V_{\tau x}\bm V_{xx}^{-1} \bm V_{x\tau})$ is independent of $\bm B$.
Thus, 
the asymptotic sampling variance of $\sqrt{N}(\hat{\tau}-\tau)$ under a re-randomization mechanism $\phi\in\Phi$ is 
\begin{equation}\label{eq:AsymptoticVariance4Phi}
\bbV_a\big(\sqrt{N}(\hat{\tau}-\tau)|\phi=1\big) = 
(1-R^2_{\infty})V_{\tau\tau,\infty} + 
    \bbE\left(\bm B^T_{\infty}\bbeta_{\infty}\bbeta^T_{\infty}\bm B_{\infty}|\bm B_{\infty} \in \mathcal{B}_{\phi,\infty}\right),
\end{equation}
where $R^2_{\infty}$ and $\bbeta_{\infty}$ are the limits of the constant scalar and vector defined below:
\begin{eqnarray}
\label{eq:R2_infty}
    R^2 &\triangleq& \bm V_{\tau x}\bm V_{xx}^{-1}\bm V_{x\tau}/V_{\tau \tau},\\
    \label{eq:beta_infty}
    \bm\beta &\triangleq& \bm V_{xx}^{-1}\bm V_{x\tau}.
\end{eqnarray}
Clearly, $\bm\beta$ and $R^2$ are the regression coefficient for individual causal effect and covariates, and the squared multiple correlation, respectively.


In practice, as suggested by \cite{Li2018}, we can measure the efficiency improvement by a re-randomization procedure $\phi$ relative to complete randomization via the \emph{Percent Reduction in Asymptotic Sampling Variance} (PRIASV) of $\sqrt{N}(\hat{\tau}-\tau)$ defined below:
\begin{equation}\label{eq:PRIAV}
    \text{PRIASV}_\phi \triangleq 100 \times \left[1-\frac{\bbV_a\big(\sqrt{N}(\hat{\tau}-\tau)\mid \phi=1\big)}{\bbV_a\big(\sqrt{N}(\hat{\tau}-\tau)\big)}\right]
    = 100 \times \left(1-r_\phi\right)\times R^2_{\infty},
\end{equation}
where
\begin{equation}\label{eq:r_phi}
r_\phi=\frac{\bbE\left(\bm B^T_{\infty}\bbeta_{\infty}\bbeta_{\infty}^T\bm B_{\infty}|\bm B_{\infty} \in \mathcal{B}_{\phi,\infty}\right)}{\bbeta^T_{\infty} \bm V_{xx,\infty}\bbeta_{\infty}} = \frac{\bbE\left(\bm B^T_{\infty}\bbeta_{\infty}\bbeta^T_{\infty}\bm B_{\infty}|\bm B_{\infty} \in \mathcal{B}_{\phi,\infty}\right)}{\bbE\left(\bm B^T_{\infty}\bbeta_{\infty}\bbeta_{\infty}^T\bm B_{\infty}\right)}.
\end{equation}

\section{Bayesian perspective for re-randomization} \label{sec:ReB}
In this study, we aim to propose novel re-randomization mechanisms that are more efficient than the classic ReM-based ones.
The key idea roots in the observation that all ReM-based re-randomization mechanisms, including ReM, ReM$_T$, PCA-ReM and Ridge-ReM, fail to take full consideration of the heterogeneity of different covariates with respect to the causal effect, while there is a clear insight that covariates associated with causal effect more closely should enjoy higher priority to be balanced.
Although ReM$_T$ considers covariate heterogeneity by partitioning the covariates into tiers and giving higher priority to balancing covariates in the leading tiers, such a strategy is sub-optimal because it treats covariates in the same tier equally.
We have better choices if we can quantify the heterogeneity of covariates with respect to causal effect more precisely.

\subsection{Re-randomization under the oracle}
To overcome the limitation of the ReM-based re-randomization mechanisms, we seek a principled way to design more efficient re-randomization mechanisms.
For any $\phi \in \Phi$, we can define its asymptotic acceptance rate in the sequence of finite populations as 
\begin{equation*}
    \gamma_{\phi} =  \lim_{N \rightarrow \infty} \bbP\big(\sqrt{N}\bm D \in \mathcal{B}_{\phi}\big)=\bbP\big(\bm B_{\infty} \in \mathcal{B}_{\phi,\infty}\big).
\end{equation*}
For any $\alpha \in (0,1)$, let 
\begin{equation*}
    \Phi_{\alpha} = \{\phi\in\Phi:\gamma_{\phi}=\alpha\}
\end{equation*}
be the subset of $\Phi$ that covers all re-randomization mechanisms with asymptotic acceptance rate $\alpha$. 
In principle, it is ideal to find the most efficient re-randomization mechanism in $\Phi_{\alpha}$ that minimizes the asymptotic conditional sampling variance $\bbV_a\big(\sqrt{N}(\hat{\tau}-\tau)|\phi=1\big)$, i.e,
\begin{equation*}
\phi^*=\arg\min_{\phi\in\Phi_{\alpha}}\bbV_a\big(\sqrt{N}(\hat{\tau}-\tau)|\phi=1\big).
\end{equation*}
The following theorem gives the solution to this optimization problem.

\begin{theorem}\label{thm:ReO}
Suppose Condition \ref{cond1} holds and $\bm\beta$ is known. 
The following $\bm\beta$-specific re-randomization procedure
\begin{equation*}
\phi_{\bm\beta}\big(\sqrt{N}\bm D, \bm V_{xx}\big) = I\big(d_{\bm\beta} \leq \sigma^2_{\bm\beta}\cdot\xi_{\alpha,1}\big)
\end{equation*}
satisfies $\phi_{\bm\beta} \in \Phi_{\alpha}$ and 
\[\begin{split}
    \bbV_a\big(\big.\sqrt{N}(\hat{\tau}-\tau)\big|\phi_{\bm\beta}=1\big) \leq \bbV_a\big(\big.\sqrt{N}(\hat{\tau}-\tau)\big|\phi=1\big),\ \forall \phi\in\Phi_{\alpha},
\end{split}\]
where $\xi_{\alpha,1}$ is the $\alpha$-quantile of the $\chi^2_1$ distribution,
\begin{eqnarray*}
d_{\bbeta} &=& \big(\sqrt{N}\bm D\big)^T \bm\beta\bm\beta^T \big(\sqrt{N}\bD\big)= N\bm D^T \bm\beta\bm\beta^T \bm D,\\
\text{and}\ \ 
\sigma^2_{\bm\beta}&=& \bm\beta^T\bm V_{xx} \bm\beta = \bV_{\tau x}\bV_{xx}^{-1}\bV_{x\tau}.
\end{eqnarray*}
\end{theorem}

Theorem~\ref{thm:ReO} shows that $\phi_{\bm\beta}$ is the optimal re-randomization mechanism in $\Phi_{\alpha}$ when $\bm\beta$ is known, and $\phi_{\bm\beta}$ leads to a ribbon-shaped acceptance region in the space of $\bm D$, i.e.,
\begin{equation*}
\mathcal{B}_{\phi_{\bm\beta}}=\left\{
    \sqrt{N}\bm D: \big(\sqrt{N}\bm\beta^T\bm D\big)^2 \leq \bm\beta^T \bm V_{xx} \bm\beta \cdot \xi_{\alpha,1}\right\},
\end{equation*}
whose orientation is orthogonal to the direction of $\bm\beta$. In addition, we note that 
\[\begin{split}
    \bm\beta^T \bm D = \frac{1}{N_t} \sum_{i=1}^N \bm Z_i(\bm\beta)W_i - \frac{1}{N_c} \sum_{i=1}^N \bm Z_i(\bm\beta)(1-W_i) = \bar{Z}_t(\bm\beta) - \bar{Z}_c(\bm\beta),
\end{split}\]
where $Z_i(\bm\beta) = \bm X_i^T \bm\beta$ is the index variable obtained by taking the inner product of $\bm X_i$ and $\bm\beta$. 
Therefore, controlling $d_{\bm\beta} = N\bm D^T \bm\beta\bm\beta^T \bm D$ as described in $\phi_{\bm\beta}$ is equivalent to controlling the imbalance of the index variable $Z_i(\bm\beta) = \bm X_i^T \bm\beta$ in the Euclidean space. 
Thus, when $\bm\beta$ is known, we can do dimension reduction first by projecting the original high-dimensional covariates into an one-dimensional sub-space spanned by $\bm\beta$, and then achieve more efficient covariate balance after the dimension reduction. 
Hereinafter, we refer to $\phi_{\bm\beta}$ as the \emph{re-randomization under the oracle} (ReO), as it is the most efficient re-randomization mechanism achievable in the asymptotic perspective when $\bm\beta$ is precisely known. 

The theorem below gives the asymptotic distribution and PRIASV of $\sqrt{N}(\hat{\tau}-\tau)$ under $\phi_{\bm\beta}$ relative to complete randomization.

\begin{theorem}\label{PRIAV-ReO}
Under the oracle $\phi_{\bm\beta}$ and Condition \ref{cond1}, 
\begin{equation}\label{eq:AsymptoticDistributionReO}
\sqrt{N}(\hat{\tau}-\tau) \mid \phi_{\bbeta}=1 \overset{\cdot}{\sim} \sqrt{V_{\tau\tau}}\big(\sqrt{1-R^2}\cdot \varepsilon_0 + \sqrt{R^2}\cdot L_{1,\xi_{\alpha,1}}\big),
\end{equation}
where $\varepsilon_0\sim \mathcal{N}(0,1)$ is independent of $L_{1,\xi_{\alpha,1}} \sim (D|D^2 \leq \xi_{\alpha,1})$ with $D \sim \mathcal{N}(0,1)$. 
Moreover, 
\begin{equation}\label{eq:PRIAV-ReO}
    \text{PRIASV}_{ReO} 
    = 
    100 \times (1-v_{\alpha,1})R^2_{\infty},
\end{equation}
where $v_{\alpha,1} = \bbE\left(\chi^2_1|\chi^2_1\leq \xi_{\alpha,1}\right)$. 
Compared with ReM, we have 
\begin{equation}\label{compare}
    \frac{\text{PRIASV}_{ReM}}{\text{PRIASV}_{ReO}} = \frac{1-v_{\alpha,p}}{1-v_{\alpha,1}} = O\left(\frac{1}{\sqrt{p}}\right).
\end{equation}
\end{theorem}


\begin{figure}
\begin{center}
\includegraphics[width=3in]{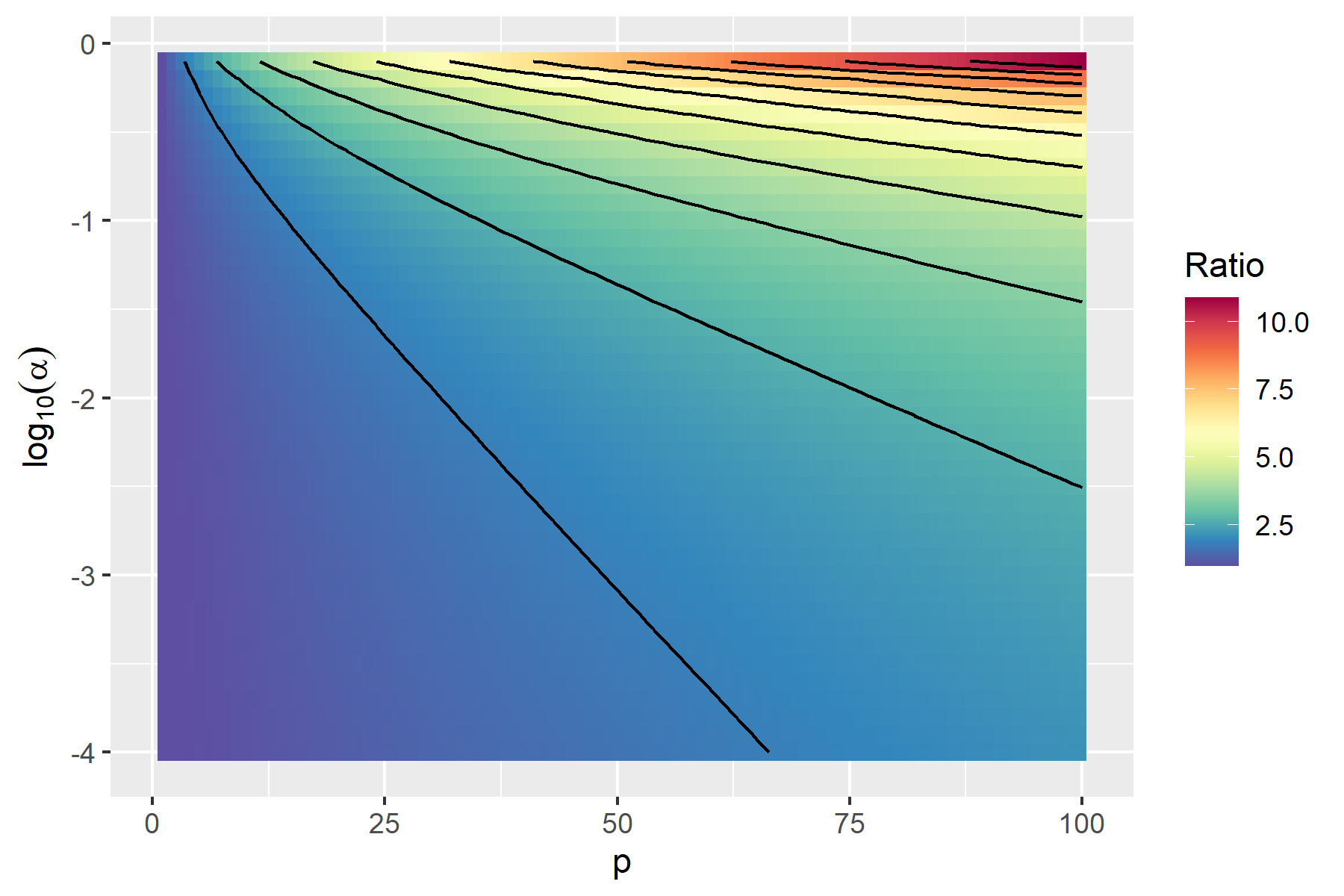}
\end{center}
\caption{Ratio of $\text{PRIASV}_{ReO}$ and $\text{PRIASV}_{ReM}$ for varying covariate dimension $p$ and logarithm acceptance probability $\alpha$.
\label{fig:PRIAV_ReO/ReM}}
\end{figure}

Figure \ref{fig:PRIAV_ReO/ReM} illustrates the superiority of ReO over ReM in terms of PRIASV by a heat map of $\text{PRIASV}_{ReO}/\text{PRIASV}_{ReM}$ as a function of covariate dimension $p$ and acceptance probability $\alpha$ based on Eq.~\eqref{compare}. 
From the figure, we observe an increasing performance advantage for ReO over ReM with the increase of covariate dimension $p$.

\subsection{Bayesian Criterion for Re-randomization}
A critical limitation of ReO, however, is that we must know $\bbeta$ for its implementation, which is often not realistic.
In this subsection, we retreat to a more realistic case where prior information about $\bbeta$, instead of $\bbeta$ itself, can be obtained, and try to establish a Bayesian criterion for re-randomization.

Suppose $\bbeta_\infty$, which is the limit of $\bbeta$, is associated with a prior distribution $\pi$.
Under the Bayesian framework, we quantify the performance of a re-randomization mechanism $\phi$ under prior $\pi$ by averaging its $\bbeta_\infty$-specific asymptotic sampling variance with respect to $\pi$. 
Formally, 
{ 
define the prior-integrated asymptotic sampling variance
\begin{equation*}\label{prior-integrated}
\begin{split}
\bbV_{\pi,a}\big(\sqrt{N}(\hat{\tau}-\tau)\mid\phi=1\big)
&= \bbE_\pi\left[\bbV_a\big(\sqrt{N}(\hat{\tau}-\tau)|\phi=1\big)\right]\\
&\vspace{1cm} = 
\bbE_{\pi}\Big[(1-R^2_{\infty})V_{\tau\tau,\infty} + \bbE\left(\bm B_{\infty}^T\bbeta_{\infty}\bbeta_{\infty}^T\bm B_{\infty}|\bm B_{\infty} \in \mathcal{B}_{\phi,\infty}\right) \Big]
\end{split}
\end{equation*}}
as the objective function for evaluating a re-randomization mechanism $\phi\in\Phi$ under the Bayesian setting.
We aim to search $\Phi_{\alpha}$ for the solution of the optimization problem below:
\begin{equation*}
\phi^*=\arg\min_{\phi\in\Phi_{\alpha}}\bbV_{\pi,a}\big(\sqrt{N}(\hat{\tau}-\tau)|\phi=1\big).
\end{equation*}
The following theorem gives the solution to this optimization problem.

\begin{theorem}\label{thm:ReB}
Suppose Condition \ref{cond1} is satisfied and $\bbeta_{\infty}$ follows prior distribution $\pi$.  Let $\bm\mu_{\pi}$ and $\bSigma_{\pi}$ be the mean vector and covariance matrix of $\pi$
respectively, and $\bLambda_{\pi}=\bm\mu_{\pi}\bm\mu_{\pi}^T + \bSigma_{\pi}$ be the corresponding characteristic matrix. Define the following prior-induced distance
\begin{equation}
    d_{\pi} = \big(\sqrt{N}\bD\big)^T \bm\Lambda_{\pi} \big(\sqrt{N}\bD\big)
\end{equation}
as the measure of covariate imbalance. The Bayesian re-randomization mechanism 
\begin{equation}
    \phi_{\pi}\big(\sqrt{N}\bm D, \bm V_{xx}\big) = I\big(d_{\pi}\leq \xi_{\alpha,\blambda}\big) 
\end{equation}
satisfies $\phi_{\pi} \in \Phi_{\alpha}$, and 
\[\begin{split}
    \bbV_{\pi,a}\big(\sqrt{N}(\hat{\tau}-\tau)\mid\phi_{\pi}=1\big) \leq \bbV_{\pi,a}\big(\sqrt{N}(\hat{\tau}-\tau)\mid\phi=1\big),\ \forall \phi \in \Phi_{\alpha},
\end{split}\]
where $\xi_{\alpha,\blambda}$ is $\alpha$-quantile of $\chi^2_{\blambda} = \sum_{j=1}^p \lambda_j Z_j^2$ with $\blambda =(\lambda_1,\ldots,\lambda_p)$ being vector of eigenvalues of matrix $\bm P = \big(\bm V_{xx}^{1/2}\big)^T\bm\Lambda_{\pi}\bm V_{xx}^{1/2}$,  $Z_1,...,Z_p$ being i.i.d. random variables from standard Gaussian distribution, and $\bm V_{xx}^{1/2}$ being the Cholesky square root of $\bm V_{xx}$ satisfying $\bm V_{xx}^{1/2}\big(\bm V_{xx}^{1/2}\big)^T=\bm V_{xx}$.
\end{theorem}

\begin{figure}
\begin{center}
\includegraphics[width=2.5in]{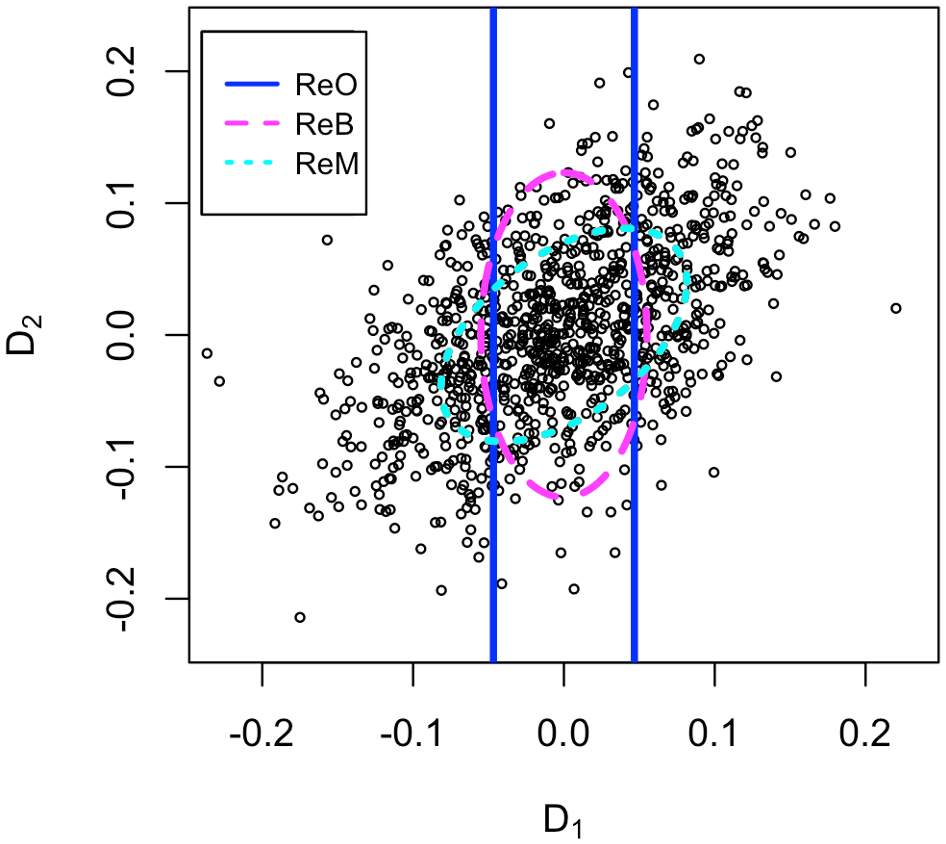}
\end{center}
\caption{Acceptance region of $\bD=(D_1,D_2)$ under different criteria for re-randomization with only 2 covariates. $\bbeta=c(1,0)^T$ in ReO and $\bm\Lambda_\pi=\bbeta\bbeta^T+\mbox{diag}\{0.25,0.25\}$ in ReB when acceptance probability $\alpha=0.5$.
\label{fig:illustrative}}
\end{figure}

The theorem tells us that $\phi_{\pi}$, the optimal re-randomization procedure in $\Phi_{\alpha}$ under the Bayesian criterion, achieves the smallest prior-integrated asymptotic sampling variance and leads to an elliptical acceptance region, whose shape is determined by $\bm\Lambda_{\pi}$, i.e., the characteristic matrix of the prior distribution $\pi$ (see Figure \ref{fig:illustrative} for the graphical illustration). 
Analogous to $d_{\bm\beta} = N\bm D^T \bm\beta\bm\beta^T\bm D$ in ReO, $d_{\pi} = N\bD^T \bm\Lambda_{\pi} \bD$ plays a similar role in achieving covariance balance by projecting the high dimensional $\bm D$ into a lower dimensional space. 
Hereinafter, we refer to $\phi_{\pi}$ as the \emph{re-randomization under the Bayesian criterion} (ReB). 

Theorem \ref{thm:ReB} shows that the Bayesian criterion for re-randomization leads to the acceptance rule $\phi_\pi=I(N\bD^T\bLambda_\pi\bD\leq\xi_{\alpha,\blambda})$, which is completely determined by $\bLambda_\pi=\bmu_\pi\bmu_\pi^T+\bSigma_\pi$, the characteristic matrix of prior distribution $\pi$. 
By specifying $\bmu_\pi$ and $\bSigma_\pi$ to different values, we find that Bayesian re-randomization takes many other re-randomization criteria as special cases.
According to Table \ref{tab:connection}, ReB degenerates to ReO, ReM, Ridge-ReM and PCA-ReM under different specifications of $\bmu_\pi$ and $\bSigma_\pi$.
Thus, ReB provides us a unified framework to understand various re-randomization criteria in the literature, and highlight their differences from the Bayesian perspectives.


\begin{table}[h]
\footnotesize
\centering
\caption{ReB leads to other re-randomization criteria under specific prior specifications}
\label{tab:connection}
    \setlength{\tabcolsep}{3.5mm}{
    \begin{tabular}{ccccc}
\midrule
$\bmu_{\pi}$ & $\bSigma_{\pi}$ & $\bLambda_{\pi}$ & $d_{\pi}$ & Criteria \\
\hline 
\multicolumn{5}{c}{}\vspace{-0.4cm}\\
$\bbeta$ & $\bm 0_{p\times p}$ & $\bbeta\bbeta^T$ & $N\bD^T\bbeta\bbeta^T\bD$ & ReO\\
$\bm 0_p$ & $\bV_{xx}^{-1}=(N\bSigma_\bD)^{-1}$ & $(N\bSigma_\bD)^{-1}$ & $\bD^T\bSigma_\bD^{-1}\bD$ & ReM\\
$\bm 0_p$ & $N^{-1}\big(\bSigma_\bD+\lambda\bI_{p}\big)^{-1}$ & $N^{-1}\big(\bSigma_\bD+\lambda\bI_{p}\big)^{-1}$ & $\bD^T\big(\bSigma_\bD+\lambda\bI_{p}\big)^{-1}\bD$ & Ridge-ReM\\
$\bm 0_p$ & $\big(N\bSigma_\bD^{(k)}\big)^{-1}$ & $\big(N\bSigma_\bD^{(k)}\big)^{-1}$ & $\bD^T\big(\bSigma_\bD^{(k)}\big)^{-1}\bD$ & PCA-ReM\\
\midrule
\end{tabular}}
\end{table}


An interesting property of ReB is that its acceptance region shows invariance to any scale transformation of prior distribution $\pi$.
Intuitively, ReB enjoys such a property because only the direction, not the magnitude, of $\bbeta$ matters in determining the relative importance of covariates, and thus the acceptance region.
The following corollary based on Thm.~\ref{thm:ReB} summarizes this phenomenon explicitly.
The detailed proof can be found in the Supplementary Material.

\begin{corollary}\label{invariance}
Suppose that $\phi_{\pi_1}$ and $\phi_{\pi_2}$ are two ReB procedures in $\Phi_{\alpha}$ with different prior specifications $\pi_1$ and $\pi_2$, respectively. 
We have 
\[\begin{split}
    \phi_{\pi_2} = \phi_{\pi_1}\ \text{as long as } \bm\Lambda_{\pi_2} = r\bm\Lambda_{\pi_1}\ \text{for some }r>0.
\end{split}\]
\end{corollary}

\subsection{Frequentist properties of ReB}
Although ReB is established from the Bayesian perspective, we can study its performance with comparison to other re-randomization mechanisms, e.g., ReM, from the frequentist perspective. 
According to \eqref{eq:AsymptoticVariance4Phi}, 
the relative efficiency of $\phi_\pi$ with respect to complete randomization can be quantified in term of PRIASV of $\sqrt{N}(\hat\tau-\tau)$:
\begin{equation}\label{PRIAV_ReB}\text{PRIASV}_{ReB}=
100\times(1-r_{\phi_{\pi}})R^2_{\infty},
\end{equation}
where 
\begin{equation}\label{R2_pi}
r_{\phi_{\pi}}=v_{\alpha,\pi} \triangleq \frac{\bbE\big(\bm B_{\infty}^T\bbeta_{\infty}\bbeta_{\infty}^T\bm B_{\infty}\mid\bm B_{\infty} \in \mathcal{B}_{\phi_{\pi},\infty} \big)}{\bbE\big(\bm B^T_{\infty}\bbeta_{\infty}\bbeta_{\infty}^T\bm B_{\infty}\big)}.
\end{equation}

When $\bLambda_{\pi}$ is
of full rank, the following theorem gives the asymptotic distribution as well as the explicit form of $v_{\alpha,\pi}$ for $\sqrt{N}(\hat{\tau}-\tau)$ under ReB.
\begin{theorem}\label{thm:AsymptoticDistributionReBfullrank}
{Under the Bayesian setting with $\pi$ as the prior distribution of $\bbeta_\infty$ and Condition \ref{cond1},
when 
$\bLambda_{\pi}$
is symmetric and positive definite with Cholesky decomposition $\bLambda_{\pi} = \bM\bM^T$, then  
$\phi_{\pi}$ leads to 
\begin{equation}\label{eq:AsymptoticDistributionReB}
\sqrt{N}(\hat{\tau}-\tau) \mid \phi_{\pi}=1\overset{\cdot}{\sim} \sqrt{V_{\tau\tau}(1-R^2)}\cdot \varepsilon_0 + \bbeta^T(\bM^{-1})^T \bm Q \mid \bm Q^T\bm Q \leq \xi_{\alpha,\blambda},
\end{equation}
where $\bm Q \sim \mathcal{N}(\bm 0,\bM^T\bV_{xx}\bM)$, and $\bbeta^T(\bM^{-1})^T \bm Q \sim \mathcal{N}(0,\bV_{\tau x}\bV_{xx}\bV_{x\tau})$. Moreover, 
\begin{equation}\label{eq:v_alpha_pi}
    v_{\alpha,\pi} = \frac{\bbeta^T_{\infty} \big(\bm M^{-1}\big)^T\bm\Gamma^T_{\infty}
\bm C_{\infty}\bm\Gamma_{\infty} \bm M^{-1}\bbeta_{\infty}}{\bV_{\tau x,\infty}\bV_{xx,\infty}\bV_{x\tau,\infty}}=\frac{\bbeta^T_{\infty} \big(\bm M^{-1}\big)^T\bm\Gamma^T_{\infty}
\bm C_{\infty}\bm\Gamma_{\infty} \bm M^{-1}\bbeta_{\infty}}{\bbeta^T_{\infty} \big(\bm M^{-1}\big)^T\bm\Gamma^T_{\infty}
\bm \Lambda_{\infty}\bm\Gamma_{\infty} \bm M^{-1}\bbeta_{\infty}},
\end{equation}
where $\bm \Lambda = diag\{\lambda_{1},\ldots,\lambda_{p}\}$, 
$\bm C = diag\{c_{1},\ldots,c_{p}\}$ with 
$c_{i} = \bbE\left(\lambda_{i}Z_i^2\mid \sum_{j=1}^p \lambda_{j} Z_j^2 \leq \xi_{\alpha,\blambda}\right)$ for $i=1,\ldots,p$, $\blambda= (\lambda_{1},\ldots,\lambda_{p})$, and 
$Z_1,...,Z_p$ being i.i.d. standard normal distributed random variables.
$\bm\Gamma$ is an orthogonal matrix such that $\bm M^T \bm V_{xx} \bm M =\bm\Gamma^T \bm \Lambda \bm\Gamma$. $\bGamma_{\infty}, \bm C_{\infty}$ and $\bLambda_{\infty}$ are limits of $\bGamma, \bm C$ and $\bLambda$ in the sequence of finite populations.}

 
\end{theorem}

Theorem \ref{thm:AsymptoticDistributionReBfullrank} gives the explicit expression of $v_{\alpha,\pi}$ when ${\bLambda}_\pi$ is of full rank, which is easy to hold in practice. The value of $v_{\alpha,\pi}$ depends on the eigenvalues $\blambda$ of $\big(\bV_{xx}^{1/2}\big)^T{\bLambda}_\pi\bV_{xx}^{1/2}$ via $\bbE\big(Z_i^2\mid \sum_{j=1}^p \lambda_{j} Z_j^2 \leq \xi_{\alpha,\blambda}\big), i=1,\ldots,p.$ This expectation will degenerate to $v_{\alpha,p} = \bbE\big(Z_i^2\mid \sum_{j=1}^p Z_j^2 \leq \xi_{\alpha,p}\big)$ in ReM, where ${\bLambda}_\pi=\bm V_{xx}^{-1}$ and thus $\lambda_{1}=\cdots=\lambda_{p}=1$. 


To build a geometry intuition on the effect of the precision of the prior on the performance of ReB, we further consider the case where the prior distribution $\pi$ degenerates to a point mass. 
{ The theorem below shows a simpler form of the asymptotic distribution and PRIASV of $\sqrt{N}(\hat{\tau}-\tau)$ in this scenario.

\begin{theorem}\label{thm:ReB2sigma0}
Under the Bayesian setting where the prior distribution $\pi$ degenerates to a point mass at $\bmu_{\pi}$, which means that $\bm\Sigma_{\pi}=\bm 0$ and ${\bLambda}_{\pi}=\bmu_{\pi}\bmu_{\pi}^T$, we have
\begin{equation}\label{eq:AsymptoticDistributionReB2}
\sqrt{N}(\hat{\tau}-\tau) \mid \phi_{\pi}=1 \overset{\cdot}{\sim}
\sqrt{V_{\tau\tau}}\left(\sqrt{1-R^2_{\pi}}\cdot \varepsilon_0 + \sqrt{R^2_{\pi}}\cdot L_{1,\xi_{\alpha,1}}\right),
\end{equation}
where 
\begin{equation}\label{eq:R2pi}
R^2_{\pi} = \frac{(\bmu_{\pi}^T\bV_{xx}\bbeta)^2}{(\bmu_{\pi}^T\bV_{xx}\bmu_{\pi})(\bbeta^T\bV_{xx}\bbeta)}\cdot R^2 = (\cos\theta)^2\cdot R^2
\end{equation}
with $\theta$ being the angle between $\big(\bV_{xx}^{1/2}\big)^T\bmu_{\pi}$ and $\big(\bV_{xx}^{1/2}\big)^T\bbeta$. Denote the limit of $\theta$ as $\theta_{\infty}$, then
\begin{equation}\label{eq:vapi4sigma0}
    v_{\alpha,\pi}  = 1-(1-v_{\alpha,1})(\cos\theta_{\infty})^2.
\end{equation}
\end{theorem}}

Intuitively, when ${\bLambda}_{\pi}=\bmu_{\pi}\bmu_{\pi}^T$, $\bmu_{\pi}$ becomes our prior guess for $\bbeta$.
Thus, Thm.~\ref{thm:ReB2sigma0} tells us that the asymptotic performance of ReB only depends on the limiting angle between $\bmu_{\pi}$ and $\bbeta$ after an affine transformation. 
Apparently, a smaller $\theta_{\infty}$ would lead to a re-randomization procedure that behaves more like ReO, and thus enjoys a better chance to outperform ReM.
The corollary below gives the sufficient and necessary condition for ReB (under the degenerated prior) to outperform ReM in terms of the admissible value range of $\theta_{\infty}$.

\begin{corollary}\label{cor:outperformReM}
Under the same setting as in Thm.~\ref{thm:ReB2sigma0} with $\theta_{\infty}$ standing for the limit of the angle between $\big(\bV_{xx}^{1/2}\big)^T\bmu_{\pi}$ and $\big(\bV_{xx}^{1/2}\big)^T\bbeta$. 
ReB $\phi_\pi$ outperforms ReM $\phi_M$ if and only if  \begin{equation*}
(\cos\theta_{\infty})^2 > \frac{1-v_{\alpha,p}}{1-v_{\alpha,1}}.
\end{equation*}
\end{corollary}

\begin{figure}
\begin{center}
\includegraphics[width=3in]{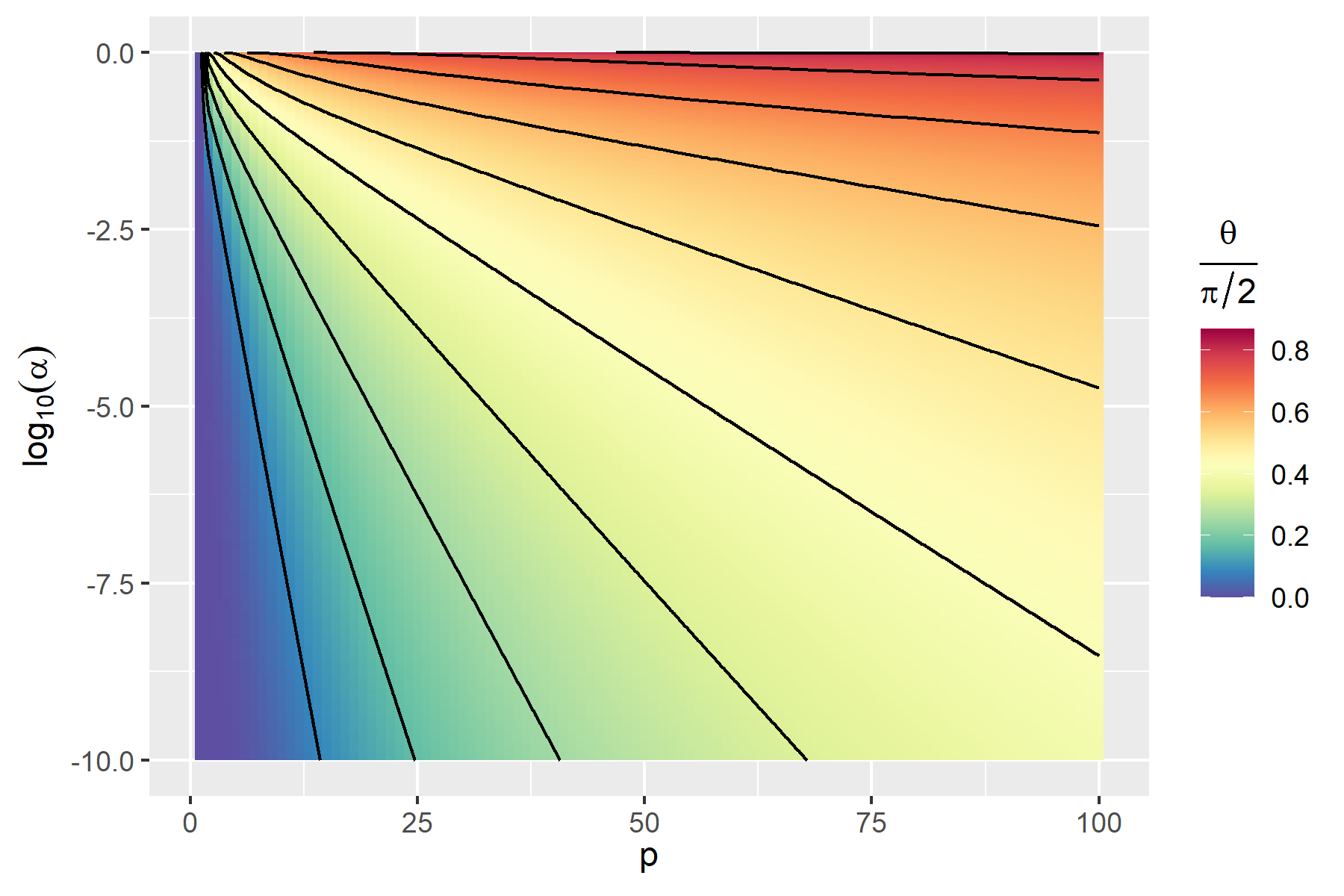}
\end{center}
\caption{Largest limiting angle between prior vector $\bmu_{\pi}$ and $\bbeta$ after transformation according to $\bV_{xx}^{1/2}$ that guarantees out-performance of ReB over ReM for varying covariate dimension $p$ and logarithm acceptance probability $\alpha$.
\label{fig:theta}}
\end{figure}

Corollary \ref{cor:outperformReM} implies that once the probability mass of the prior distribution $\pi$ concentrates in a cone centered in the specific direction along $\bbeta$ (after the transformation according to $\bV_{xx}^{1/2}$), ReB would benefit from the prior information and outperform ReM.
Figure \ref{fig:theta} illustrates how the ratio between the opening angle of the cone (when restricted on $[0,\pi/2]$) and $\pi/2$ changes with $\alpha$ and $p$ when $\bV_{xx}$ is an identity matrix.
From the figure, we can see that as the dimensionality $p$ gets larger, the admissible limiting value range of $\theta$ enlarges to a wider region spanning from 0 to nearly $\pi/2$, indicating that the superiority of ReB over ReM in practice becomes less sensitive to the mis-specification of $\bbeta$ as $p$ gets larger.
Such a result gives us the confidence that ReB would be a useful re-randomization procedure in practice as long as our prior knowledge about $\bbeta$ is not that misleading.

\section{Implementing ReB via a two-stage experiment}\label{sec:Two-stage}

The theoretical analysis in the previous section shows that ReB is a more efficient re-randomization procedure than ReM when the prior distribution $\pi$ is informative to highlight $\bbeta$.
In case that such an informative prior is not available in advance, we still can implement ReB if a two-stage experiment is feasible, where the first stage involves a small pilot randomized experiment  with $N_1$ randomly selected units to learn an informative prior for the unknown $\bbeta$, while the second stage implements a larger-scale re-randomized experiment with other $N_2$ randomly selected units via ReB guided by the learned prior in pursuit of a more efficient causal estimator. 

In the pilot experiment in the first stage, we implement a balanced completely randomized experiment to get a preliminary causal effect estimator 
$\hat{\tau}_1 = \bar{Y}_T^{(1)}-\bar{Y}_C^{(1)}$ satisfying 
\begin{equation*}
\bbE\left(\hat{\tau}_1\right) = \tau_1,\ 
    \bbV_a\big(\sqrt{N_1}(\hat{\tau}_1-\tau_1)\big) = \bbV_a\big(\sqrt{N}(\hat{\tau}-\tau)\big),
\end{equation*}
where $\bar{Y}_T^{(1)}$ and $\bar{Y}_C^{(1)}$ are average responses in the two arms of the first-stage experiment, $\tau_1$ and $\tau$ are average treatment effects in the first stage and the complete study, respectively,
and $\hat{\tau}$ is the difference-in-means estimator of $\tau$ when we conduct a complete randomization on the total $N=N_1+N_2$ units.

{ According to Eq's.~\eqref{eq:V} and \eqref{eq:beta_infty}, 
$$\bbeta = r_0 \cdot \bS_{\bX}^{-2}\bS^2_{\bX,Y(1)} + r_1 \cdot \bS_{\bX}^{-2}\bS^2_{\bX,Y(0)}=r_0\cdot\bbeta_1 + r_1\cdot\bbeta_0,$$
where $\bbeta_1$ and $\bbeta_0$ are projection coefficient vector of $\bY(1)$ and $\bY(0)$ on $\bX$, respectively. We can estimate $\bbeta_1$ and $\bbeta_0$ with observed outcomes and covariates in the first stage experiment, denoted as $\hat{\bbeta}_1$ and $\hat{\bbeta}_0$. Assisted with linear model, we obtain a prior variance of $\hat{\bbeta}_1$ and $\hat{\bbeta}_0$, denoted as $\bV_1$ and $\bV_0$. Then an informative prior distribution of $\bbeta$ is 
\begin{equation}\label{eq:prior}
\pi_{N_1} = \mathcal{N}\big(r_0\cdot\hat{\bbeta}_1 + r_1\cdot\hat{\bbeta}_0, r_0^2\cdot\bV_1 + r_1^2\cdot\bV_0\big).
\end{equation}}
In the second stage, we implement ReB on the other $N_2$ units with the $\pi_{N_1}$ derived in the first stage as the prior distribution of $\bbeta$, leading to a Bayesian re-randomization procedure referred to as $\phi_{\pi_{N_1}}$ 
with the balance criterion below: 
\begin{equation*}
\phi_{\pi_{N_1}}\big(\sqrt{N_2}\bD^{(2)},\bV_{xx}^{(2)}\big) = I\big(N_2\cdot (\bD^{(2)})^T\bLambda_{\pi_{N_1}}\bD^{(2)}\leq  \xi_{\alpha,\blambda_{\pi_{N_1}}}\big),
\end{equation*}
where $\bm D^{(2)}$ is the difference-in-means of covariates in the second-stage experiment, $\bm V_{xx}^{(2)}$ is the sampling covariance of $\sqrt{ N_2}\cdot\bm X^{(2)}$, $\bm X^{(2)}$ is the covariate matrix in the second stage
and
$\blambda_{\pi_{N_1}}$ is the vector of eigenvalues of matrix $\bm P_{N_1,{N_2}} = \Big(\big(\bV_{xx}^{(2)}\big)^{1/2}\Big)^T \bLambda_{\pi_{N_1}} \big(\bV_{xx}^{(2)}\big)^{1/2}$. 
This procedure leads to another causal effect estimator 
$\hat{\tau}_2 = \bar{Y}_T^{(2)}-\bar{Y}_C^{(2)}$ satisfying 
\begin{equation*}
    \bbE\left(\hat{\tau}_2\right) = \tau_2,\ 
    \bbV_a\big(\sqrt{N_2}(\hat{\tau}_2-\tau_2)\big) = 
    \left[1-(1-v_{\alpha,\pi_{N_1}})R^2_{\infty}\right]\bbV_a\big(\sqrt{N}(\hat{\tau}-\tau)\big),
\end{equation*}
where $\bar{Y}_T^{(2)}$ and $\bar{Y}_C^{(2)}$ are average responses in the two arms of the second-stage experiment, and $\tau_2$ is the average treatment effects in the second stage.

{ Denote the proportion of sample size in first stage as $\varrho$.
Since $\tau = \varrho\cdot\tau_1 + (1-\varrho)\cdot\tau_2$, 
\begin{equation*}
\hat{\tau} = \varrho\cdot\hat{\tau}_1 + (1-\varrho)\cdot\hat{\tau}_2
\end{equation*}
is an unbiased estimator of $\tau$.} Denote the above 
two-stage re-randomization procedure as $\phi_{BCRD\text{-}ReB}$. Similarly, BCRD in the first stage can be replaced by ReM, resulting in an alternative two-stage
procedure referred to as $\phi_{ReM\text{-}ReB}$.
The following theoren gives the PRIASV of $\hat{\tau}$ under BCRD-ReB and ReM-ReB.

{ \begin{theorem}\label{2stageReO-variance}
Suppose Condition \ref{cond1} is satisfied, acceptance rate in the second stage of BCRD-ReB is $\alpha_1$ and the acceptance rates in the two stages of ReM-ReB are $\alpha_2$ and $\alpha_3$, respectively.
When $N_1$ and $N_2$ both go to infinity, $\varrho = N_1/N$ converges to a constant $\varrho^*$, 
the PRIASV of $\hat{\tau}$ under BCRD-ReB is 
\begin{equation}\label{eq:asymptotic-variance2}
\text{PRIASV}_{BCRD\text{-}ReB}=
100\times (1-\varrho^*)(1-v_{\alpha_1,1})R^2_{\infty}.
\end{equation}
Similarly, the PRIASV of  $\Tilde{\tau}_{N_1,N_2}$ under ReM-ReB is 
\begin{equation}\label{eq:asymptotic-variance3}
\text{PRIASV}_{ReM\text{-}ReB}=
100\times \left[\varrho^*(1-v_{\alpha_2,p}) + (1-\varrho^*)(1-v_{\alpha_3,1})\right]R^2_{\infty}.
\end{equation}
\end{theorem}}

It is easy to see that both two-stage ReB procedures are more efficient than complete randomization. ReM-ReB is more efficient than BCRD-ReB when the acceptance rates in the second stage of both methods are the same, i.e., $\alpha_1=\alpha_3$. 
The following corollary is a simple application of the above analysis and it shows that if the proportion of sample size in first stage is small, BCRD-ReB and ReM-ReB with $\alpha_1=\alpha_3=\alpha$ are asymptotically equivalent to ReO with acceptance rate $\alpha$.

\begin{corollary}\label{N1infinity}
Under the same settings in Theorem \ref{2stageReO-variance}, if the acceptance rates in ReO, namely $\alpha$, is the same as acceptance rate in the second stage of BCRD-ReB and ReM-ReB, i.e., $\alpha=\alpha_1=\alpha_3$, then
\begin{equation*}
    \text{PRIASV}_{BCRD\text{-}ReB}=\text{PRIASV}_{ReM\text{-}ReB}=\text{PRIASV}_{ReO},
\end{equation*}
as $\varrho = N_1/N \rightarrow \varrho^* =  0$.
\end{corollary}


{ Additionally, Eq.~\eqref{eq:prior} allows defining a prior for $\bbeta$ in the two-stage procedure using a linear model. A simpler approach is to degenerate $\pi_{N_1}$ into a point mass on $r_0\cdot\hat{\bbeta}_1 + r_1\cdot\hat{\bbeta}_0$, using only the first moment. We refer to the procedure employing this degenerated prior as BCRD-ReO and ReM-ReO, as their second-stage criterion closely resembles that of ReO.}

\section{Simulation studies}\label{sec:Simulation}

In this section, we validate the theoretical analyses in previous sections and compare various re-randomization procedures, including ReM, ReO, ReB and two-stage ReB, via simulation.
We generate simulated datasets with a linear model for potential outcomes:
{ \[\begin{split}
Y_i(1) &= a + \bbeta_1^T \bX_i + \epsilon_{1,i},\ \ i=1,\ldots,N,\\
Y_i(0) &= \bbeta_0^T \bX_i + \epsilon_{0,i},\ \ i=1,\ldots,N.
\end{split}\]
Here, $a = 5$, covariates $\bX \sim \mathcal{N}(\bm 0_p,(1-\rho)\bI_p + \rho \bm 1_p \bm 1_p^T)$, 
$\epsilon_1,\epsilon_0 \sim \mathcal{N}(0,\sigma^2_{\bepsilon})$, and 
regression coefficients $\bbeta_1 \sim \mathcal{N}( 2\cdot \bm 1_p,2\sigma^2_{\bbeta}\cdot \bm I_p)$, $\bbeta_0 \sim \mathcal{N}(\bm 1_p,2\sigma^2_{\bbeta}\cdot \bm I_p)$. We assume balanced treatment assignment with $r_1=r_0=0.5$, resulting in the projection coefficient of treatment effect on difference-in-means of covariates $\bbeta = r_1 \cdot \bbeta_0 + r_0 \cdot \bbeta_1 \sim \mathcal{N}(1.5 \cdot \bm 1_p,\sigma^2_{\bbeta}\cdot \bm I_p)$. Denote the distribution of $\bbeta$ as $\pi(\bbeta)$.} We specify $p\in \{2,5,10,20\}$, $\rho\in \{0,0.2,0.5\}$, $\sigma^2_\bbeta\in\{0,0.1,1\}$. For each combination of $(p,\rho,\sigma^2_\bbeta)$, we porperly set $\sigma^2_\bepsilon$ to tune the average value of $R^2$ in the datasets referred to as $\bar{R}^2\in\{0.2, 0.5, 0.8\}$, and choose sample size $N\in\{100,200,600,1200\}$, leading to a configuration space $\cS$ containing $4\times 3\times 3\times 3\times 4=432$ combinations of the 5 controlling factors $(p,\rho,\sigma^2_\bbeta,\bar{R}^2,N)$.
Table \ref{tab:Simu_Factors} summarizes these factors, including a scheme factor for different re-randomization procedures. 
Each configuration in $\cS$ generates 100 independent datasets, totaling 43,200 simulated datasets.

\begin{table}
\footnotesize
    \centering
    \caption{Six major factors in the simulation study.}
    \setlength{\tabcolsep}{5.5mm}{
    \begin{tabular}{cll}
    \midrule
    Factor & Levels & Description  \\
    \hline
    $p$ & \{2, 5, 10, 20\} & Dimensionality of $\bX$\\
    $\rho$ & \{0, 0.2, 0.5\} & Correlation coefficient of $\bX$\\
    $\sigma^2_\bbeta$ & $\{0, 0.1, 1\}$ & Variance of $\pi(\bbeta)$\\
    $\bar{R}^2$ & \{0.2, 0.5, 0.8\} & Noise level of regression model\\
    $N$ & \{100, 200, 600, 1200\}  & Total sample size\\
    $\mbox{Scheme}$ & \{ReM, ReO, ReB\} & Re-randomization procedures\\
    \midrule
    \end{tabular}
    }
    \label{tab:Simu_Factors}
\end{table}


\subsection{ReO and ReB versus ReM}\label{sec:onestage}
First, we study the performance of ReO and ReB versus ReM in reducing the estimation variance of $\hat{\tau}$ compared to Balanced Complete Randomized Design (BCRD). These schemes are applied to our simulated datasets for quantitative evaluation. We maintain a fixed acceptance probability of $\alpha=0.05$, while assuming known covariate dimensionality $p$, sample size $N$, and correlation $\rho$.
In ReO, $\bbeta$ is considered known, while in ReB, the investigator has information about the sampling distribution $\pi(\bbeta)$ but not the specific $\bbeta$ for each dataset, and ReM solely relies on simulated data without any prior information about $\bbeta$ or its distribution.

For the $j$-th simulated dataset $\cD_{ij}$ under the $i$-th simulation configuration in $\cS$, the estimation variance $\bbV(\hat\tau\mid\phi=1)$ of a re-randomization procedure $\phi$ is estimated based on 1000 accepted allocations from $\phi$ (referred to as $\widehat\bbV_{ij}(\hat\tau\mid\phi=1)$).
Based on the estimated $\widehat\bbV_{ij}(\hat\tau\mid\phi=1)$, PRIV achieved by $\phi$ with respect to BCRD can be obtained for $\cD_{ij}$ (denoted as PRIV$_{ij}(\phi)$), and the average performance of a re-randomization procedure $\phi$ in simulation setting $i\in\cS$ can be summarized by
\begin{equation}\label{eq:PRIV_i_phi}
    \mbox{PRIV}_{i}(\phi)=\frac{1}{100}\sum_{j=1}^{100} \mbox{PRIV}_{ij}(\phi),
\end{equation}
which is equivalent to averaging the PRIV by prior of $\bbeta$. 

To assess the impact of the six factors in Table \ref{tab:Simu_Factors} on PRIV of $\hat\tau$, we conducted an ANOVA for the estimated $\{$PRIV$_{ij}(\phi)\}$ considering these factors and their interactions.
Table \ref{tab:Sim_ANOVA} presents the ANOVA outcomes with the 5 most influential effects in explaining the variance of PRIV$_{ij}(\phi)$.
The term ``residual'' encompasses all remaining effects beyond these top 5 influential factors.
Notably, the top 5 factors account for nearly all PRIV$_{ij}(\phi)$ variance, with the top 2 factors explaining over 95\% of the total variance.
Specifically, $\bar{R}^2$ has the most significant impact on PRIV variance, followed by scheme, covariate dimension $p$, and their interactions, highlighting the re-randomization procedure's critical role.
Moreover, sample size $N$ does not significantly affect PRIV.
These findings are consistent with the theoretical results presented in Section \ref{sec:ReB}.


\begin{table}[h]
    \centering
    \footnotesize
    \caption{The 5 most influential effects for the ANOVA simulation results.}
    \setlength{\tabcolsep}{6.5 mm}{
    \begin{tabular}{lcccc}
    \midrule
    Sources & DF & MS & CP & $F$-value\\
    \hline
    $\bar{R}^2$ & 2 & 3415.91 & 90.96 & 506659.51\\
$\mbox{Scheme}$ & 2 & 206.21 & 96.45 & 30585.59 \\
$p$ & 3 & 59.88 & 98.04 & 8881.31 \\
$\mbox{Scheme} \times p$ & 6 & 32.32 & 98.91 & 4793.10\\
$\mbox{Scheme} \times \bar{R}^2$ & 4 & 21.20 & 99.47 & 3143.95\\
\hline
Residual & 129582 & 19.91 & 100 & \\
    \midrule
    \multicolumn{5}{l}{\footnotesize \makecell[l]{DF: ``degrees of freedom'', MS: ``mean square'', CP: ``cumulative percentage of MS''}}
    \end{tabular}
    }
    \label{tab:Sim_ANOVA}
\end{table}

\begin{table}[ht]
    \centering
    \footnotesize
    \caption{Average PRIV achieved by different re-randomization procedures when $N=200$ and $\sigma^2_\bbeta=1$.}
    \begin{tabular}{cccccccccc}
    \midrule
    & \multicolumn{3}{c}{$\bar{R}^2=0.2$}&\multicolumn{3}{c}{$\bar{R}^2=0.5$}&\multicolumn{3}{c}{$\bar{R}^2=0.8$}\\
    \cmidrule(lr){2-4}
    \cmidrule(lr){5-7}
    \cmidrule(lr){8-10}
    Scheme & $\rho=0$ & 0.2 & 0.5 & 0 & 0.2 & 0.5 & 0 & 0.2 & 0.5 \\\hline
    \multicolumn{10}{c}{}\vspace{-0.2cm}\\
    & \multicolumn{9}{c}{$p=2$}\\\cmidrule(lr){2-10}
    ReO & 19.36 & 19.61 & 20.15 & 49.50 & 49.71 & 49.75 & 79.74 & 79.75 & 79.78 \\ 
    ReB  & 19.27 & 19.32 & 19.26 & 48.65 & 49.04 & 48.53 & 78.40 & 78.71 & 78.23 \\ 
    ReM & 19.03 & 18.98 & 19.05 & 48.07 & 48.24 & 48.03 & 77.57 & 77.60 & 77.56 \\ 
    \cmidrule(lr){2-10}
    \multicolumn{10}{c}{}\vspace{-0.3cm}\\
    & \multicolumn{9}{c}{$p=5$}\\\cmidrule(lr){2-10}
    ReO & 20.12 & 20.45 & 19.90 & 49.71 & 49.69 & 49.91 & 79.69 & 79.83 & 79.80 \\ 
    ReB & 17.18 & 17.95 & 16.44 & 46.07 & 47.29 & 44.40 & 74.56 & 76.85 & 72.17 \\ 
    ReM  & 15.53 & 15.53 & 15.50 & 41.18 & 40.94 & 41.16 & 66.75 & 66.81 & 66.86 \\ 
    \cmidrule(lr){2-10}
    \multicolumn{10}{c}{}\vspace{-0.3cm}\\
    & \multicolumn{9}{c}{$p=10$}\\\cmidrule(lr){2-10}
    ReO & 19.54 & 19.85 & 19.04 & 49.14 & 49.11 & 49.15 & 79.59 & 79.60 & 79.66 \\ 
    ReB  & 14.29 & 14.92 & 12.89 & 43.74 & 45.71 & 39.86 & 73.34 & 76.45 & 67.18 \\ 
    ReM  & 10.21 & 10.21 & 10.28 & 31.98 & 32.03 & 32.03 & 54.03 & 53.99 & 54.17 \\ 
    \cmidrule(lr){2-10}
    \multicolumn{10}{c}{}\vspace{-0.3cm}\\
    & \multicolumn{9}{c}{$p=20$}\\\cmidrule(lr){2-10}
    ReO  & 19.35 & 19.88 & 20.36 & 50.30 & 49.70 & 49.99 & 79.91 & 80.04 & 79.84 \\ 
    ReB  & 10.84 & 11.23 & 8.80 & 42.36 & 43.61 & 35.46 & 73.43 & 76.16 & 62.13 \\ 
    ReM  & 6.66 & 6.68 & 6.66 & 24.61 & 24.45 & 24.46 & 42.02 & 42.02 & 41.95 \\ 
    \midrule
    \end{tabular}
    \label{tab:SimReB}
\end{table}

Table \ref{tab:SimReB} provides a detailed comparison of the average PRIV achieved by the three competing re-randomization procedures under various simulation complications. 
Given the insignificance of sample size $N$ in the ANOVA and the greater challenge posed by larger $\sigma^2_\bbeta$ for Bayesian re-randomization procedures, we focus on reporting results for settings where $N=200$ and $\sigma^2_\bbeta=1$.
Notable findings include ReO consistently outperforming others and ReB surpassing classic ReM, with its advantage increasing as $p$ grows.



\begin{figure}[t]
\begin{center}
\includegraphics[width=3.5in]{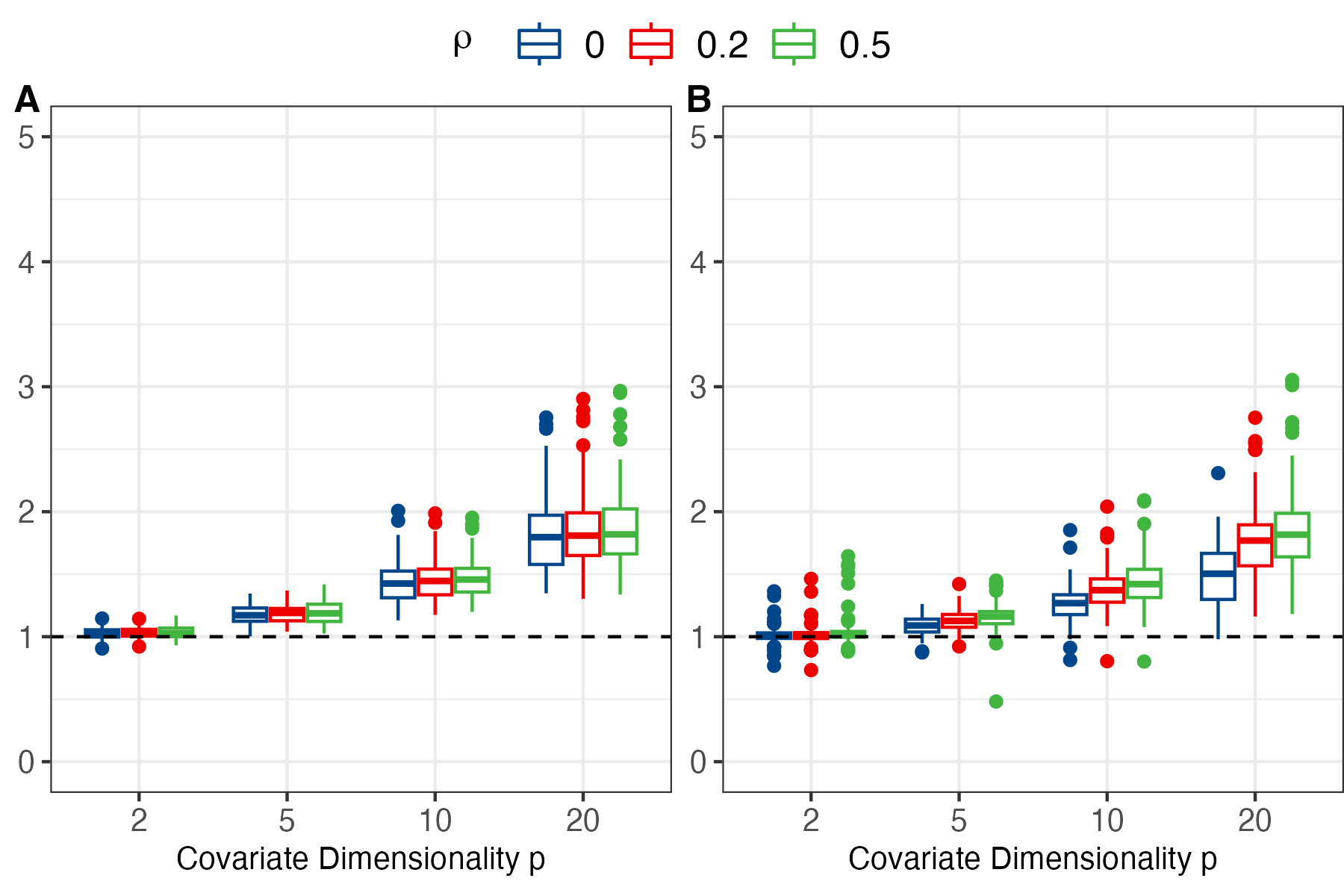}
\end{center}
\caption{The PRIV ratio of ReB over ReM when sample size $N$ is fixed at 600 and $\bar{R}^2$ fixed at 0.5 for different $p$ and $\rho$ with (A) $\sigma^2_{\bbeta} = 0.1$ and (B) $\sigma^2_{\bbeta} = 1$.
\label{fig:Sim_Boxplot}}
\end{figure}


In Figure \ref{fig:Sim_Boxplot}, we examine how ReB's advantage over ReM varies with different factors using the PRIV ratio of ReB to ReM. We maintain a fixed sample size of 600 and a constant $\bar{R}^2$ of 0.5 while varying $p$ and $\rho$. Two scenarios are considered for $\sigma^2_{\bbeta}$: 0.1 and 1, and each box-plot summarizes results from 100 repeated experiments.
The figure shows that the majority of PRIV ratios are above 1 in all scenarios, even with larger $\sigma^2_{\bbeta}$, and exceptions become less common as $\rho$ and $p$ increase.

{Figure B.1} in the Supplement shows average PRIV for ReO and ReM across various covariate dimensions ($p$) with different $\bar{R}^2$ values at $\rho=0$, $\sigma^2_\bbeta=0.1$ and $N=600$.
Both methods show substantial PRIV improvements, and ReM's average PRIV decreases with higher $p$, while ReO remains stable. These findings align with Theorem \ref{PRIAV-ReO}'s predictions. Moreover, additional studies on acceptance probability precision are detailed in { Section B.2} of the Supplementary Material. As sample size $N$ grows, acceptance probabilities for the three procedures converge to theoretical values, with ReO and ReB outperforming ReM.

\subsection{Two-stage ReB versus ReM and ReO}
In this subsection, we evaluate the performance of two-stage ReB procedures, including BCRD-ReB, ReM-ReB, BCRD-ReO, and ReM-ReO, in comparison to ReO and ReM. Our analysis focuses on simulated datasets with $\rho=0$, $\sigma^2_\bbeta=0$, $\bar{R}^2 = 0.5$, a first-stage sample proportion $r \in {0.2,0.3,0.4}$, and a minimum sample size of $N\geq200$, excluding cases with $N=100$ due to potential two-stage ReB method limitations in such scenarios.
For a fair comparison, ReM's acceptance probability is set at $\alpha=0.05$, and other re-randomization methods' acceptance probabilities are adjusted numerically to ensure comparable computational costs, i.e., the running time of sampling 1000 assignments. Each simulated dataset provides PRIV estimates based on 1000 accepted allocations, and we calculate the average PRIV for each simulation configuration. Detailed acceptance rates and running times are documented in { Section B.3} of the Supplementary Material.

Table \ref{tab:2stageReB} shows average PRIV for different methods in various simulation scenarios, highlighting the best RPIV among ReM and two-stage procedures in bold. 
Our observations are as follows:
In most cases, ReM-ReB outperforms other methods, except when the first-stage sample size is insufficient relative to the covariate dimension ($p$) or when $p$ is small. Two-stage methods exhibit a more pronounced advantage over ReM when dealing with high-dimensional covariates. However, in scenarios with small sample sizes, BCRD-ReO and ReM-ReO perform notably worse than BCRD-ReB and ReM-ReB. This suggests that introducing variance through ReB improves causal effect estimation, helping mitigate point estimator bias when sample sizes are limited.

\begin{table}[ht]
    \centering
    \footnotesize
    \caption{Average PRIV achieved by different re-randomization procedures when $\bar{R}^2=0.5$.}
    \label{tab:2stageReB}
    \setlength{\tabcolsep}{2mm}{
\begin{tabular}{cccccccccc}
    \midrule
    & \multicolumn{3}{c}{$N=200$}&\multicolumn{3}{c}{$N=600$}&\multicolumn{3}{c}{$N=1200$}\\
    \cmidrule(lr){2-4}
    \cmidrule(lr){5-7}
    \cmidrule(lr){8-10}
    Scheme & $r=0.2$ & 0.3 & 0.4 & 0.2 & 0.3 & 0.4 & 0.2 & 0.3 & 0.4 \\\hline
    \multicolumn{10}{c}{}\vspace{-0.2cm}\\
    & \multicolumn{9}{c}{$p=2$}\\\cmidrule(lr){2-10}
\cellcolor[HTML]{D9D9D9} ReM & \cellcolor[HTML]{D9D9D9}\bf 48.08 &\cellcolor[HTML]{D9D9D9}  {48.08}& \cellcolor[HTML]{D9D9D9} \bf  48.08 & \cellcolor[HTML]{D9D9D9} 48.90 & \cellcolor[HTML]{D9D9D9} {48.90}
& \cellcolor[HTML]{D9D9D9} 48.90 & \cellcolor[HTML]{D9D9D9} 48.55 & \cellcolor[HTML]{D9D9D9} 48.55 & \cellcolor[HTML]{D9D9D9} 48.55 \\ 
 ReM-ReB & 48.01 & \bf 48.15 & 47.87 & \bf 49.21 & \bf 49.43 & \bf 49.33 &\bf  49.35 & \bf 49.17 & \bf 48.94 \\ 
  ReM-ReO & 47.08 & 46.94 & 47.59 & 49.03 & 48.82 & 48.95 & 49.11 & 49.07 & 49.24 \\ 
  BCRD-ReB & 38.94 & 34.39 & 29.57 & 40.04 & 34.20 & 30.98 & 39.63 & 34.84 & 29.64 \\ 
  BCRD-ReO & 36.95 & 32.78 & 28.52 & 39.80 & 34.20 & 30.79 & 39.55 & 34.40 & 29.72 \\
    \cmidrule(lr){2-10}
    \multicolumn{10}{c}{}\vspace{-0.2cm}\\
    & \multicolumn{9}{c}{$p=5$}\\\cmidrule(lr){2-10}
     \cellcolor[HTML]{D9D9D9} ReM &\cellcolor[HTML]{D9D9D9} 41.27 & \cellcolor[HTML]{D9D9D9} 41.27 & \cellcolor[HTML]{D9D9D9} 41.27 & \cellcolor[HTML]{D9D9D9} 41.94 & \cellcolor[HTML]{D9D9D9} 41.94 & \cellcolor[HTML]{D9D9D9} 41.94 & \cellcolor[HTML]{D9D9D9} 42.56 & \cellcolor[HTML]{D9D9D9} 42.56 & \cellcolor[HTML]{D9D9D9} 42.56 \\ 
    ReM-ReB & \textbf{42.65} & \bf 43.66 &\bf  43.11 & \bf 46.01 & \bf 46.08 & \bf 45.32 & \bf 47.00 & \bf 46.97 & \bf 46.45 \\ 
  ReM-ReO & 37.92 & 41.16 & 41.73 & 45.22 & 45.22 & 44.49 & 46.81 & 46.76 & 46.07 \\ 
  BCRD-ReB & 37.36 & 33.12 & 28.74 & 38.68 & 34.16 & 29.63 & 39.78 & 35.38 & 30.93 \\ 
  BCRD-ReO & 31.44 & 29.37 & 26.46 & 36.37 & 33.62 & 29.16 & 39.32 & 34.97 & 30.11 \\
    \cmidrule(lr){2-10}
    \multicolumn{10}{c}{}\vspace{-0.2cm}\\
    & \multicolumn{9}{c}{$p=10$}\\\cmidrule(lr){2-10}
    \cellcolor[HTML]{D9D9D9} ReM & \cellcolor[HTML]{D9D9D9}\textbf{32.74} &\cellcolor[HTML]{D9D9D9} 32.74 &\cellcolor[HTML]{D9D9D9} 32.74 &\cellcolor[HTML]{D9D9D9} 33.83 &\cellcolor[HTML]{D9D9D9} 33.83 &\cellcolor[HTML]{D9D9D9} 33.83 &\cellcolor[HTML]{D9D9D9} 34.48 &\cellcolor[HTML]{D9D9D9} 34.48 &\cellcolor[HTML]{D9D9D9} 34.48 \\ 
    ReM-ReB & 31.75 & \textbf{34.60} & \textbf{35.78} & \textbf{41.86} & \bf 41.24 & \bf 40.50 & \bf 44.52 & \bf 43.56 &\bf  43.07 \\ 
  ReM-ReO & 21.18 & 28.35 & 32.29 & 40.02 & 40.45 & 39.59 & 43.67 & 43.10 & 42.75 \\ 
  BCRD-ReB & 28.67 & 28.13 & 25.66 & 35.99 & 31.77 & 28.25 & 38.58 & 34.07 & 29.64 \\ 
  BCRD-ReO & 17.20 & 21.62 & 20.90 & 33.38 & 30.16 & 27.08 & 37.26 & 33.66 & 28.69 \\
    \cmidrule(lr){2-10}
    \multicolumn{10}{c}{}\vspace{-0.2cm}\\
    & \multicolumn{9}{c}{$p=20$}\\\cmidrule(lr){2-10}
  \cellcolor[HTML]{D9D9D9} ReM & \cellcolor[HTML]{D9D9D9}\textbf{23.59} &\cellcolor[HTML]{D9D9D9} \textbf{23.59} &\cellcolor[HTML]{D9D9D9} 23.59 &\cellcolor[HTML]{D9D9D9} 26.41 &\cellcolor[HTML]{D9D9D9} 26.41 &\cellcolor[HTML]{D9D9D9} 26.41 &\cellcolor[HTML]{D9D9D9} 26.26 &\cellcolor[HTML]{D9D9D9} 26.26 &\cellcolor[HTML]{D9D9D9} 26.26 \\ 
    ReM-ReB& $--$ & 22.32 & \textbf{25.05} & \textbf{34.50} & \textbf{35.39} & \textbf{34.43} & \textbf{40.11} & \textbf{38.69} & 37.42 \\ 
    ReM-ReO & $--$ &14.10 & 20.45 & 29.13 & 32.63 & 33.44 & 38.45 & 38.10 & \textbf{37.54} \\ 
    BCRD-ReB & $--$ & 18.42 & 18.64 & 29.71 & 28.73 & 26.14 & 34.91 & 31.56 & 27.01 \\ 
    BCRD-ReO & $--$ & 9.82 & 12.69 & 24.80 & 25.57 & 23.49 & 32.44 & 30.38 & 26.74 \\ 
    \midrule
    \end{tabular}}
\end{table}

\begin{figure}[h]
\begin{center}
\includegraphics[width=4in]{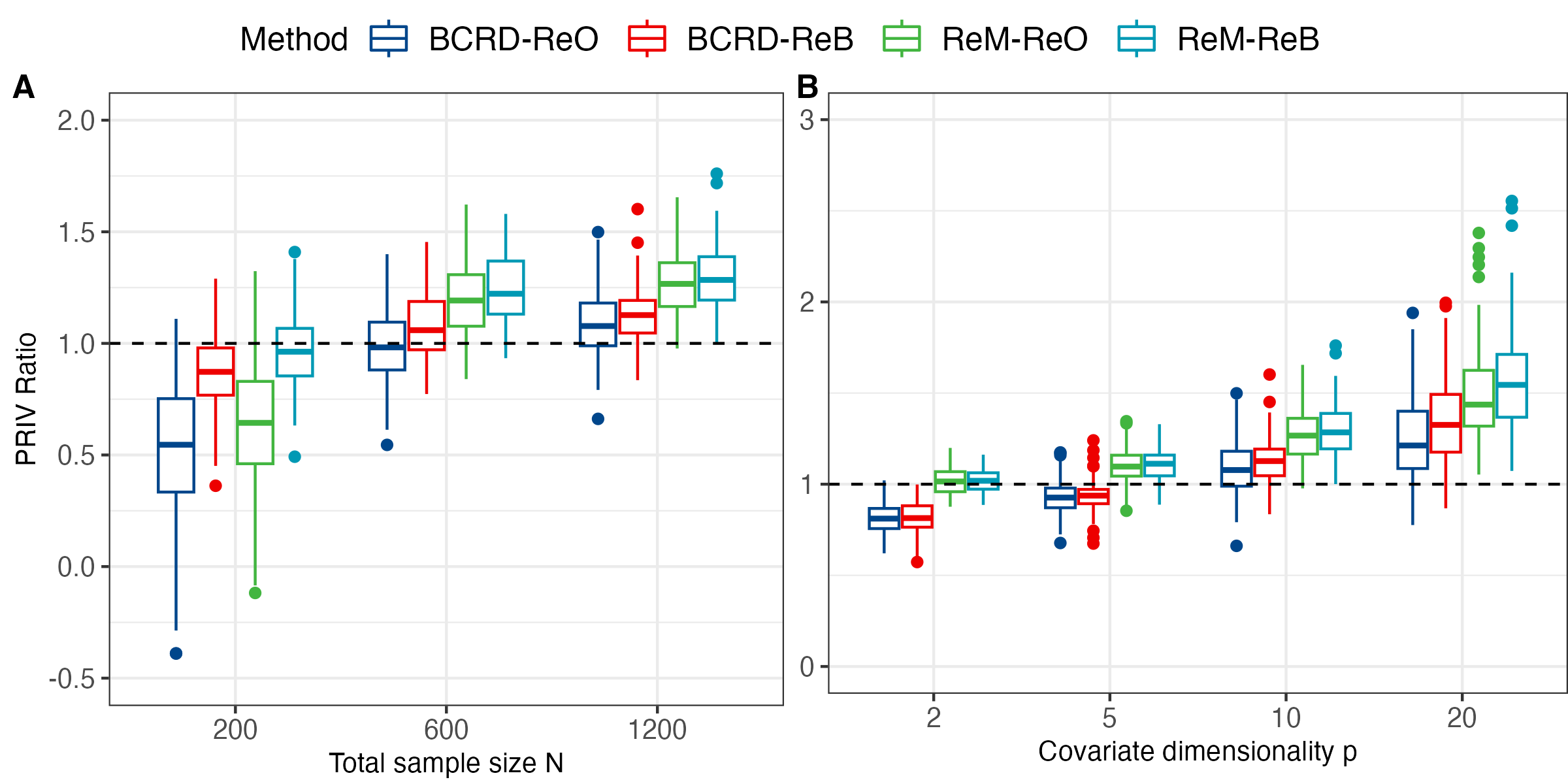}
\end{center}
\caption{Performance of BCRD-ReO, BCRD-ReB, ReM-ReO and ReM-ReB compared with ReM with $r=0.2$. 
Sub-figure (A) explores the impact of factor $N$ with a fixed $p=10$, while sub-figure (B) investigates factor $p$ with a fixed $N=1200$.
\label{fig:linear}}
\end{figure}


To compare the performance of two-stage ReB procedures to ReM, Figure \ref{fig:linear} shows how different factors affect the PRIV ratio of BCRD-ReO, BCRD-ReB, ReM-ReO, and ReM-ReB relative to ReM. With $r=0.2$ fixed, sub-figure (A) explores the impact of factor $N$ with a fixed $p=10$, while sub-figure (B) investigates factor $p$ with a fixed $N=1200$. The figure highlights that most PRIV ratios of ReM-ReB over ReM are greater than 1, with exceptions diminishing as $N$ and $p$ increase. In contrast, the other two-stage methods perform less effectively, particularly when sample size or covariate dimensionality is small.

In { Section B.4} of the Supplement, we examine how varying sample proportions in the first stage affect two-stage ReB procedures. We find that ReM-ReB with around 20\% of samples in the first stage consistently demonstrates the best performance and robustness. Similar results also hold in more challenging nonlinear scenarios, as shown in { Section B.5} of the Supplementary Material. 
In summary, simulations in both linear and nonlinear settings support the superiority of the two-stage ReB approach over ReM, particularly when the overall sample size is reasonably large (as indicated by Theorem \ref{2stageReO-variance}). Among the different two-stage ReB methods, ReM-ReB is generally the preferred choice in most cases.

\section{Real Data Application}\label{sec:RealDataAnalysis}

In this section, we evaluate the performance of Bayesian re-randomization procedures in a real-world application using a strategy based on \cite{Hill2011}. We employ covariates from the Infant Health and Development Program (IHDP), a study investigating the impact of intensive child care and home visits on low-birth-weight and premature infants. The IHDP dataset consists of 747 units and 25 pretreatment variables, including 6 continuous standardized covariates and 19 binary covariates. Here, we modify response surfaces based on \cite{Hill2011} by adding more noise to response surface A. Specifically, we use the following form for potential outcomes:
$$\bm Y(0) \sim N(\bm X\bbeta,3)\quad\text{and}\quad \bm Y(1) \sim N(\bm X\bbeta+4,3),$$
where $\bm X$ is an expanded covariate matrix (747$\times$26) with first column being a vector of ones, and
$\bbeta$ is a 26-dimensional random coefficient vector with elements sampled from a discrete support $\{0,1,2,3,4\}$ based on probability $(0.5,0.2,0.15,0.1,0.05)$. 

\begin{figure}[t]
\begin{center}
\includegraphics[width=4in]{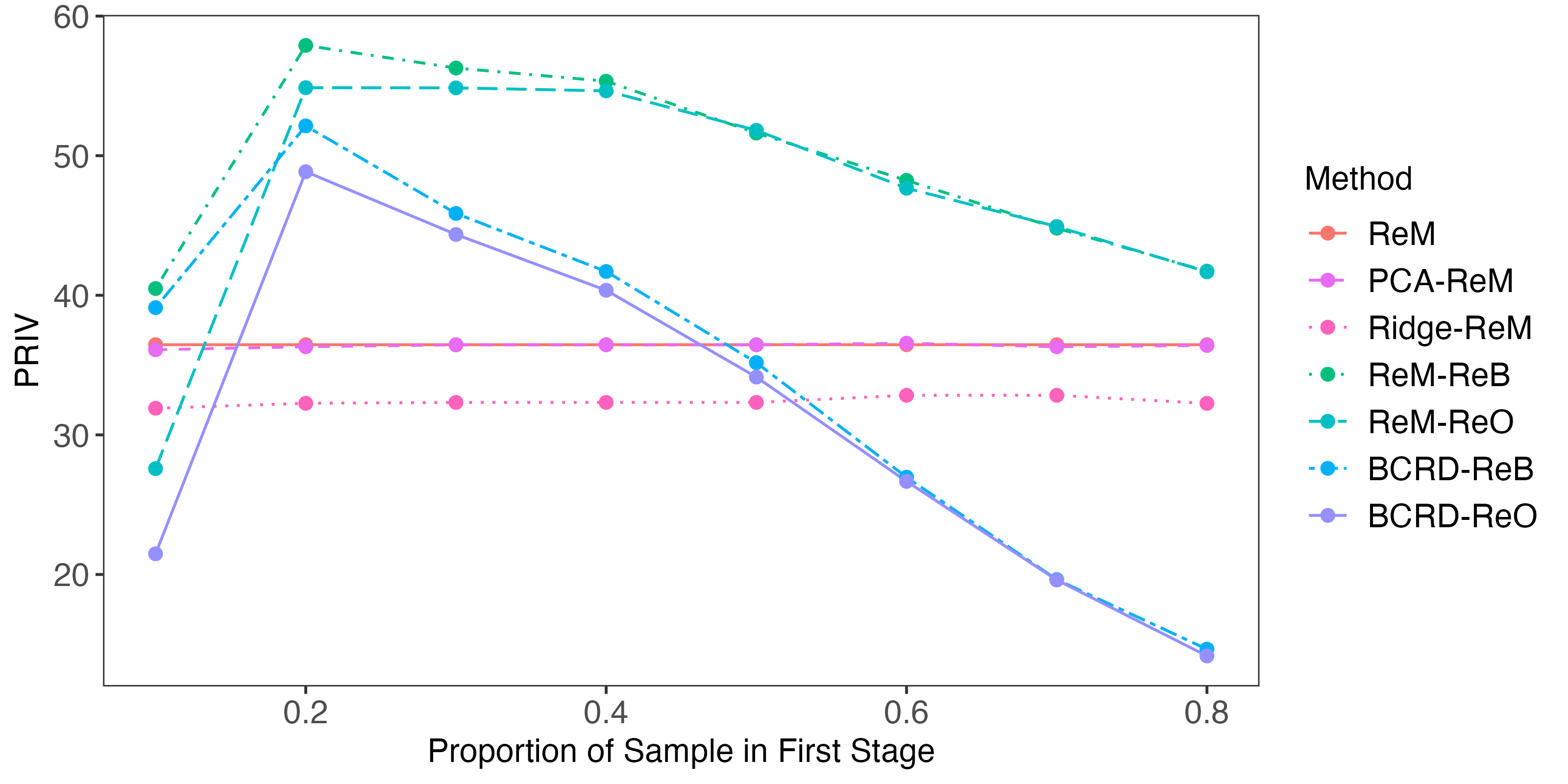}
\end{center}
\caption{PRIV of ReM, BCRD-ReB, BCRD-ReO, ReM-ReB and ReM-ReO with different proportion of sample in first stage $r$ in real application.
\label{fig:priv}}
\end{figure}

To keep a balanced design, we exclude the last unit from the dataset, and allocate the remaining 746 units into two groups of equal sample size (i.e., $N_0=N_1=373)$. Given the popularity of PCA-ReM and Ridge-ReM in high-dimensional scenarios, we assess the performance of two-stage ReB procedures (ReM-ReB, ReM-ReO, BCRD-ReB, and BCRD-ReO) alongside four different randomization schemes: BCRD, ReM, PCA-ReM, and Ridge-ReM. We maintain ReM's acceptance probability at $\alpha=0.05$ while numerically tuning the acceptance probabilities of other re-randomization procedures to match ReM's computational cost during the design phase. In PCA-ReM, we select principal components capturing 95\% of the variance. The running time ratio of these re-randomization procedures relative to ReM is detailed in { Section B.6} of the Supplement. For the two-stage procedures, we vary the proportion of samples in the first stage from 10\% to 80\% to assess their robustness. We calculate PRIV for rerandomization procedures over BCRD based on 1000 accepted allocations for each method over 100 iterations. The average PRIV for these methods is presented in Figure \ref{fig:priv}.

Based on the figure, we can see the following facts.
ReM-ReB and ReM-ReO outperform ReM, PCA-ReM, and Ridge-ReM, even when allocating a substantial proportion of samples to the first stage. Notably, Ridge-ReM fares poorly under equivalent computational constraints. Additionally, ReM-ReB and ReM-ReO exhibit greater robustness to changes in first-stage sample proportions compared to BCRD-ReB and BCRD-ReO. Moreover, BCRD-ReO and ReM-ReO may struggle with very small first-stage sample sizes. As a practical guideline, we recommend using ReM-ReB with a 20\% sample proportion in the first stage for scenarios similar to this setting.

\section{Conclusion and Discussions}\label{sec:Discussion}

In this study, we re-consider re-randomization from the Bayesian perspective. 
Assuming that the importance vector of covariates $\bbeta$ is uncertain with respect to a prior distribution $\pi$, a Bayesian criterion is established to derive a more efficient re-randomization procedure, referred to as ReB, by minimizing the prior-integrated conditional asymptotic variance of the casual effect estimator.
We find that ReB only depends on the second moment of prior distribution $\pi$, and enjoys a better efficiency than the ReM-based re-randomization procedures in balancing high-dimensional covariates and reducing the estimation variance of the difference-in-mean estimator for causal effect, as long as the prior distribution is informative to highlight the true importance vector.
We also note that ReB is invariant to any scale transformation of the prior distribution of $\bbeta$, indicating that re-scaling either $\pi$ or $\bX$ would not change the re-randomization procedure under the Bayesian criterion. 
Taking ReM, PCA-ReM, Ridge-ReM as its special cases under different specifications of the prior distribution $\pi$, ReB provides us a unified framework to understand and interpret various re-randomization procedures.
More importantly, ReB extends the idea behind ReM$_T$, and establishes a principled way to utilize prior information about $\bbeta$ in re-randomization.

Additionally, the Bayesian nature of ReB also provides an additional avenue for designing more efficient experiments in sequential trials.
The success of the two-stage ReB gives us an important message: once we can observe the responses of some units before the arrival of other experimental units, we would be able to design a more efficient randomized experiment by following the spirit of ReB.
Now, suppose a typical scenario where experiment units arrive in a sequential fashion with their potential outcomes observed along the experiment process so that we can update our knowledge about $\bbeta$ gradually. 
In such cases, we would be able to adjust the measurement for covariate imbalance simultaneously, and thus improve the overall balance of covariates more efficiently by carefully allocating treatment assignments to upcoming units based on the adjusted imbalance measurement induced by the Bayesian principle for re-randomization.
Compared to the classic approaches for sequential design, which typically do not take into consideration the responses of units received along the experiment process, the above alternative approach is apparently more attractive.
Due to the limited space of this paper, however, we reserve this exciting topic for future research.

\section{Funding}
This work was supported by National Science Foundation of China under Grant No. 11931001.

\bibliographystyle{apalike}
\bibliography{Rerand}

\newpage
\appendix

\begin{center}
    \textbf{\centering \Large Supplementary Material}
\end{center}
\setcounter{equation}{0}
\renewcommand{\theequation}{A.\arabic{equation}}
\setcounter{figure}{0}
\renewcommand{\thefigure}{B.\arabic{figure}}
\setcounter{table}{0}
\renewcommand{\thetable}{B.\arabic{table}}

In this document, Appendix \ref{sec:proof} contains technical proofs, and  
Appendix \ref{sec:simulations} shows additional simulation results.

\section{Technical Proofs}\label{sec:proof}

\subsection{Useful Lemmas for Theorem \ref{thm:ReO}}

To prove Theorem \ref{thm:ReO}, we propose the following two lemmas.
\begin{lemma}\label{lma:GeneralSolution4OptimizationProblem}
Let $\bm Z$ be a random vector defined on a probability space $(\Omega,\mathcal{F},\mathcal{P})$, $g(\bm Z)$ be a measurable function of $\bm Z$ satisfying $\bbE|g(\bm Z)|<\infty$, and  
$$\Psi_{\alpha}=\left\{\phi(\bm Z)=0\text{ or }1: \bbP(\phi(\bm Z)=1)=\alpha\right\}$$
be the set of re-randomization mechanisms for $\bm Z$ with acceptance rate $\alpha$. 
Define 
\[\begin{split}
    \phi_{g,\alpha}(\bm Z) \triangleq I\left(g(\bm Z)\leq a\right) \text{ with } \bbP(\phi_{g,\alpha}(\bm Z)=1) = \bbP(g(\bm Z)\leq a)=\alpha
\end{split}\]
as re-randomization mechanism that defines the acceptance region based on the contour line of $g(\cdot)$. 
We have:
\[\begin{split}
    \phi_{g,\alpha} = \arg \min_{\phi\in{\Psi_{\alpha}}} \bbE\big(g(\bm Z)|\phi(\bm Z)=1\big).
\end{split}\]
\end{lemma}

\begin{lemma}\label{lma:AssymptoticDistribution4d_beta}
Suppose Condition \ref{cond1} holds and $\bm\beta$ is known. 
Under complete randomization, we have
\[\begin{split}d_{\bm\beta}/\sigma^2_{\bm\beta} \xrightarrow{d} \chi^2_1\ \text{ as }N \rightarrow \infty,
\end{split}\]
where $d_{\bbeta}$
and $\sigma^2_{\bm\beta}$ are defined in Theorem \ref{thm:ReO}.
\end{lemma}

\subsection{Proof of Lemma \ref{lma:GeneralSolution4OptimizationProblem}}
Assume that there exists another $\phi^*(\bm Z) \in \Psi_\alpha$ that satisfies $\phi^* \neq \phi_{g,\alpha}$ and 
$$\bbE(g(\bZ)\mid\phi^*(\bm Z)=1)
< 
\bbE(g(\bZ)\mid\phi_{g,\alpha}(\bm Z)=1).$$
Let $A \triangleq \left\{\omega: \phi_{g,\alpha}(\bm Z(w))=1\right\} \in \mathcal{F}$, $A^* \triangleq \left\{\omega: \phi^*(\bm Z(w))=1\right\} \in \mathcal{F}$, then $\bZ$ is uniformly distributed on $A^*$ and $A$ under $\phi^*$ and $\phi$, and we have 
$$\int_{A^*} g(\bm Z) dF_{\bZ} < \int_{A} g(\bm Z) dF_{\bZ}.$$
Considering that
\begin{eqnarray*}
\int_{A^*} g(\bm Z) dF_{\bZ}&=&\int_{A^* \cap A} g(\bm Z) dF_{\bZ} + \int_{A^*\cap A^c} g(\bm Z) dF_{\bZ},\\
\int_{A} g(\bm Z) dF_{\bZ}&=&\int_{A^* \cap A} g(\bm Z) dF_{\bZ} + \int_{A\cap A^{*c}} g(\bm Z) dF_{\bZ},
\end{eqnarray*}
and $\bbE|g(\bm Z)| < \infty$, we have 
$$\int_{A^*\cap A^c} g(\bm Z) dF_{\bZ} < \int_{A\cap A^{*c}} g(\bm Z) dF_{\bZ}.$$

On the other hand, because $A = \left\{\omega: g(\bm Z(w))\leq a\right\}$, we have
\begin{eqnarray*}
\int_{A^*\cap A^c} g(\bm Z) dF_{\bZ}&>&a\cdot \mathbb{P}(A^*\cap A^c),\\
\int_{A\cap A^{*c}} g(\bm Z) dF_{\bZ}&\leq&a\cdot \mathbb{P}(A\cap A^{*c}).
\end{eqnarray*}
Considering that the acceptance rates of $\phi_g$ and $\phi^*$ are the same, we have 
$$\mathbb{P}(A^*\cap A^c)=\mathbb{P}(A\cap A^{*c}).$$
Therefore, 
\begin{eqnarray*}
\int_{A^*\cap A^c} g(\bm Z) dF_{\bZ}>
\int_{A\cap A^{*c}} g(\bm Z) dF_{\bZ},
\end{eqnarray*}
and we reach a contradiction.
\QEDA


\subsection{Proof of Lemma \ref{lma:AssymptoticDistribution4d_beta}}
According to the finite population central limit theorem, under regularity conditions in Condition \ref{cond1}, the large sampling distribution, over all randomizations, of $\sqrt{N}\bD$ is asymptotically Gaussian:
\[\begin{split}
    \sqrt{N}\bD \overset{\cdot}{\sim} \bm B,\ \bm B \sim \mathcal{N}\left(\bm 0, \bV_{xx}\right),
\end{split}\]
where $\overset{\cdot}{\sim}$ means two random variables converge to the same distribution weakly. Hence,
\[\begin{split}
    \sqrt{N}\bD^T\bbeta \overset{\cdot}{\sim} \bm B^T\bbeta,\text{ where } \bm B^T\bbeta \sim \mathcal{N}\left(\bm 0, \bbeta^T\bV_{xx}\bbeta\right).
\end{split}\]
Continuous mapping theorem and Slutsky's theorem imply that 
\[\begin{split}
    \frac{N\bD^T\bbeta\bbeta^T\bD}{\bbeta^T\bV_{xx}\bbeta} \overset{\cdot}{\sim} \frac{\bm B^T\bbeta\bbeta^T \bm B}{\bbeta^T\bV_{xx}\bbeta}, \text{ where }\frac{\bm B^T\bbeta\bbeta^T \bm B}{\bbeta^T\bV_{xx}\bbeta} \sim \chi^2_1.
\end{split}\]
Therefore, 
\[\begin{split}d_{\bm\beta}/\sigma^2_{\bm\beta} \xrightarrow{d} \chi^2_1,
\end{split}\]
where $\sigma^2_{\bm\beta} = \bm\beta^T\bV_{xx}\bm\beta$.
\QEDA


\subsection{Proof of Theorem \ref{thm:ReO}}

First of all,
Lemma \ref{lma:AssymptoticDistribution4d_beta} implies that the asymptotic acceptance rate of $\phi_{\bm\beta}$ is 
\[\begin{split}\gamma_{\phi_{\bm\beta}} = \lim_{N\rightarrow \infty} \bbP\big(d_{\bm\beta}/\sigma^2_{\bm\beta} \leq \xi_{\alpha,1}\big) = \alpha.
\end{split}\]
Hence, $\phi_{\bm\beta} \in \Phi_{\alpha}$. 
Since the first term in \eqref{eq:AsymptoticVariance4Phi}, i.e., $(1-R^2_{\infty})V_{\tau\tau,\infty}$, is a constant that does not depend on the specification of the re-randomization mechanism $\phi$, and the asymptotic sampling variance in \eqref{eq:AsymptoticVariance4Phi} is determined by the interaction of two factors $\bm\beta_{\infty}$ and $\bB_{\infty}$ in the second term.
Therefore,  
minimizing $\bbV_a\big(\sqrt{N}(\hat{\tau}-\tau)|\phi=1\big)$ is equivalent to minimizing 
\[\begin{split}
\bbE\left(\bm B^T_{\infty}\bm\beta_{\infty}\bm\beta_{\infty}^T\bm B_{\infty}|\bm B_{\infty} \in \mathcal{B}_{\phi,\infty}\right)
\end{split}\]
within $\phi \in \Phi_{\alpha}$. 
Direct application of Lemma \ref{lma:GeneralSolution4OptimizationProblem} implies the asymptotic acceptance region of the optimal criterion should take the form of 
\begin{equation}\label{criterion_ReO}
\mathcal{B}_{\phi,\infty} = \big\{\bm \mu: \bm \mu^T\bm\beta_{\infty}\bm\beta_{\infty}^T\bm \mu \leq a\big\},
\end{equation}
where the threshold $a$ is determined by solving 
\[\begin{split}
    \bbP\left(\bm B_{\infty}^T\bm\beta_{\infty}\bm\beta_{\infty}^T\bm B_{\infty} \leq a\right) = \alpha.
\end{split}\]
Since $\bB_{\infty} \sim \mathcal{N}(\bZero,\bV_{xx,\infty})$, we have $a = \bm\beta_{\infty}^T \bm V_{xx,\infty}\bm\beta_{\infty} \cdot \xi_{\alpha,1}$. 
As the acceptance region of $\phi_{\bm\beta}$ is 
\[\begin{split}
    \mathcal{B}_{\phi_{\bm\beta}}=\left\{\bm\mu: \bm \mu^T\bm\beta\bm\beta^T\bm \mu \leq \bm\beta^T \bm V_{xx}\bm\beta \cdot \xi_{\alpha,1}\right\},
\end{split}\]
whose limit is consistent with the form of (\ref{criterion_ReO}). Therefore, $$\phi_{\bbeta} = \argmin_{\phi \in \Phi_{\alpha}} \bbV_a\big(\sqrt{N}(\hat{\tau}-\tau)|\phi=1\big).$$

\QEDA

\subsection{Proof of Theorem \ref{PRIAV-ReO}}
Under Condition \ref{cond1}, 
\[\begin{split}
\sqrt{N}(\hat{\tau}-\tau) \mid \phi_{\bbeta}=1 &\overset{\cdot}{\sim} \epsilon + \bm V_{\tau x}\bm V_{xx}^{-1}\bm B\mid \bm B \in \mathcal{B}_{\phi_{\bbeta}}\\&\sim \epsilon + \bbeta^T \bB \mid (\bbeta^T \bB)^2 \leq \sigma^2_{\bbeta}\cdot \xi_{\alpha,1},
\end{split}\]
where $\epsilon \sim \mathcal{N}(0,V_{\tau\tau}-\bm V_{\tau x}\bm V_{xx}^{-1} \bm V_{x\tau})$ is independent of $\bm B\sim \mathcal{N}(0,\bV_{xx})$, and $\sigma^2_{\bbeta} = \bbV(\bbeta^T \bB) = \bm V_{\tau x}\bm V_{xx}^{-1} \bm V_{x\tau}$. Let $\tilde{\bB} = \sigma^{-1}_{\bbeta}\bbeta^T \bB$, we have $\tilde{\bB} \sim \mathcal{N}(0,1)$, and 
\[\begin{split}
\bbeta^T \bB \mid (\bbeta^T \bB)^2 \leq \sigma^2_{\bbeta}\cdot \xi_{\alpha,1} &\sim \sigma_{\bbeta}\cdot\tilde{\bB} \mid \tilde{\bB}^2 \leq \xi_{\alpha,1}.
\end{split}\]
Since $\sigma_{\bbeta}^2 = V_{\tau\tau}\cdot R^2$, we have 
\[\begin{split}
\sqrt{N}(\hat{\tau}-\tau) \mid \phi_{\bbeta}=1 &\overset{\cdot}{\sim} \sqrt{V_{\tau\tau}(1-R^2)}\cdot\varepsilon_0 + \sqrt{V_{\tau\tau}\cdot R^2}\cdot
(D \mid D^2 \leq \xi_{\alpha,1}),
\end{split}\]
where $\varepsilon_0,D \sim \mathcal{N}(0,1)$ are independent.
Moreover, 
by Eq.~\eqref{eq:r_phi}, 
\[\begin{split}
r_{\phi_{\bbeta}}= \frac{\bbE\left(\bm B^T_{\infty}\bbeta_{\infty}\bbeta^T_{\infty}\bm B_{\infty}|\bm B_{\infty} \in \mathcal{B}_{\phi_{\bbeta},\infty}\right)}{\bbE\left(\bm B^T_{\infty}\bbeta_{\infty}\bbeta_{\infty}^T\bm B_{\infty}\right)} = \frac{\bbE\left(\big(\bbeta^T_{\infty}\bm B_{\infty}\big)^2|\big(\bbeta^T_{\infty}\bm B_{\infty}\big)^2 \leq 
\bbeta^T_{\infty} \bm V_{xx,\infty} \bbeta_{\infty} \cdot \xi_{\alpha,1}
\right)}{\bbE\left(\big(\bbeta^T_{\infty}\bm B_{\infty}\big)^2\right)},
\end{split}\]
where 
$\left(\bbeta_{\infty}^T\bm B_{\infty}\right)^2 \sim \bbeta^T_{\infty} \bm V_{xx,\infty} \bbeta_{\infty} \cdot \chi_1^2$.
Hence, $r_{\phi_{\bbeta}} =
    v_{\alpha,1}$ and
$$\text{PRIASV}_{ReO}=100\times(1-v_{\alpha,1})R^2_{\infty}.$$
Recall that ReM achieves 
\[\begin{split}
\sqrt{N}(\hat{\tau}-\tau) \mid \phi_{\bbeta}=1 \overset{\cdot}{\sim} \sqrt{V_{\tau\tau}}\big(\sqrt{1-R^2}\cdot \varepsilon_0 + \sqrt{R^2}\cdot L_{1,\xi_{\alpha,p}}\big),
\end{split}\]
and 
\[\begin{split}
    \text{PRIASV}_{ReM} =  100 \times (1-v_{\alpha,p})R^2_{\infty},
\end{split}\]
where $v_{\alpha,p} = \bbE\left(\chi^2_p|\chi^2_p\leq \xi_{\alpha,p}\right)/p$. 
Therefore,
$$\frac{\text{PRIASV}_{ReM}}{\text{PRIASV}_{ReO}}=\frac{1-v_{\alpha,p}}{1-v_{\alpha,1}}.$$
From the definition of $v_{\alpha,p}$, we have
$$v_{\alpha,p} = \frac{\bbP(\chi^2_{p+2}\leq \xi_{\alpha,p})}{\bbP(\chi^2_p\leq \xi_{\alpha,p})}
     =\frac{1}{\alpha}\cdot\frac{\int_0^{\xi_{\alpha,p}}{x^{\frac{p}{2}}e^{-\frac{x}{2}}}dx}{2^{\frac{p+2}{2}}\cdot\Gamma(\frac{p+2}{2})}.$$
Based on the integration by parts, 
\begin{eqnarray*}
\int_0^{\xi_{\alpha,p}}x^{\frac{p}{2}}e^{-\frac{x}{2}}dx
&=&-2\left[x^\frac{p}{2}e^{-\frac{x}{2}}|_{x=\xi_{\alpha,p}}-\frac{p}{2}\int_0^{\xi_{\alpha,p}}x^{\frac{p}{2}-1}e^{-\frac{x}{2}}dx\right]\\
&=&-2\cdot x^\frac{p}{2}e^{-\frac{x}{2}}|_{x=\xi_{\alpha,p}}+p\cdot\frac{\int_0^{\xi_{\alpha,p}}x^{\frac{p}{2}-1}e^{-\frac{x}{2}}dx}{2^{\frac{p}{2}}\cdot\Gamma(\frac{p}{2})}\cdot 2^{\frac{p}{2}}\cdot\Gamma(\frac{p}{2})\\
&=&-2\cdot x^\frac{p}{2}e^{-\frac{x}{2}}|_{x=\xi_{\alpha,p}}+p\cdot\bbP(\xi^2_p\leq\xi_{\alpha,p})\cdot 2^{\frac{p}{2}}\cdot\Gamma(\frac{p}{2})\\
&=&-2\cdot x^\frac{p}{2}e^{-\frac{x}{2}}|_{x=\xi_{\alpha,p}}+p\cdot\alpha\cdot 2^{\frac{p}{2}}\cdot\Gamma(\frac{p}{2}),
\end{eqnarray*}
indicating that
\begin{eqnarray*}
v_{\alpha,p}&=&\frac{1}{\alpha}\cdot\frac{-2\cdot x^\frac{p}{2}e^{-\frac{x}{2}}|_{x=\xi_{\alpha,p}}+p\cdot\alpha\cdot 2^{\frac{p}{2}}\cdot\Gamma(\frac{p}{2})}{2^{\frac{p+2}{2}}\cdot\Gamma(\frac{p+2}{2})}\nonumber\\
&=&-\frac{x^\frac{p}{2}e^{-\frac{x}{2}}|_{x=\xi_{\alpha,p}}}{\alpha\cdot 2^{\frac{p}{2}}\cdot\Gamma(\frac{p+2}{2})}+\frac{p\cdot\alpha\cdot 2^{\frac{p}{2}}\cdot\Gamma(\frac{p}{2})}{\alpha\cdot 2^{\frac{p+2}{2}}\cdot\Gamma(\frac{p+2}{2})}\nonumber\\
&=&-\frac{x^\frac{p}{2}e^{-\frac{x}{2}}|_{x=\xi_{\alpha,p}}}{\alpha\cdot 2^{\frac{p}{2}}\cdot\Gamma(\frac{p+2}{2})}+1,
\end{eqnarray*}
and thus,
$$1-v_{\alpha,p}=\frac{x^\frac{p}{2}e^{-\frac{x}{2}}|_{x=\xi_{\alpha,p}}}{\alpha\cdot 2^{\frac{p}{2}}\cdot\Gamma(\frac{p+2}{2})}.$$
Considering that $\chi_p^2$ is close to normal distribution $N(p,2p)$ when $p$ is large, we have
$$\xi_{\alpha,p}= p+\sqrt{2p}\cdot\Phi^{-1}(\alpha)+o(1),$$
and thus,
\begin{eqnarray*}
\frac{x^\frac{p}{2}\cdot e^{-\frac{x}{2}}|_{x=\xi_{\alpha,p}}}{p^{\frac{p}{2}}\cdot e^{-\frac{p}{2}}}
&=&\left(1+\frac{\sqrt{2}\cdot\Phi^{-1}(\alpha)}{\sqrt{p}}+o\left(p^{-1}\right)\right)^{\frac{p}{2}}\cdot\exp\left\{-\frac{\sqrt{p}\cdot\Phi^{-1}(\alpha)+o(1)}{\sqrt{2}}\right\}\\
&\rightarrow&\exp\left\{\frac{\sqrt{p}\cdot\Phi^{-1}(\alpha)}{\sqrt{2}}\right\}\cdot\exp\left\{-\frac{\sqrt{p}\cdot\Phi^{-1}(\alpha)}{\sqrt{2}}\right\}=1,
\end{eqnarray*}
i.e., $x^\frac{p}{2}\cdot e^{-\frac{x}{2}}|_{x=\xi_{\alpha,p}}=O\left(p^{\frac{p}{2}}\cdot e^{-\frac{p}{2}}\right)$. 
On the other hand, because
$$\Gamma\left(\frac{p}{2}+1\right)=
\begin{cases}
\frac{p}{2}!, & p\mbox{ is even},\\
\frac{p!\sqrt{\pi}}{2^p(\frac{p-1}{2})!}, & p\mbox{ is odd},
\end{cases}
$$
and $p!=O(p^{p+\frac{1}{2}}\cdot e^{-p})$ based on the Stirling equation, we have 
$$\Gamma\left(\frac{p}{2}+1\right)=O\left(\left(\frac{p}{2}\right)^{\frac{p+1}{2}}\cdot e^{-\frac{p}{2}-1}\right).$$
And, thus
$$1-v_{\alpha,p}
=\frac{x^\frac{p}{2}e^{-\frac{x}{2}}|_{x=\xi_{\alpha,p}}}{\alpha\cdot 2^{\frac{p}{2}}\cdot\Gamma(\frac{p+2}{2})}
=\frac{O\left(p^{\frac{p}{2}}\cdot e^{-\frac{p}{2}}\right)}{\alpha\cdot 2^{\frac{p}{2}}\cdot O\left(\big(\frac{p}{2}\big)^{\frac{p+1}{2}}\cdot e^{-\frac{p}{2}-1}\right)}=O\left(\frac{1}{\sqrt{p}}\right).$$
Because $(1-v_{\alpha,1})$ is constant, we finally have
$$\frac{\text{PRIASV}_{ReM}}{\text{PRIASV}_{ReO}}=\frac{1-v_{\alpha,p}}{1-v_{\alpha,1}}=O\left(\frac{1}{\sqrt{p}}\right).$$
\QEDA

\subsection{Useful Lemmas for Theorem \ref{thm:ReB}}
To prove Theorem \ref{thm:ReB}, we propose the following two lemmas.
\begin{lemma}\label{lma:AssymptoticDistribution4d_pi}
Under Condition \ref{cond1} and complete randomization, we have 
\begin{equation}\label{eq:dpi}
    d_{\pi} \overset{\cdot}{\sim} \chi^2_{\blambda} = \sum_{j=1}^p \lambda_j Z_j^2,
\end{equation}
where 
$Z_1,...,Z_p$ are i.i.d. random variables from standard Gaussian distribution, and vector $\blambda =(\lambda_1,\ldots,\lambda_p)$ contains eigenvalues of matrix $\bm P = \big(\bm V_{xx}^{1/2}\big)^T\bm\Lambda_{\pi}\bm V_{xx}^{1/2}$ with $\bm V_{xx}^{1/2}$ being the Cholesky square root of $\bm V_{xx}$ satisfying $\bm V_{xx}^{1/2}\big(\bm V_{xx}^{1/2}\big)^T=\bm V_{xx}$.
\end{lemma}

\begin{lemma}\label{lemma3}
Suppose random variables $X_n$ and $Y_n$ converge weakly to the same distribution as $X$, i.e.,
\[\begin{split}
    X_n \xrightarrow{d} X,\ Y_n \xrightarrow{d} X,
\end{split}\]
and the distribution function $F_X$ is continuous. Let $\xi_{Y_n,\alpha}$ be the $\alpha$-quantile of the distribution of $Y_n$, then 
\[\begin{split}
    \bbP\left(X_n\leq \xi_{Y_n,\alpha}\right) \rightarrow \alpha.
\end{split}\]
\end{lemma}

\subsection{Proof of Lemma \ref{lma:AssymptoticDistribution4d_pi}}
Suppose the Cholesky decomposition of $\bSigma_{\bD}$ is $\bSigma_{\bD} = \bSigma_{\bD}^{1/2}\big(\bSigma_{\bD}^{1/2}\big)^T$ and let
$\Tilde{\bZ} = \bSigma_{\bD}^{-1/2}\bD$. Then we have $\bD = \bSigma_{\bD}^{1/2}\Tilde{\bZ}$ and 
$$d_\pi=N\bD^T\bLambda_\pi\bD=N\Tilde{\bZ}^T\left(\big(\bSigma_{\bD}^{1/2}\big)^T\bLambda_\pi\bSigma_\bD^{1/2}\right)\Tilde{\bZ}=\Tilde{\bZ}^T\bm P\Tilde{\bZ},$$
where $\bm P=N\big(\bSigma_{\bD}^{1/2}\big)^T\bLambda_\pi\bSigma_\bD^{1/2}=\big(\bV_{xx}^{1/2}\big)^T\bLambda_\pi\bV_{xx}^{1/2}$ is a symmetric positive-definite real matrix with the following eigenvalue decomposition
$$\bm P=\bGamma diag\{\lambda_1,\cdots,\lambda_p\}\bGamma^T\quad\mbox{with}\quad\bGamma^T\bGamma=\bI_p.$$
Central limit theorem of finite population implies that $\sqrt{N}\bD$ is asymptotically Gaussian under Condition \ref{cond1}: 
\[\begin{split}
    \sqrt{N}\bD \overset{\cdot}{\sim} \bm B,\ \bm B \sim \mathcal{N}(\bm 0, \bm V_{xx}).
\end{split}\]
Thus, 
$$\Tilde{\bZ} = \bSigma_{\bD}^{-1/2}\bD \overset{\cdot}{\sim} \bV_{xx}^{-1/2}\bm B \sim \mathcal{N}(\bm 0, \bm I),\ \ \bGamma^T\Tilde{\bZ} \sim \mathcal{N}(\bm 0, \bm I).$$
Therefore,
$$d_\pi=\Tilde{\bZ}^T\bm P\Tilde{\bZ} \overset{\cdot}{\sim} \sum_{j=1}^p\lambda_j Z_j^2,$$
where $Z_j$'s are i.i.d. standard normal random variables and weights $\lambda_j$'s are eigenvalues of matrix $\bm P$. \QEDA

\subsection{Proof of Lemma \ref{lemma3}}
Let $\xi_{X,a}$ be the $\alpha$-quantile of distribution of random variable $X$, $F_{X_n}(\cdot),F_{Y_n}(\cdot),F_{X}(\cdot)$ be the distribution function of $X_n$, $Y_n$ and $X$ respectively. Polya's theorem implies 
\[\begin{split}
    &\lim_{n\rightarrow\infty} \sup_{x\in\mathcal{R}} \left|F_{X_n}(x) - F_X(x)\right|=0,\\
    &\lim_{n\rightarrow\infty} \sup_{x\in\mathcal{R}} \left|F_{Y_n}(x) - F_X(x)\right|=0.
\end{split}\]
Hence, 
$$\left|\bbP(X_n \leq \xi_{Y_n,\alpha}) - \bbP(X \leq \xi_{Y_n,\alpha})\right|\rightarrow 0.$$
Since
\[\begin{split}
    \left|\bbP(X \leq \xi_{Y_n,\alpha}) - \bbP(X \leq \xi_{X ,\alpha})\right| &= \left|F_X(\xi_{Y_n,\alpha}) -F_X(\xi_{X,\alpha}) \right|\\ &\leq \left|F_X(\xi_{Y_n,\alpha}) -F_{Y_n}(\xi_{Y_n,\alpha}) \right|+\left|F_{Y_n}(\xi_{Y_n,\alpha}) -F_X(\xi_{X,\alpha}) \right|\\ &= \left|F_X(\xi_{Y_n,\alpha}) -F_{Y_n}(\xi_{Y_n,\alpha}) \right| \rightarrow 0,
\end{split}\]
we have 
\[\begin{split}
    \left|\bbP(X_n \leq \xi_{Y_n,\alpha}) - \bbP(X \leq \xi_{X,\alpha})\right| &\leq \left|\bbP(X_n \leq \xi_{Y_n,\alpha}) - \bbP(X \leq \xi_{Y_n,\alpha})\right|\\&+
    \left|\bbP(X \leq \xi_{Y_n,\alpha}) - \bbP(X \leq \xi_{X,\alpha})\right| \rightarrow 0.
\end{split}\]
That is, 
\[\begin{split}
    \bbP(X_n \leq \xi_{Y_n,\alpha}) \rightarrow \alpha.
\end{split}\]
\QEDA

\subsection{Proof of Theorem \ref{thm:ReB}}
First of all, Lemma \ref{lma:AssymptoticDistribution4d_pi} and Lemma \ref{lemma3} imply that the asymptotic acceptance rate of $\phi_{\pi}$ is 
\[\begin{split}
\gamma_{\phi_{\pi}} = \lim_{N\rightarrow \infty} \bbP\big(d_{\pi} \leq \xi_{\alpha,\blambda}\big) = \alpha.
\end{split}\]
Hence, $\phi_{\pi} \in \Phi_{\alpha}$. 
Considering that $\bbE_{\pi}\left[(1-R^2_{\infty})V_{\tau\tau,\infty}\right]$ is independent of $\phi$, minimizing $\bbV_{\pi,a}\big(\sqrt{N}(\hat{\tau}-\tau)|\phi=1\big)$ is equivalent to minimizing
\begin{equation}\label{eq:ObjectiveFunction4ReB}
    \begin{split}
\bbE_{\pi}\Big[\bbE\left(\bm B_{\infty}^T\bbeta_{\infty}\bbeta_{\infty}^T\bm B_{\infty}|\bm B_{\infty} \in \mathcal{B}_{\phi,\infty}\right) \Big]
    &= \bbE\left[\bm B_{\infty}^T\bbE_{\pi}\left(\bbeta_{\infty}\bbeta_{\infty}^T\right)\bm B_{\infty}\mid\bm B_{\infty} \in \mathcal{B}_{\phi,\infty}\right]\\
    &=\bbE\left[\bm B_{\infty}^T\bLambda_{\pi}\bm B_{\infty}\mid\bm B_{\infty} \in \mathcal{B}_{\phi,\infty}\right],
    \end{split}
\end{equation}
where
$$\bm\Lambda_{\pi} = \bbE_\pi\left(\bbeta_\infty\bbeta_\infty^T\right) = \bm\mu_{\pi}\bm\mu_{\pi}^T + \bm\Sigma_{\pi}$$
is the second moment of $\pi$ and is referred to as the characteristic matrix of the prior distribution,
with $\bm\mu_{\pi}$ and $\bm\Sigma_{\pi}$ being the expectation vector and covariance matrix of $\pi$, respectively.
Direct application of Lemma \ref{lma:GeneralSolution4OptimizationProblem} implies the asymptotic acceptance region of the optimal criterion should take the form of 
\begin{equation}\label{criterion}
\mathcal{B}_{\phi,\infty} = \big\{\bm \mu: \bm \mu^T\bLambda_{\pi}\bm \mu \leq a\big\},
\end{equation}
where the threshold $a$ is determined by solving 
\[\begin{split}
    \bbP\left(\bm B_{\infty}^T\bLambda_{\pi}\bm B_{\infty} \leq a\right) = \alpha.
\end{split}\]
Since $\bB_{\infty} \sim \mathcal{N}(\bZero,\bV_{xx,\infty})$, we have $a =  \xi_{\alpha,\blambda_{\infty}}$, where $\blambda_\infty$ is vector of eigenvalues of $\big(\bm V_{xx,\infty}^{1/2}\big)^T\bm\Lambda_{\pi}\bm V_{xx,\infty}^{1/2}$, and $Z_1,...,Z_p$ being i.i.d. random variables from standard Gaussian distribution.
As the acceptance region of $\phi_{\pi}$ is 
\[\begin{split}
\mathcal{B}_{\phi_{\pi}}=\left\{\bm\mu: \bm \mu^T\bLambda_{\pi}\bm \mu \leq \xi_{\alpha,\blambda}\right\},
\end{split}\]
whose limit is consistent with the form of (\ref{criterion}). Therefore, $$\phi_{\pi} = \argmin_{\phi \in \Phi_{\alpha}} \bbV_{\pi,a}\big(\sqrt{N}(\hat{\tau}-\tau)|\phi=1\big).$$
\QEDA

\subsection{Proof of Corollary \ref{invariance}}

Based on Theorem \ref{thm:ReB}, we have
\begin{eqnarray*}
&&\phi_{\pi_1}=I\left(d_{\pi_1}\leq \xi_{\alpha,\blambda_1}\right),\\
&&\phi_{\pi_2}=I\left(d_{\pi_2}\leq \xi_{\alpha,\blambda_2}\right),
\end{eqnarray*}
where $\blambda_1$ and $\blambda_1$ are vectors of eigenvalues of $\big(\bm V_{xx}^{1/2}\big)^T\bLambda_{\pi_1}\bm V_{xx}^{1/2}$ and $\big(\bm V_{xx}^{1/2}\big)^T\bLambda_{\pi_2}\bm V_{xx}^{1/2}$, respectively. Since $\bLambda_{\pi_2}=r\bLambda_{\pi_1}$, we have 
$$d_{\pi_2}=N\bD^T\bLambda_{\pi_2}\bD=rN\bD^T\bLambda_{\pi_1}\bD=r\cdot d_{\pi_1},\ \xi_{\alpha,\blambda_2} = r\cdot \xi_{\alpha,\blambda_1},$$
and thus
$$\phi_{\pi_2}=I\left(d_{\pi_2}\leq \xi_{\alpha,\blambda_2}\right)=I\left(r\cdot d_{\pi_1}\leq r \cdot \xi_{\alpha,\blambda_1}\right) = \phi_{\pi_1}.$$ The proof is complete. \QEDA

\subsection{Proof of Theorem \ref{thm:AsymptoticDistributionReBfullrank}}
Under Condition \ref{cond1}, 
\[\begin{split}
\sqrt{N}(\hat{\tau}-\tau) \mid \phi_{\pi}=1 &\overset{\cdot}{\sim} \epsilon + \bm V_{\tau x}\bm V_{xx}^{-1}\bm B\mid \bm B \in \mathcal{B}_{\phi_{\pi}}\\&\sim \epsilon + \bbeta^T \bB \mid \bB^T\bLambda_{\pi}\bB \leq \xi_{\alpha,\blambda},
\end{split}\]
where $\epsilon \sim \mathcal{N}(0,V_{\tau\tau}-\bm V_{\tau x}\bm V_{xx}^{-1} \bm V_{x\tau})$ is independent of $\bm B\sim \mathcal{N}(0,\bV_{xx})$. Since $\bLambda_{\pi} = \bM\bM^T$ is of full rank, then $\bM$ is reversible.
Hence, 
\[\begin{split}
\sqrt{N}(\hat{\tau}-\tau) \mid \phi_{\pi}=1 &\overset{\cdot}{\sim} \epsilon + \bbeta^T(\bM^{-1})^T \bm Q \mid \bm Q^T\bm Q \leq \xi_{\alpha,\blambda},
\end{split}\]
where $\bm Q = \bM^T\bB\sim \mathcal{N}(\bm 0,\bM^T\bV_{xx}\bM)$, and $\bbeta^T(\bM^{-1})^T \bm Q \sim \mathcal{N}(0,\bm V_{\tau x}\bm V_{xx}^{-1} \bm V_{x\tau})$.
Therefore,
\[\begin{split}
\sqrt{N}(\hat{\tau}-\tau) \mid \phi_{\pi}=1 &\overset{\cdot}{\sim} \sqrt{V_{\tau\tau}(1-R^2)}\cdot \varepsilon_0 + \bbeta^T(\bM^{-1})^T \bm Q | \bm Q^T\bm Q \leq \xi_{\alpha,\blambda}.
\end{split}\]

By Eq.~\eqref{R2_pi}, 
$$r_{\phi_{\pi}} \triangleq \frac{\bbE\big(\bm B_{\infty}^T\bbeta_{\infty}\bbeta_{\infty}^T\bm B_{\infty}\mid\bm B_{\infty} \in \mathcal{B}_{\phi_{\pi},\infty} \big)}{\bbE\big(\bm B^T_{\infty}\bbeta_{\infty}\bbeta_{\infty}^T\bm B_{\infty}\big)} = \lim_{N\rightarrow\infty} 
\frac{\bbE\big(\bm B^T\bbeta\bbeta^T\bm B\mid\bm B \in \mathcal{B}_{\phi_{\pi}} \big)}{\bbE\big(\bm B^T\bbeta\bbeta^T\bm B\big)},$$
where $\bm B \sim N(\bm 0, \bm V_{xx})$ and 
\[\begin{split}
    \mathcal{B}_{\phi_{\pi}} = \left\{\bmu:\bmu^T \bLambda_{\pi}\bmu \leq \xi_{\alpha,\blambda}\right\},
\end{split}\]
where $\blambda = (\lambda_{1},...,\lambda_{p})^T$ is the vector of eigenvalues of  $\bm P=\big(\bV_{xx}^{1/2}\big)^T\bLambda_\pi\bV_{xx}^{1/2}$ and $\bm B^T\bLambda_{\pi}\bm B \sim \chi^2_{\blambda}$. 
For $\bm Q =  \bm M^T \bm B$, we have
\[\begin{split}
\bbE\left(\bm B^T\bbeta\bbeta^T\bm B\mid \bm B \in \mathcal{B}_{\phi_{\pi}}\right) &= 
\bbE\left(\bbeta^T\bm B\bm B^T\bbeta\mid\bm B^T\bLambda_{\pi}\bm B \leq \xi_{\alpha,\blambda}\right)\\
&= \bbeta^T \left(\bm M^{-1}\right)^T
\bbE\left(\bm Q\bm Q^T\mid\bm Q^T\bm Q \leq \xi_{\alpha,\blambda}\right) \bm M^{-1}\bbeta,
\end{split}\]
where $\bm Q^T\bm Q \sim \chi^2_{\blambda}$.
Since $\bm Q \sim N\left(\bm0,\bm M^T \bm V_{xx} \bm M\right)$ and $\bm M^T \bm V_{xx} \bm M$ has the same eigenvalues as $\bm P$, it can be expressed as 
\[\begin{split}
    \bm M^T \bm V_{xx} \bm M = \bm\Gamma^T diag\{\lambda_{1},...,\lambda_{p}\} \bm\Gamma,
\end{split}\]
where $\bm\Gamma$ is an orthogonal matrix. Let $\bm \Lambda = diag\{\lambda_{1},...,\lambda_{p}\}$,
then $\Tilde{\bm Q} = \bm\Gamma\bm Q\sim N\left(\bm0,\bm \Lambda\right)$ and 
\[\begin{split}
\bbE\big(\bm B^T\bbeta\bbeta^T\bm B\mid \bm B \in \mathcal{B}_{\phi_{\pi}}\big)= 
\bbeta^T \left(\bm M^{-1}\right)^T\bm\Gamma^T
\bbE\big(\Tilde{\bm Q}\Tilde{\bm Q}^T\mid\Tilde{\bm Q}^T\Tilde{\bm Q} \leq \xi_{\alpha,\blambda}\big)\bm\Gamma \bm M^{-1}\bbeta.
\end{split}\]
Suppose $\bm \Tilde{\bm Q} = \left(q_1,...,q_p\right)^T$, for $i \not= j$, symmetry implies
\[\begin{split}
    \bbE\big(q_iq_j\mid\Tilde{\bm Q}^T\Tilde{\bm Q} \leq \xi_{\alpha,\blambda}\big) &= \bbE\left(\bbE\big(q_iq_j\mid q_j,\Tilde{\bm Q}^T\Tilde{\bm Q} \leq \xi_{\alpha,\blambda}\big)\mid\Tilde{\bm Q}^T\Tilde{\bm Q} \leq \xi_{\alpha,\blambda}\right)\\
    &= \bbE\left(q_j\bbE\big(q_i\mid q_j,\Tilde{\bm Q}^T\Tilde{\bm Q} \leq \xi_{\alpha,\blambda}\big)\mid\Tilde{\bm Q}^T\Tilde{\bm Q} \leq \xi_{\alpha,\blambda}\right)\\
    &= \bbE\big(q_j\times 0\mid\Tilde{\bm Q}^T\Tilde{\bm Q} \leq \xi_{\alpha,\blambda}\big)=0.
\end{split}\]
Let $c_i=\bbE\left(\lambda_{i} Z_i^2\mid \sum_{j=1}^p \lambda_{j} Z_j^2 \leq \xi_{\alpha,\blambda}\right)$, where $i=1,\ldots,p$ and $Z_1,\ldots,Z_p$ are i.i.d. standard normal distributed random variables, then
\[\begin{split}
\bbE\big(\Tilde{\bm Q}\Tilde{\bm Q}^T\mid\Tilde{\bm Q}^T\Tilde{\bm Q} \leq \xi_{\alpha,\blambda}\big) &= diag\left\{\bbE\big(q_i^2\mid \Tilde{\bm Q}^T\Tilde{\bm Q} \leq \xi_{\alpha,\blambda}\right)\big\}_{i=1}^p\\
&= diag\left\{\bbE\left(\lambda_{i} Z_i^2\mid \sum_{j=1}^p \lambda_{j} Z_j^2 \leq \xi_{\alpha,\blambda}\right)\right\}_{i=1}^p\\
&= \bm C = diag\{c_1,...,c_p\}.
\end{split}\]
Therefore, 
\[\begin{split}
 \bbeta^T \big(\bm M^{-1}\big)^T\bm\Gamma^T
\bbE\big(\Tilde{\bm Q}\Tilde{\bm Q}^T\mid\Tilde{\bm Q}^T\Tilde{\bm Q} \leq \xi_{\alpha,\blambda}\big)\bm\Gamma \bm M^{-1}\bbeta= \bbeta^T \big(\bm M^{-1}\big)^T\bm\Gamma^T
\bm C\bm\Gamma \bm M^{-1}\bbeta,
\end{split}\]
and thus, 
\[\begin{split}
    r_{\phi_{\pi}} &=\lim_{N\rightarrow\infty}  \frac{\bbE\big(\bm B^T\bbeta\bbeta^T\bm B\mid\bm B \in \mathcal{B}_{\phi_{\pi}}\big)}{\bbeta^T \bm V_{xx} \bbeta} = \lim_{N\rightarrow\infty} \frac{\bbeta^T \big(\bm M^{-1}\big)^T\bm\Gamma^T
\bm C\bm\Gamma \bm M^{-1}\bbeta}{\bbeta^T \bm V_{xx}\bbeta}.
\end{split}\]
Since $\bbeta^T \bm V_{xx}\bbeta = \bbeta^T \big(\bm M^{-1}\big)^T\bm\Gamma^T
\bm \Lambda\bm\Gamma\bm M^{-1}\bbeta$, we complete the proof.
\QEDA

\subsection{Proof of Theorem \ref{thm:ReB2sigma0}}
Under Condition \ref{cond1}, 
\[\begin{split}
\sqrt{N}(\hat{\tau}-\tau) \mid \phi_{\pi}=1 &\overset{\cdot}{\sim} \epsilon + \bm V_{\tau x}\bm V_{xx}^{-1}\bm B\mid \bm B \in \mathcal{B}_{\phi_{\pi}}\\&\sim \epsilon + \bbeta^T \bB \mid \bB^T\bLambda_{\pi}\bB \leq \xi_{\alpha,\blambda},
\end{split}\]
where $\epsilon \sim \mathcal{N}(0,V_{\tau\tau}-\bm V_{\tau x}\bm V_{xx}^{-1} \bm V_{x\tau})$ is independent of $\bm B\sim \mathcal{N}(0,\bV_{xx})$. Since $\bLambda_{\pi} = \bmu_{\pi}\bmu_{\pi}^T$, we have 
\[\begin{split}
\bbeta^T \bB \mid \bB^T\bLambda_{\pi}\bB \leq \xi_{\alpha,\blambda} &\sim \epsilon_{\pi} +(\bbeta^T\bV_{xx}\bmu_{\pi})(\bmu_{\pi}^T\bV_{xx}\bmu_{\pi})^{-1}\bmu_{\pi}^T\bB\mid \bB^T\bmu_{\pi}\bmu_{\pi}^T\bB \leq \xi_{\alpha,\blambda},
\end{split}\]
where $\epsilon_{\pi} \sim \mathcal{N}(0,\bm V_{\tau x}\bm V_{xx}^{-1} \bm V_{x\tau}-(\bbeta^T\bV_{xx}\bmu_{\pi})(\bmu_{\pi}^T\bV_{xx}\bmu_{\pi})^{-1}(\bmu_{\pi}^T\bV_{xx}\bbeta))$ is independent of $\bmu_{\pi}^T\bB$, and $\xi_{\alpha,\blambda} = \bmu_{\pi}^T\bV_{xx}\bmu_{\pi}\cdot \xi_{\alpha,1}$ is the $\alpha$-quantile of $\bmu_{\pi}^T\bV_{xx}\bmu_{\pi}\cdot \chi^2_1$ distribution. 
Denote
\[\begin{split}
    R^2_{\pi} = \frac{(\bbeta^T\bV_{xx}\bmu_{\pi})(\bmu_{\pi}^T\bV_{xx}\bmu_{\pi})^{-1}(\bmu_{\pi}^T\bV_{xx}\bbeta)}{V_{\tau\tau}},
\end{split}\] we have 
\[\begin{split}
\sqrt{N}(\hat{\tau}-\tau) \mid \phi_{\pi}=1 &\overset{\cdot}{\sim} \epsilon + \epsilon_{\pi} +(\bbeta^T\bV_{xx}\bmu_{\pi})(\bmu_{\pi}^T\bV_{xx}\bmu_{\pi})^{-1}\bmu_{\pi}^T\bB\mid \bB^T\bmu_{\pi}\bmu_{\pi}^T\bB \leq \xi_{\alpha,\blambda}\\
&\sim \sqrt{V_{\tau\tau}(1-R^2_{\pi})}\cdot \varepsilon_0 + \sqrt{V_{\tau\tau}\cdot R^2_{\pi}}\cdot L_{1,\xi_{\alpha,1}}.
\end{split}\]
Since $V_{\tau\tau}\cdot R^2 = \bbeta^T\bV_{xx}\bbeta$, 
\[\begin{split}
R^2_{\pi} = \frac{(\bbeta^T\bV_{xx}\bmu_{\pi})(\bmu_{\pi}^T\bV_{xx}\bmu_{\pi})^{-1}(\bmu_{\pi}^T\bV_{xx}\bbeta)}{\bbeta^T\bV_{xx}\bbeta}R^2
= \frac{(\bmu_{\pi}^T \bV_{xx} \bbeta)^2}{(\bmu_{\pi}^T \bV_{xx} \bmu_{\pi})(\bbeta^T \bV_{xx} \bbeta)}R^2.
\end{split}\]
Let $\theta$ be the angle between $\big(\bV_{xx}^{1/2}\big)^T\bmu_{\pi}$ and $\big(\bV_{xx}^{1/2}\big)^T\bbeta$, then we have 
\[\begin{split}
R^2_{\pi} 
= (\cos\theta)^2\cdot R^2.
\end{split}\]

Moreover, denote the limit of $\theta$ and $R^2_{\pi}$ as $\theta_{\infty}$ and $R^2_{\pi,\infty} = (\cos\theta_{\infty})^2\cdot R^2_{\infty}$, respectively, then 
\[\begin{split}
\frac{\bbV_a\big(\sqrt{N}(\hat{\tau}-\tau) \mid \phi_{\pi}=1\big)}{\bbV_a\big(\sqrt{N}(\hat{\tau}-\tau)\big)} &= \frac{V_{\tau\tau,\infty}(1-R^2_{\pi,\infty}) + V_{\tau\tau,\infty}\cdot R^2_{\pi,\infty} \cdot v_{\alpha,1}}{V_{\tau\tau,\infty}} = 1-(1-v_{\alpha,1})R^2_{\pi,\infty}.
\end{split}\]
By definition in Eq.~\eqref{eq:PRIAV}, 
\[\begin{split}
    r_{\phi_{\pi}} &=1 - (1-v_{\alpha,1})R^2_{\pi,\infty}/R^2_{\infty}=1 - (1-v_{\alpha,1})
    (\cos\theta_{\infty})^2.
\end{split}\]
\QEDA

\subsection{Proof of Theorem \ref{2stageReO-variance}}
Take BCRD-ReB as an example and we suppose its acceptance probability in second stage be $\alpha_1$. Since $\hat{\tau} = \varrho\cdot\hat{\tau}_1 + (1-\varrho)\cdot\hat{\tau}_2$, we have 
\[\begin{split}
\bbV\left(\hat{\tau}\mid \phi_{BCRD\text{-}ReB}=1\right) &= \varrho^2\cdot \bbV\left(\hat{\tau}_1\right)+(1-\varrho)^{2}
\cdot \bbV\left(\hat{\tau}_2\mid \phi_{\pi_{N_1}=1}\right).
\end{split}\]
Hence, for $N=N_1+N_2$, we have 
\begin{equation}\label{eq:asympvariancetotal}
\begin{split}
    &\indent\bbV_a\left(\sqrt{N}(\hat{\tau}-\tau)\mid \phi_{BCRD\text{-}ReB}=1\right)\\ &=
\lim_{N_1,N_2 \rightarrow \infty}\left[
\frac{\varrho^2}{N_1/N}
\cdot \bbV\left(\sqrt{N_1}(\hat{\tau}_1-\tau_1)\right)+\frac{(1-\varrho)^{2}}{N_2/N}
\cdot \bbV\left(\sqrt{N_2}(\hat{\tau}_2-\tau_2)\mid \phi_{\pi_{N_1}=1}\right)\right]\\ &= \lim_{N_1,N_2\rightarrow\infty}\left[
\varrho
\cdot \bbV\left(\sqrt{N_1}(\hat{\tau}_1-\tau_1)\right)+(1-\varrho)
\cdot \bbV\left(\sqrt{N_2}(\hat{\tau}_2-\tau_2)\mid \phi_{\pi_{N_1}=1}\right)\right].
\end{split}
\end{equation}
Denote $\hat{\tau}^{(2)}$ as the treatment effect estimator when we conduct complete randomization in the second stage. For fixed $N_1$, as $N_2 \rightarrow \infty$, we have
\[\begin{split}
    \lim_{N_2\rightarrow \infty}\bbV\left(\sqrt{N_2}(\hat{\tau}_2-\tau_2)\mid \phi_{\pi_{N_1}=1}\right) &= \lim_{N_2\rightarrow \infty}\left[1-(1-v_{\alpha_1,\pi_{N_1}})R^2_{N_2}\right]\cdot \bbV\left(\sqrt{N_2}(\hat{\tau}^{(2)}-\tau_2)\right)\\&=\left[1-(1-v_{\alpha_1,\pi_{N_1}})R^2_{\infty}\right]\cdot\lim_{N\rightarrow \infty}\bbV\left(\sqrt{N}(\hat{\tau}-\tau)\right).
\end{split}\]
Moreover, under Condition \ref{cond1}, $\pi_{N_1}$ converges to the point mass at $\bbeta_{\infty}$
as $N_1\rightarrow\infty$, and hence, $v_{\alpha_1,\pi_{N_1}} \rightarrow v_{\alpha_1,1}$.
Therefore, 
\[\begin{split}
    \lim_{N_1,N_2\rightarrow \infty}\bbV\left(\sqrt{N_2}(\hat{\tau}_2-\tau_2)\mid \phi_{\pi_{N_1}=1}\right) &=\left[1-(1-v_{\alpha_1,1})R^2_{\infty}\right]\cdot\lim_{N\rightarrow \infty}\bbV\left(\sqrt{N}(\hat{\tau}-\tau)\right).
\end{split}\]
Since
\[\begin{split}
\lim_{N_1\rightarrow\infty}
\bbV\left(\sqrt{N_1}(\hat{\tau}_1-\tau_1)\right)=\lim_{N\rightarrow\infty}
\bbV\left(\sqrt{N}(\hat{\tau}-\tau)\right),
\end{split}\]
for $\varrho\rightarrow\varrho^*$, we have 
\[\begin{split} 
\frac{\bbV_a\left(\sqrt{N}(\hat{\tau}-\tau)\mid \phi_{BCRD\text{-}ReB}=1\right)}{ \bbV_a\left(\sqrt{N}(\hat{\tau}-\tau)\right)}
&=
\varrho^*
+(1-\varrho^*)
\cdot \left[1-(1-v_{\alpha_1,1})R^2_{\infty}\right].
\end{split}\] 
Therefore, as $N_1,N_2 \rightarrow \infty$, we have
\[\begin{split}
    \text{PRIASV}_{BCRD-ReB} &= 100\times \left\{1-\left\{\varrho^*
+(1-\varrho^*)
\cdot \left[1-(1-v_{\alpha_1,1})R^2_{\infty}\right]\right\}\right\}
  \\ &
  =100\times (1-\varrho^*)(1-v_{\alpha_1,1})R^2_{\infty}.
\end{split}\]
PRIASV of ReM-ReB can be similarly derived.
\QEDA

\newpage
\section{Additional simulation results}\label{sec:simulations}
\subsection{Comparison of ReO and ReM}
\begin{figure}[h]
\begin{center}
\includegraphics[width=4in]{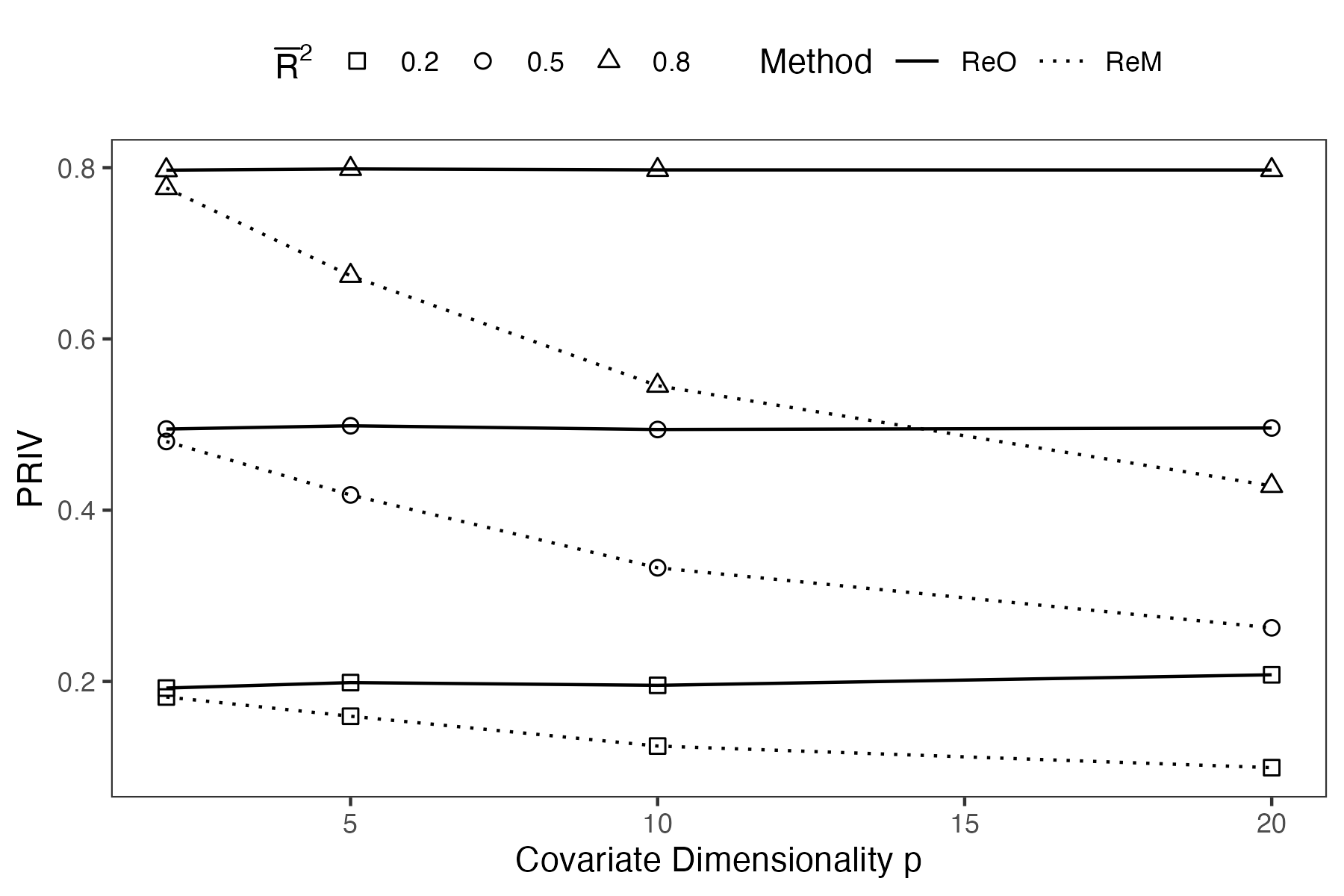}
\end{center}
\caption{ReO versus ReM in terms of average PRIV for different $p$ and $\bar{R}^2$ when $N=600$ and $\sigma^2_\bbeta=0.1$.\label{fig:Sim1ReO}}
\end{figure}

To verify Theorem \ref{PRIAV-ReO} and compare ReO and ReM, Figure \ref{fig:Sim1ReO} shows average PRIV for both methods across various covariate dimensions ($p$) with different $\bar{R}^2$ values at $\rho=0$, $\sigma^2_\bbeta=0.1$ and $N=600$. As $\bar{R}^2$ increases, both methods show substantial PRIV improvements. ReM's average PRIV decreases with higher $p$, while ReO remains stable at fixed $\bar{R}^2$. 

\subsection{Validation of acceptance probability}\label{sec:validation}

\begin{figure}[h]
    \centering
    \includegraphics[scale=0.8]{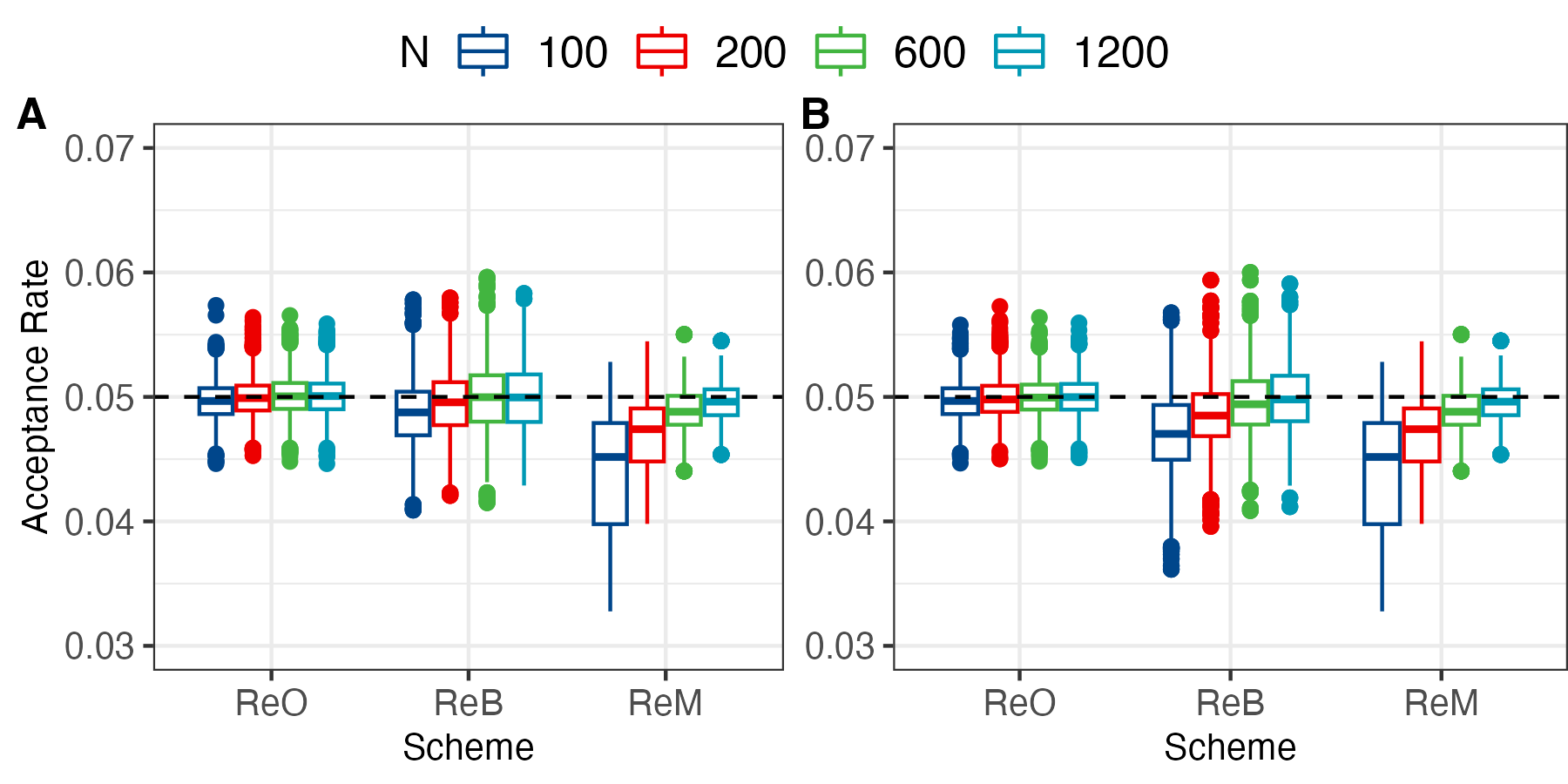}
    \caption{The realized acceptance probability of ReO, ReB and ReM when (A) $\sigma_{\bbeta}^2=0.1$ and (B) $\sigma_{\bbeta}^2=1$. The black dotted line represents value 0.05.}
    \label{fig:Acceptance_rate}
\end{figure}

To verify the realized acceptance probability of the three re-randomization schemes, running time of re-randomizing for each method in each setting is recorded. Figure \ref{fig:Acceptance_rate} shows the reciprocal of the running time, which is equivalent to acceptance rate for each method. We can see that as sample size increases, the acceptance probability will get closer to 0.05, which is exactly what we expect.

\subsection{Tuned acceptance rates for different re-randomization procedures and their corresponding running time under linear settings}\label{sec:linear_time}

Table \ref{tab:ARlinear} and Table \ref{tab:RTlinear} show the numerically tuned acceptance rates of different re-randomization procedures and their corresponding average running time of sampling 1000 assignments under linear settings.  
We can see that the computational cost of these methods are comparative under their acceptance rates and thus, the comparisons of PRIV are fair. 

\begin{table}[ht]
    \centering
    \footnotesize
\caption{Acceptance rates of different re-randomization procedures under linear settings.}
\label{tab:ARlinear}
    \setlength{\tabcolsep}{2.5mm}{
\begin{tabular}{cccccccccc}
    \midrule
    & \multicolumn{3}{c}{$N=200$}&\multicolumn{3}{c}{$N=600$}&\multicolumn{3}{c}{$N=1200$}\\
    \cmidrule(lr){2-4}
    \cmidrule(lr){5-7}
    \cmidrule(lr){8-10}
    Scheme & $r=0.2$ & 0.3 & 0.4 & 0.2 & 0.3 & 0.4 & 0.2 & 0.3 & 0.4 \\\hline
    \multicolumn{10}{c}{}\vspace{-0.2cm}\\
    & \multicolumn{9}{c}{$p=2$}\\\cmidrule(lr){2-10}
    ReM & 0.050 & 0.050 & 0.050 & 0.050 & 0.050 & 0.050 & 0.050 & 0.050 & 0.050 \\
    ReM-ReB & 0.086 & 0.089 & 0.096 & 0.073 & 0.071 & 0.073 & 0.061 & 0.063 & 0.066 \\   
    ReM-ReO & 0.082 & 0.086 & 0.087 & 0.069 & 0.071 & 0.071 & 0.063 & 0.062 & 0.062 \\  
    BCRD-ReB & 0.052 & 0.047 & 0.045 & 0.042 & 0.040 & 0.045 & 0.046 & 0.038 & 0.034 \\ 
    BCRD-ReO & 0.053 & 0.048 & 0.049 & 0.046 & 0.040 & 0.040 & 0.043 & 0.040 & 0.035 \\ 
    \cmidrule(lr){2-10}
    \multicolumn{10}{c}{}\vspace{-0.3cm}\\
    & \multicolumn{9}{c}{$p=5$}\\\cmidrule(lr){2-10}
    ReM & 0.050 & 0.050 & 0.050 & 0.050 & 0.050 & 0.050 & 0.050 & 0.050 & 0.050 \\
    ReM-ReB & 0.093 & 0.087 & 0.092 & 0.069 & 0.069 & 0.071 & 0.063 & 0.068 & 0.058 \\  
    ReM-ReO & 0.089 & 0.076 & 0.086 & 0.069 & 0.071 & 0.068 & 0.063 & 0.063 & 0.068 \\ 
    BCRD-ReB & 0.049 & 0.047 & 0.044 & 0.048 & 0.045 & 0.040 & 0.044 & 0.045 & 0.031 \\ 
    BCRD-ReO & 0.048 & 0.045 & 0.044 & 0.044 & 0.041 & 0.040 & 0.041 & 0.039 & 0.033 \\  
    \cmidrule(lr){2-10}
    \multicolumn{10}{c}{}\vspace{-0.3cm}\\
    & \multicolumn{9}{c}{$p=10$}\\\cmidrule(lr){2-10} 
    ReM & 0.050 & 0.050 & 0.050 & 0.050 & 0.050 & 0.050 & 0.050 & 0.050 & 0.050 \\
    ReM-ReB & 0.094 & 0.087 & 0.089 & 0.068 & 0.074 & 0.069 & 0.059 & 0.061 & 0.058 \\ 
    ReM-ReO & 0.092 & 0.085 & 0.082 & 0.067 & 0.071 & 0.067 & 0.059 & 0.060 & 0.059 \\ 
    BCRD-ReB & 0.052 & 0.047 & 0.045 & 0.048 & 0.079 & 0.040 & 0.041 & 0.039 & 0.039 \\ 
    BCRD-ReO & 0.047 & 0.044 & 0.039 & 0.043 & 0.041 & 0.038 & 0.041 & 0.038 & 0.036 \\  
    \cmidrule(lr){2-10}
    \multicolumn{10}{c}{}\vspace{-0.3cm}\\
    & \multicolumn{9}{c}{$p=20$}\\\cmidrule(lr){2-10}
    ReM & 0.050 & 0.050 & 0.050 & 0.050 & 0.050 & 0.050 & 0.050 & 0.050 & 0.050 \\
    ReM-ReB & $--$ & 0.090 & 0.087 & 0.063 & 0.067 & 0.066 & 0.058 & 0.056 & 0.059 \\ 
    ReM-ReO & $--$ & 0.088 & 0.081 & 0.065 & 0.063 & 0.068 & 0.058 & 0.056 & 0.058 \\ 
    BCRD-ReB & $--$ & 0.044 & 0.043 & 0.042 & 0.040 & 0.036 & 0.044 & 0.038 & 0.034 \\ 
    BCRD-ReO & $--$ & 0.039 & 0.037 & 0.042 & 0.038 & 0.035 & 0.041 & 0.040 & 0.033 \\  
    \midrule
    \end{tabular}
}
\end{table}

\begin{table}[ht]
    \centering
    \footnotesize
\caption{Average running time (seconds) of sampling 1000 assignments using different re-randomization procedures under linear settings with tuned acceptance rates.}
\label{tab:RTlinear}
    \setlength{\tabcolsep}{2.5mm}{
\begin{tabular}{cccccccccc}
    \midrule
    & \multicolumn{3}{c}{$N=200$}&\multicolumn{3}{c}{$N=600$}&\multicolumn{3}{c}{$N=1200$}\\
    \cmidrule(lr){2-4}
    \cmidrule(lr){5-7}
    \cmidrule(lr){8-10}
    Scheme & $r=0.2$ & 0.3 & 0.4 & 0.2 & 0.3 & 0.4 & 0.2 & 0.3 & 0.4 \\\hline
    \multicolumn{10}{c}{}\vspace{-0.2cm}\\
    & \multicolumn{9}{c}{$p=2$}\\\cmidrule(lr){2-10}
    ReM & 0.852 & 0.867 & 0.848 & 1.581 & 1.571 & 1.579 & 2.653 & 2.642 & 2.656 \\ 
  ReM-ReB & 0.894 & 0.865 & 0.799 & 1.525 & 1.591 & 1.527 & 2.714 & 2.594 & 2.557 \\ 
  ReM-ReO & 0.919 & 0.892 & 0.873 & 1.586 & 1.568 & 1.560 & 2.554 & 2.644 & 2.662 \\ 
  BCRD-ReB & 0.850 & 0.919 & 0.907 & 1.750 & 1.689 & 1.402 & 2.510 & 2.783 & 2.796 \\ 
  BCRD-ReO & 0.826 & 0.887 & 0.820 & 1.581 & 1.652 & 1.537 & 2.664 & 2.591 & 2.713 \\
    \cmidrule(lr){2-10}
    \multicolumn{10}{c}{}\vspace{-0.3cm}\\
    & \multicolumn{9}{c}{$p=5$}\\\cmidrule(lr){2-10}
    ReM & 0.923 & 0.950 & 0.917 & 1.744 & 1.721 & 1.721 & 2.897 & 2.912 & 2.908 \\ 
  ReM-ReB & 0.911 & 0.976 & 0.898 & 1.742 & 1.742 & 1.696 & 2.785 & 2.640 & 3.138 \\ 
  ReM-ReO & 0.918 & 1.090 & 0.923 & 1.724 & 1.695 & 1.746 & 2.762 & 2.833 & 2.626 \\ 
  BCRD-ReB & 0.957 & 1.018 & 0.980 & 1.653 & 1.629 & 1.670 & 2.844 & 2.729 & 3.392 \\ 
  BCRD-ReO & 0.957 & 1.004 & 0.934 & 1.783 & 1.760 & 1.654 & 2.958 & 2.900 & 3.091 \\
    \cmidrule(lr){2-10}
    \multicolumn{10}{c}{}\vspace{-0.3cm}\\
    & \multicolumn{9}{c}{$p=10$}\\\cmidrule(lr){2-10}
    ReM & 1.057 & 1.037 & 1.035 & 1.951 & 1.956 & 1.971 & 3.360 & 3.345 & 3.360 \\ 
  ReM-ReB & 1.088 & 1.063 & 1.022 & 2.002 & 1.796 & 1.977 & 3.424 & 3.337 & 3.533 \\ 
  ReM-ReO & 1.076 & 1.064 & 1.069 & 1.997 & 1.882 & 2.006 & 3.420 & 3.370 & 3.461 \\ 
  BCRD-ReB & 1.007 & 1.065 & 1.040 & 1.846 & 1.024 & 1.853 & 3.495 & 3.346 & 3.022 \\ 
  BCRD-ReO & 1.072 & 1.033 & 1.098 & 1.999 & 1.913 & 1.924 & 3.451 & 3.410 & 3.250 \\ 
    \cmidrule(lr){2-10}
    \multicolumn{10}{c}{}\vspace{-0.3cm}\\
    & \multicolumn{9}{c}{$p=20$}\\\cmidrule(lr){2-10}
    ReM  & 1.339 & 1.320 & 1.311 & 2.473 & 2.459 & 2.459 & 4.272 & 4.259 & 4.284 \\ 
  ReM-ReB & $--$& 1.350 & 1.324 & 2.652 & 2.469 & 2.445 & 4.433 & 4.445 & 4.275 \\ 
  ReM-ReO & $--$ & 1.328 & 1.343 & 2.558 & 2.543 & 2.333 & 4.326 & 4.363 & 4.329 \\ 
  BCRD-ReB & $--$ & 1.295 & 1.313 & 2.611 & 2.429 & 2.581 & 4.132 & 4.249 & 4.315 \\ 
  BCRD-ReO & $--$ & 1.333 & 1.326 & 2.492 & 2.510 & 2.468 & 4.227 & 3.908 & 4.319 \\
    \midrule
    \end{tabular}
}
\end{table}

\subsection{Performance of two-stage ReB with various sample size proportion in the first stage}
\begin{figure}[ht]
\centering
\footnotesize
\includegraphics[scale=0.55]{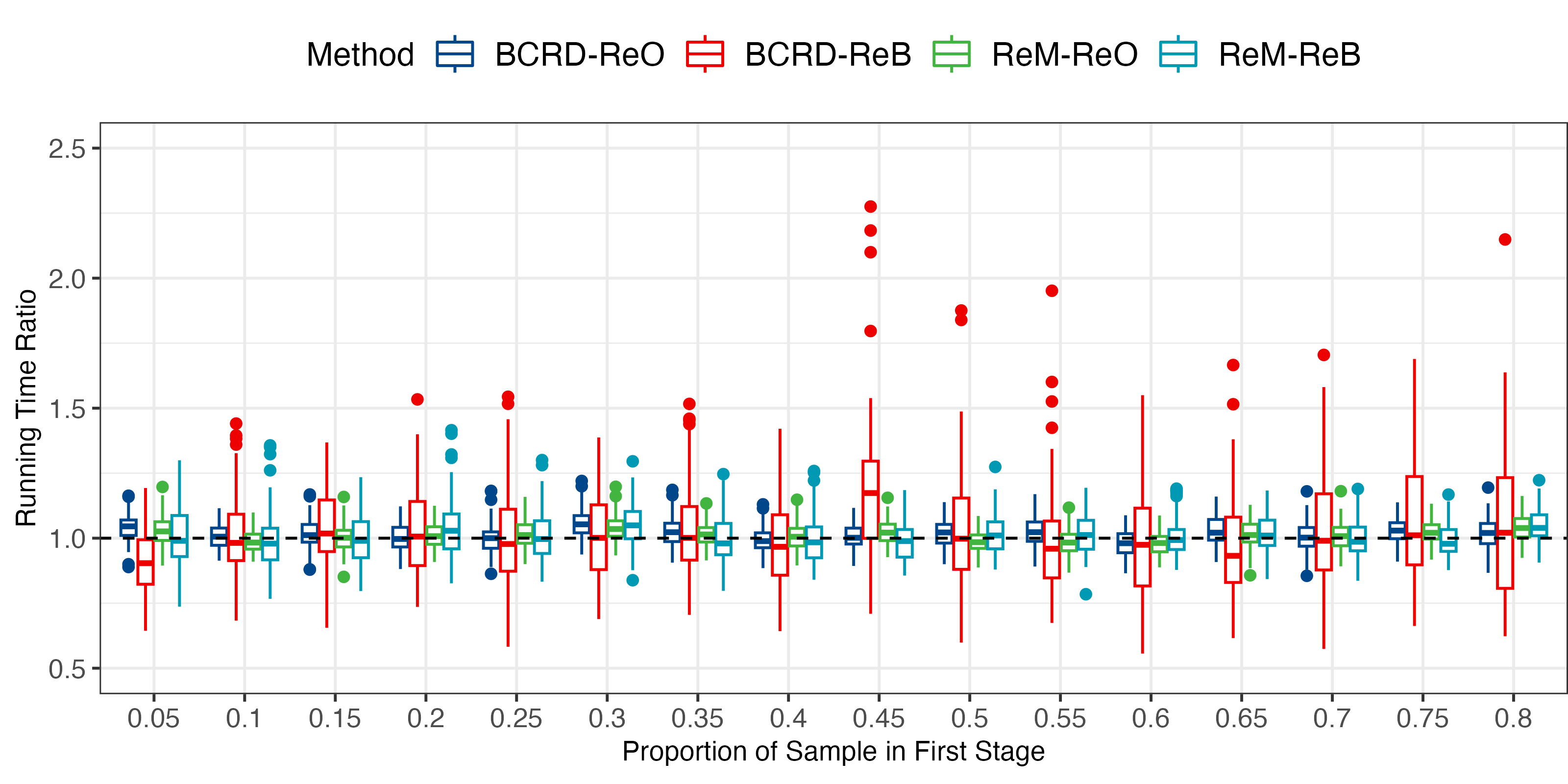}
\caption{Running time ratio of re-randomization procedures over ReM under different proportion of sample in first stage $r$ when total sample size $N=1200$ and covariate dimensionality $p=20$ in linear setting.}
\label{fig:r_time}
\end{figure}

To further study the effect of the proportion of sample in first stage on the performance of two-stage ReB procedures, we conduct an additional experiment on ReM, ReO, BCRD-ReO, BCRD-ReB, ReM-ReO and ReM-ReB with $r$ ranging from 0.05 to 0.8 when $N=1200$, $p=20$, $\bar{R}^2=0.5$ and $\rho=0$. 
We set acceptance rate of ReM as $\alpha=0.05$ and tune the proper acceptance probabilities of other re-randomization procedures numerically as before. 
The running time ratio of different re-randomization procedures over ReM is plotted in Figure \ref{fig:r_time}. { It is observed that the running time of BCRD-ReB is relatively unstable than the other two-stage ReB procedures. A conjecture is that the instability arises from the use of Monte Carlo simulation during the second stage for threshold estimation. The running time of ReM-ReB, however, remains relatively stable, possibly because it constrains the allocation in the first stage using ReM. Overall, the average running time of ReM and two-stage procedures is comparable.}

\begin{figure}[t]
\begin{center}
\includegraphics[width=4in]{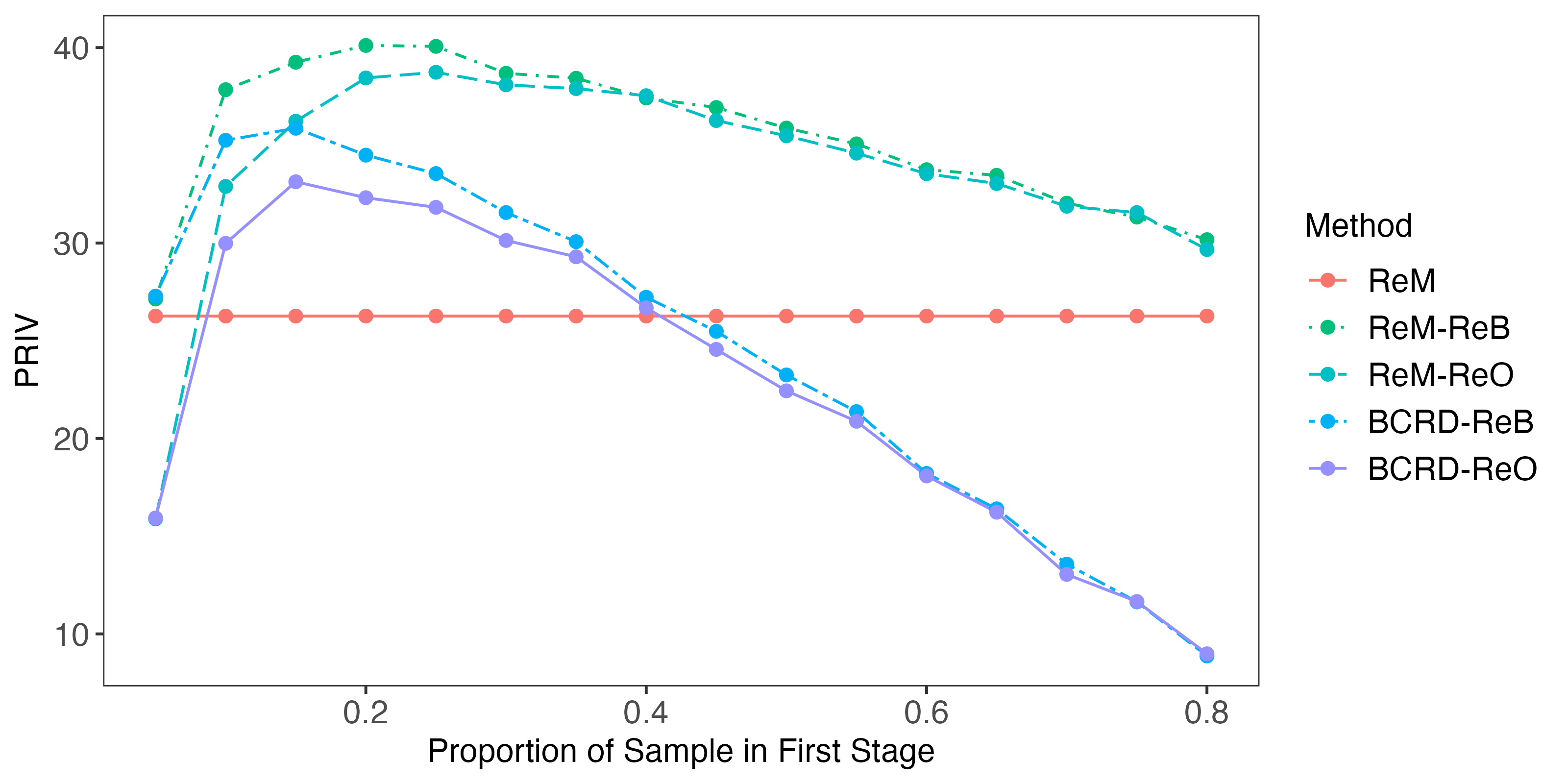}
\end{center}
\caption{Average PRIV of ReO, ReM, ReM-ReB, ReM-ReO, BCRD-ReB and BCRD-ReO, with different proportion of sample in first stage $r$ when total sample size $N=1200$ and covariate dimensionality $p=20$ in linear setting.
\label{fig:r_plot2}}
\end{figure}

Average PRIVs of these methods under different specifications of $r$ are plotted in Figure \ref{fig:r_plot2}, from which we can see the following facts immediately.
First, all two-stage methods achieve the best performance when $r$ falls into the region of $[0.1,0.3]$. 
Second, ReM-ReB exhibits the best robustness to different specifications of $r$ out of all the schemes considered.
Thus, as a rule of thumb, we recommend ReM-ReB with about 20\% of samples in the first stage as the primary method for practical usage.

\subsection{Simulation results under non-linear settings}\label{sec:nonlinear_result}
All above simulations are conducted under linear models. 
To evaluate the performance of two-stage ReB under non-linear models, we repeated the above numerical experiment with the linear regression model for potential outcomes replaced by the non-linear regression model below:
{ \[\begin{split}
 Y_i(1) &= a + \bbeta_1^T \exp\{\bX_i\} + \epsilon_{1,i},\ \ i=1,\ldots,N,\\
Y_i(0) &= \bbeta_0^T \exp\{\bX_i\} + \epsilon_{0,i},\ \ i=1,\ldots,N,
\end{split}\]}
with all the other simulation settings unchanged.
Under non-linear settings, the acceptance rate of ReM is set to be $0.05$ and the acceptance probabilities of other methods are tuned numerically as in experiments under linear settings. The detailed acceptance rates as well as their corresponding running time of sampling 1000 assignments are shown in Table \ref{tab:ARnonlinear} and Table \ref{tab:RTnonlinear}. The observations are similar to those in Section B.4.

\begin{table}[h]
    \centering
    \footnotesize
\caption{Acceptance rates of different re-randomization procedures under non-linear settings.}
\label{tab:ARnonlinear}
    \setlength{\tabcolsep}{2.5mm}{
\begin{tabular}{cccccccccc}
    \midrule
    & \multicolumn{3}{c}{$N=200$}&\multicolumn{3}{c}{$N=600$}&\multicolumn{3}{c}{$N=1200$}\\
    \cmidrule(lr){2-4}
    \cmidrule(lr){5-7}
    \cmidrule(lr){8-10}
    Scheme & $r=0.2$ & 0.3 & 0.4 & 0.2 & 0.3 & 0.4 & 0.2 & 0.3 & 0.4 \\\hline
    \multicolumn{10}{c}{}\vspace{-0.2cm}\\
    & \multicolumn{9}{c}{$p=2$}\\\cmidrule(lr){2-10}
    ReM & 0.050 & 0.050 & 0.050 & 0.050 & 0.050 & 0.050 & 0.050 & 0.050 & 0.050 \\ 
    ReM-ReB & 0.085 & 0.088 & 0.088 & 0.071 & 0.073 & 0.073 & 0.063 & 0.062 & 0.066 \\ 
    ReM-ReO & 0.096 & 0.086 & 0.083 & 0.068 & 0.071 & 0.072 & 0.063 & 0.061 & 0.061 \\ 
    BCRD-ReB & 0.055 & 0.053 & 0.049 & 0.042 & 0.040 & 0.042 & 0.044 & 0.040 & 0.033 \\
    BCRD-ReO & 0.062 & 0.047 & 0.046 & 0.044 & 0.045 & 0.043 & 0.043 & 0.039 & 0.036 \\ 
    \cmidrule(lr){2-10}
    \multicolumn{10}{c}{}\vspace{-0.3cm}\\
    & \multicolumn{9}{c}{$p=5$}\\\cmidrule(lr){2-10}
    ReM & 0.050 & 0.050 & 0.050 & 0.050 & 0.050 & 0.050 & 0.050 & 0.050 & 0.050 \\ 
    ReM-ReB & 0.096 & 0.085 & 0.096 & 0.068 & 0.069 & 0.071 & 0.061 & 0.063 & 0.063 \\ 
    ReM-ReO & 0.085 & 0.086 & 0.082 & 0.069 & 0.069 & 0.071 & 0.059 & 0.061 & 0.062 \\ 
    BCRD-ReB & 0.053 & 0.044 & 0.045 & 0.048 & 0.045 & 0.038 & 0.043 & 0.040 & 0.032 \\ 
    BCRD-ReO & 0.045 & 0.045 & 0.043 & 0.047 & 0.041 & 0.039 & 0.041 & 0.040 & 0.034 \\ 
    \cmidrule(lr){2-10}
    \multicolumn{10}{c}{}\vspace{-0.3cm}\\
    & \multicolumn{9}{c}{$p=10$}\\\cmidrule(lr){2-10}
    ReM & 0.050 & 0.050 & 0.050 & 0.050 & 0.050 & 0.050 & 0.050 & 0.050 & 0.050 \\ 
    ReM-ReB & 0.094 & 0.087 & 0.092 & 0.065 & 0.067 & 0.069 & 0.059 & 0.061 & 0.059 \\ 
    ReM-ReO & 0.094 & 0.085 & 0.087 & 0.066 & 0.072 & 0.066 & 0.059 & 0.060 & 0.058 \\ 
    BCRD-ReB & 0.052 & 0.043 & 0.040 & 0.043 & 0.041 & 0.039 & 0.045 & 0.039 & 0.033 \\ 
    BCRD-ReO & 0.048 & 0.043 & 0.040 & 0.043 & 0.045 & 0.037 & 0.042 & 0.037 & 0.033 \\ 
    \cmidrule(lr){2-10}
    \multicolumn{10}{c}{}\vspace{-0.3cm}\\
    & \multicolumn{9}{c}{$p=20$}\\\cmidrule(lr){2-10}
    ReM & 0.050 & 0.050 & 0.050 & 0.050 & 0.050 & 0.050 & 0.050 & 0.050 & 0.050 \\ 
    ReM-ReB & $--$ & 0.092 & 0.089 & 0.068 & 0.068 & 0.065 & 0.058 & 0.056 & 0.059 \\ 
    ReM-ReO & $--$ & 0.088 & 0.083 & 0.066 & 0.065 & 0.063 & 0.058 & 0.056 & 0.058 \\ 
    BCRD-ReB & $--$ & 0.044 & 0.045 & 0.046 & 0.039 & 0.036 & 0.079 & 0.038 & 0.032 \\ 
    BCRD-ReO & $--$ & 0.040 & 0.045 & 0.045 & 0.038 & 0.035 & 0.041 & 0.036 & 0.033 \\ 
    \midrule
    \end{tabular}
}
\end{table}

\begin{table}[h]
    \centering
    \footnotesize
\caption{Average running time (seconds) of sampling 1000 assignments using different re-randomization procedures under non-linear settings with tuned acceptance rates.}
\label{tab:RTnonlinear}
    \setlength{\tabcolsep}{2.5mm}{
\begin{tabular}{cccccccccc}
    \midrule
    & \multicolumn{3}{c}{$N=200$}&\multicolumn{3}{c}{$N=600$}&\multicolumn{3}{c}{$N=1200$}\\
    \cmidrule(lr){2-4}
    \cmidrule(lr){5-7}
    \cmidrule(lr){8-10}
    Scheme & $r=0.2$ & 0.3 & 0.4 & 0.2 & 0.3 & 0.4 & 0.2 & 0.3 & 0.4 \\\hline
    \multicolumn{10}{c}{}\vspace{-0.2cm}\\
    & \multicolumn{9}{c}{$p=2$}\\\cmidrule(lr){2-10}
    ReM & 0.850 & 0.850 & 0.847 & 1.591 & 1.573 & 1.614 & 2.658 & 2.643 & 2.651 \\ 
  ReM-ReB & 0.910 & 0.866 & 0.867 & 1.568 & 1.524 & 1.591 & 2.565 & 2.652 & 2.553 \\ 
  ReM-ReO & 0.790 & 0.886 & 0.900 & 1.616 & 1.551 & 1.592 & 2.551 & 2.673 & 2.695 \\ 
  BCRD-ReB & 0.814 & 0.813 & 0.836 & 1.781 & 1.677 & 1.551 & 2.610 & 2.683 & 2.853 \\ 
  BCRD-ReO & 0.713 & 0.899 & 0.869 & 1.638 & 1.483 & 1.506 & 2.665 & 2.655 & 2.677 \\ 
    \cmidrule(lr){2-10}
    \multicolumn{10}{c}{}\vspace{-0.3cm}\\
    & \multicolumn{9}{c}{$p=5$}\\\cmidrule(lr){2-10}
    ReM & 0.923 & 0.917 & 0.932 & 1.737 & 1.716 & 1.716 & 2.922 & 2.901 & 2.916 \\ 
  ReM-ReB & 0.873 & 0.957 & 0.852 & 1.742 & 1.728 & 1.718 & 2.923 & 2.844 & 2.825 \\ 
  ReM-ReO & 0.971 & 0.927 & 0.981 & 1.715 & 1.716 & 1.703 & 3.007 & 2.935 & 2.874 \\ 
  BCRD-ReB & 0.873 & 1.004 & 0.972 & 1.633 & 1.660 & 1.759 & 2.859 & 2.885 & 3.367 \\ 
  BCRD-ReO & 1.009 & 0.968 & 0.992 & 1.672 & 1.734 & 1.703 & 2.978 & 2.799 & 3.007 \\ 
    \cmidrule(lr){2-10}
    \multicolumn{10}{c}{}\vspace{-0.3cm}\\
    & \multicolumn{9}{c}{$p=10$}\\\cmidrule(lr){2-10}
    ReM & 1.052 & 1.057 & 1.046 & 1.966 & 1.959 & 1.983 & 3.334 & 3.337 & 3.333 \\ 
  ReM-ReB & 1.066 & 1.093 & 1.002 & 2.098 & 2.037 & 1.979 & 3.410 & 3.294 & 3.446 \\ 
  ReM-ReO & 1.032 & 1.087 & 1.020 & 2.040 & 1.847 & 2.049 & 3.383 & 3.368 & 3.475 \\ 
  BCRD-ReB & 1.002 & 1.168 & 1.192 & 2.044 & 1.959 & 1.943 & 3.116 & 3.333 & 3.579 \\ 
  BCRD-ReO & 1.045 & 1.104 & 1.094 & 2.035 & 1.784 & 2.005 & 3.344 & 3.460 & 3.465 \\ 
    \cmidrule(lr){2-10}
    \multicolumn{10}{c}{}\vspace{-0.3cm}\\
    & \multicolumn{9}{c}{$p=20$}\\\cmidrule(lr){2-10}
    ReM & 1.340 & 1.329 & 1.341 & 2.458 & 2.459 & 2.469 & 4.295 & 4.287 & 4.286 \\ 
  ReM-ReB & $--$ & 1.335 & 1.310 & 2.437 & 2.407 & 2.508 & 4.459 & 4.479 & 4.320 \\ 
  ReM-ReO & $--$ & 1.345 & 1.342 & 2.482 & 2.483 & 2.530 & 4.362 & 4.427 & 4.326 \\ 
  BCRD-ReB & $--$ & 1.299 & 1.275 & 2.395 & 2.550 & 2.605 & 2.313 & 4.355 & 4.543 \\ 
  BCRD-ReO & $--$ & 1.291 & 1.128 & 2.310 & 2.476 & 2.478 & 4.313 & 4.425 & 4.267 \\ 
    \midrule
    \end{tabular}
}
\end{table}

\begin{figure}[t]
\centering
\includegraphics[scale=0.6]{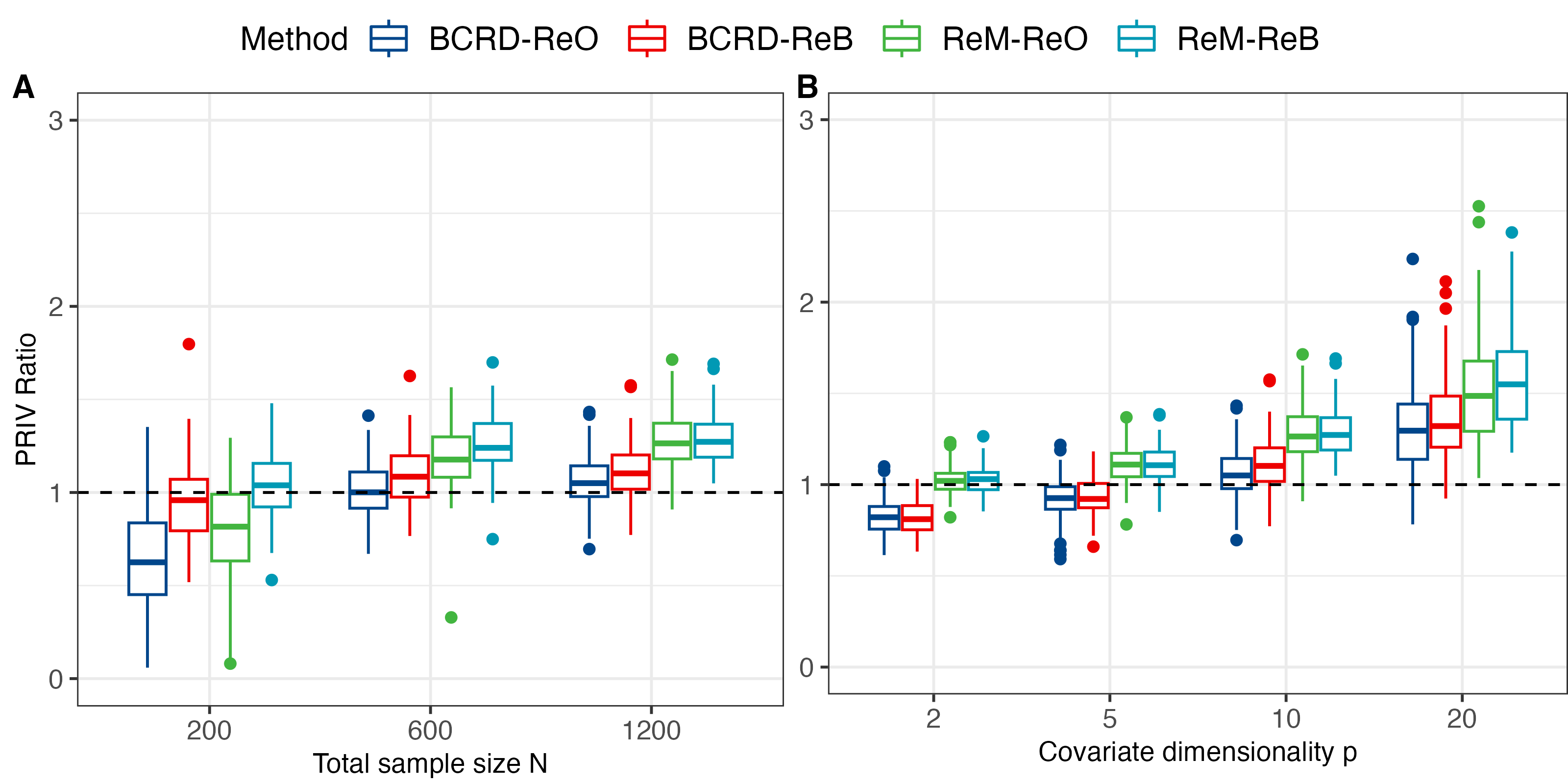}
\caption{Performance of BCRD-ReO, BCRD-ReB, ReM-ReO and ReM-ReB compared with ReM with $r=0.2$ under non-linear model. Sub-figure (A) investigates into factor 
$N$ with covariate dimensionality $p$ fixed at 
10, while sub-figure (B) investigates 
into factor $p$ with $N$ fixed at 1200.}
\label{fig:nonlinear}
\end{figure}

Table \ref{tab:nonlinear} shows the average
PRIV of different re-randomization methods with different $p, r$ and $N$. 
It implies that two-stage ReB, especially ReM-ReB, still outperforms ReM in most cases, as long as sample size is not very small, even under nonlinear model. Figure \ref{fig:nonlinear} shows the effect of $p$ on these methods with fixed $r=0.2$ and $p=10$, 
and the effect of $N$ when $r=0.2$ and $N=1200$. As sample size increases or dimension $p$ gets larger, the advantage of two-stage ReB schemes gets more obvious. Also, ReM-ReB has the best performance among the four schemes.

\begin{table}
    \centering
    \footnotesize
\caption{Average PRIV achieved by different re-randomization procedures under non-linear model.}
\label{tab:nonlinear}
    \setlength{\tabcolsep}{2.5mm}{
\begin{tabular}{cccccccccc}
    \midrule
    & \multicolumn{3}{c}{$N=200$}&\multicolumn{3}{c}{$N=600$}&\multicolumn{3}{c}{$N=1200$}\\
    \cmidrule(lr){2-4}
    \cmidrule(lr){5-7}
    \cmidrule(lr){8-10}
    Scheme & $r=0.2$ & 0.3 & 0.4 & 0.2 & 0.3 & 0.4 & 0.2 & 0.3 & 0.4 \\\hline
    \multicolumn{10}{c}{}\vspace{-0.2cm}\\
    & \multicolumn{9}{c}{$p=2$}\\\cmidrule(lr){2-10}
    \cellcolor[HTML]{D9D9D9} ReM & \cellcolor[HTML]{D9D9D9}\bf {48.28} & \cellcolor[HTML]{D9D9D9}\bf 48.28 & \cellcolor[HTML]{D9D9D9}\bf 48.28 & \cellcolor[HTML]{D9D9D9}47.53 & \cellcolor[HTML]{D9D9D9}47.53 & \cellcolor[HTML]{D9D9D9}47.53 & \cellcolor[HTML]{D9D9D9}47.28 & \cellcolor[HTML]{D9D9D9}47.28 & \cellcolor[HTML]{D9D9D9}47.28 \\ 
    ReM-ReB & 48.24 & 47.98 & 47.98 & \bf 48.28 & \bf 47.75 & 48.10 & \bf 48.23 & \bf 47.96 & \bf 48.06 \\ 
  ReM-ReO & 46.86 & 46.92 & 48.04 & 47.44 & 47.73 & \bf 48.35 & 48.19 & 47.91 & 47.85 \\ 
  BCRD-ReB & 38.74 & 34.97 & 30.40 & 39.34 & 33.54 & 29.90 & 38.65 & 33.99 & 28.57 \\ 
  BCRD-ReO & 37.17 & 33.06 & 29.23 & 38.42 & 33.14 & 29.48 & 38.72 & 33.65 & 28.88 \\ 
    \cmidrule(lr){2-10}
    \multicolumn{10}{c}{}\vspace{-0.3cm}\\
    & \multicolumn{9}{c}{$p=5$}\\\cmidrule(lr){2-10}
    \cellcolor[HTML]{D9D9D9} ReM & \cellcolor[HTML]{D9D9D9}40.55 & \cellcolor[HTML]{D9D9D9}40.55 & \cellcolor[HTML]{D9D9D9}40.55 & \cellcolor[HTML]{D9D9D9}41.24 & \cellcolor[HTML]{D9D9D9}41.24 & \cellcolor[HTML]{D9D9D9}41.24 & \cellcolor[HTML]{D9D9D9}40.86 & \cellcolor[HTML]{D9D9D9}40.86 & \cellcolor[HTML]{D9D9D9}40.86 \\
    ReM-ReB & \bf 42.97 & \bf 42.89 & \bf 42.30 & \bf 44.62 & \bf 44.82 & \bf 44.11 & \bf 45.29 & \bf 45.20 & \bf 44.57 \\ 
  ReM-ReO & 38.66 & 40.92 & 41.63 & 43.79 & 44.21 & 43.74 & 45.03 & 44.95 & 44.23 \\ 
  BCRD-ReB & 36.53 & 32.38 & 27.04 & 37.71 & 32.40 & 28.51 & 37.97 & 33.66 & 29.13 \\ 
  BCRD-ReO & 31.18 & 29.06 & 25.75 & 36.61 & 32.05 & 28.08 & 37.59 & 32.55 & 28.87 \\  
    \cmidrule(lr){2-10}
    \multicolumn{10}{c}{}\vspace{-0.3cm}\\
    & \multicolumn{9}{c}{$p=10$}\\\cmidrule(lr){2-10}
    \cellcolor[HTML]{D9D9D9} ReM & \cellcolor[HTML]{D9D9D9}32.88 & \cellcolor[HTML]{D9D9D9}32.88 & \cellcolor[HTML]{D9D9D9}32.88 & \cellcolor[HTML]{D9D9D9}32.77 & \cellcolor[HTML]{D9D9D9}32.77  & \cellcolor[HTML]{D9D9D9}32.77  & \cellcolor[HTML]{D9D9D9}33.36 & \cellcolor[HTML]{D9D9D9}33.36 & \cellcolor[HTML]{D9D9D9}33.36 \\ 
    ReM-ReB & \bf 33.69 & \bf 35.28 & \bf 36.56 & \bf 41.15 & \bf 41.01 & \bf 39.20 & \bf 42.72 & \bf 42.26 & \bf 41.36 \\ 
  ReM-ReO & 25.57 & 30.78 & 33.74 & 38.58 & 39.11 & 38.86 & 42.44 & 41.87 & 41.09 \\ 
  BCRD-ReB & 30.53 & 28.91 & 25.86 & 35.59 & 31.04 & 26.90 & 36.85 & 32.70 & 28.57 \\ 
  BCRD-ReO & 20.64 & 23.56 & 21.80 & 33.11 & 29.80 & 26.71 & 35.33 & 31.95 & 27.99 \\ 
    \cmidrule(lr){2-10}
    \multicolumn{10}{c}{}\vspace{-0.3cm}\\
    & \multicolumn{9}{c}{$p=20$}\\\cmidrule(lr){2-10}
    \cellcolor[HTML]{D9D9D9} ReM & \cellcolor[HTML]{D9D9D9}\textbf{23.93} & \cellcolor[HTML]{D9D9D9}{23.93} & \cellcolor[HTML]{D9D9D9}23.93 & \cellcolor[HTML]{D9D9D9}24.08 & \cellcolor[HTML]{D9D9D9}24.08 & \cellcolor[HTML]{D9D9D9}24.08 & \cellcolor[HTML]{D9D9D9}25.66 & \cellcolor[HTML]{D9D9D9}25.66 & \cellcolor[HTML]{D9D9D9}25.66 \\ 
      ReM-ReB & $--$ & \bf 24.50 & \bf 27.46 & \bf 34.30 & \bf 34.40 & \bf 33.88 & \bf 39.75 & \bf 38.68 & \bf 37.01 \\ 
  ReM-ReO & $--$ & 17.66 & 24.07 & 31.05 & 32.52 & 32.63 & 38.16 & 37.60 & 36.39 \\ 
  BCRD-ReB & $--$ & 19.66 & 20.33 & 30.20 & 27.76 & 24.50 & 34.23 & 30.99 & 27.27 \\ 
  BCRD-ReO & $--$ & 11.64 & 14.60 & 25.60 & 25.36 & 23.16 & 33.21 & 30.21 & 26.90 \\ 
    \midrule
    \end{tabular}
}
\end{table}

\begin{figure}[t]
    \centering
\includegraphics[scale=0.55]{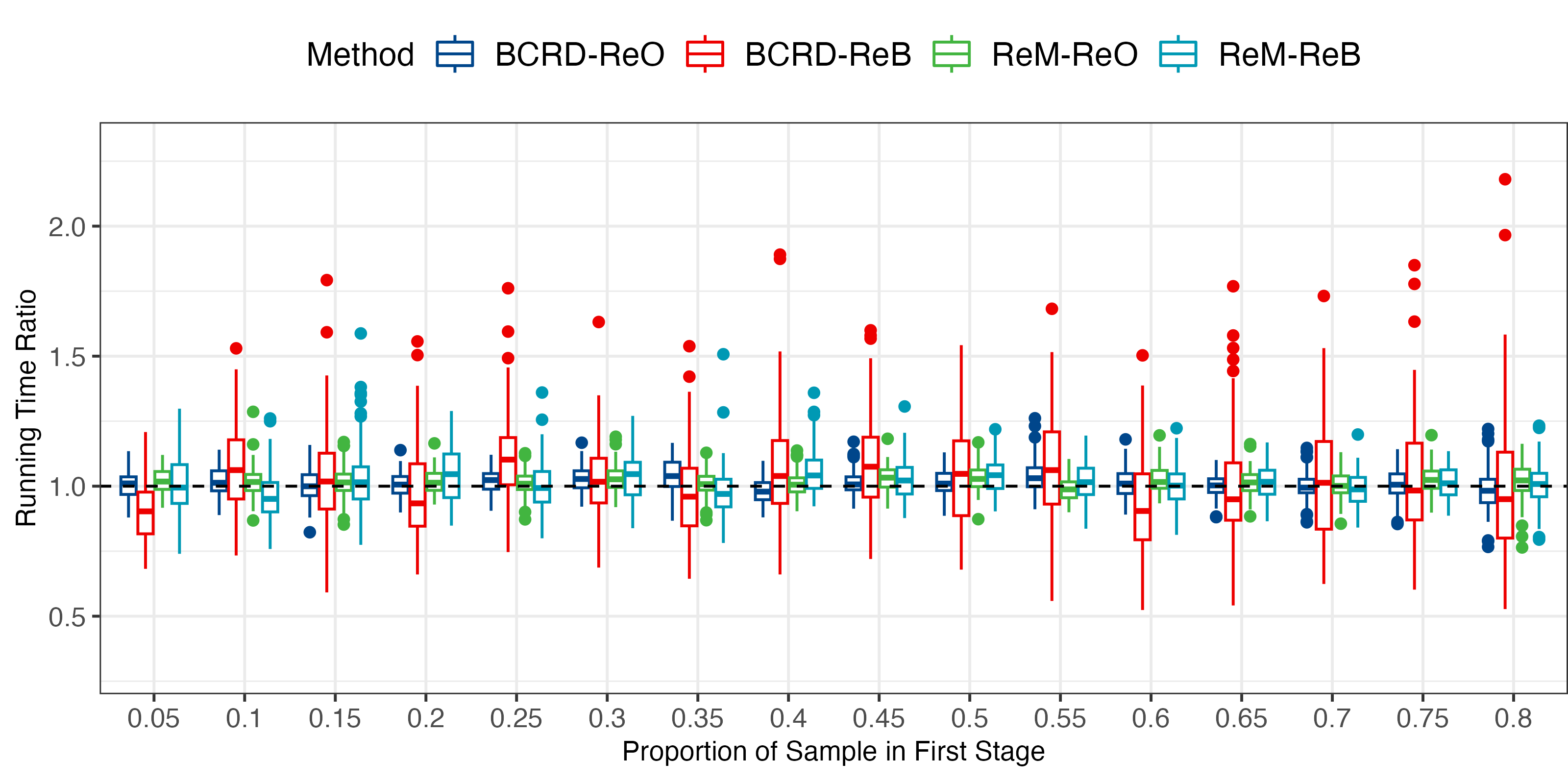}
    \caption{Running time ratio of re-randomization procedures over ReM under different proportion of sample in first stage $r$ when total sample size $N=1200$ and covariate dimensionality $p=20$ under non-linear model.}
    \label{fig:nonlinear_r_time}
\end{figure}

\begin{figure}[t]
\centering
\includegraphics[scale=0.55]{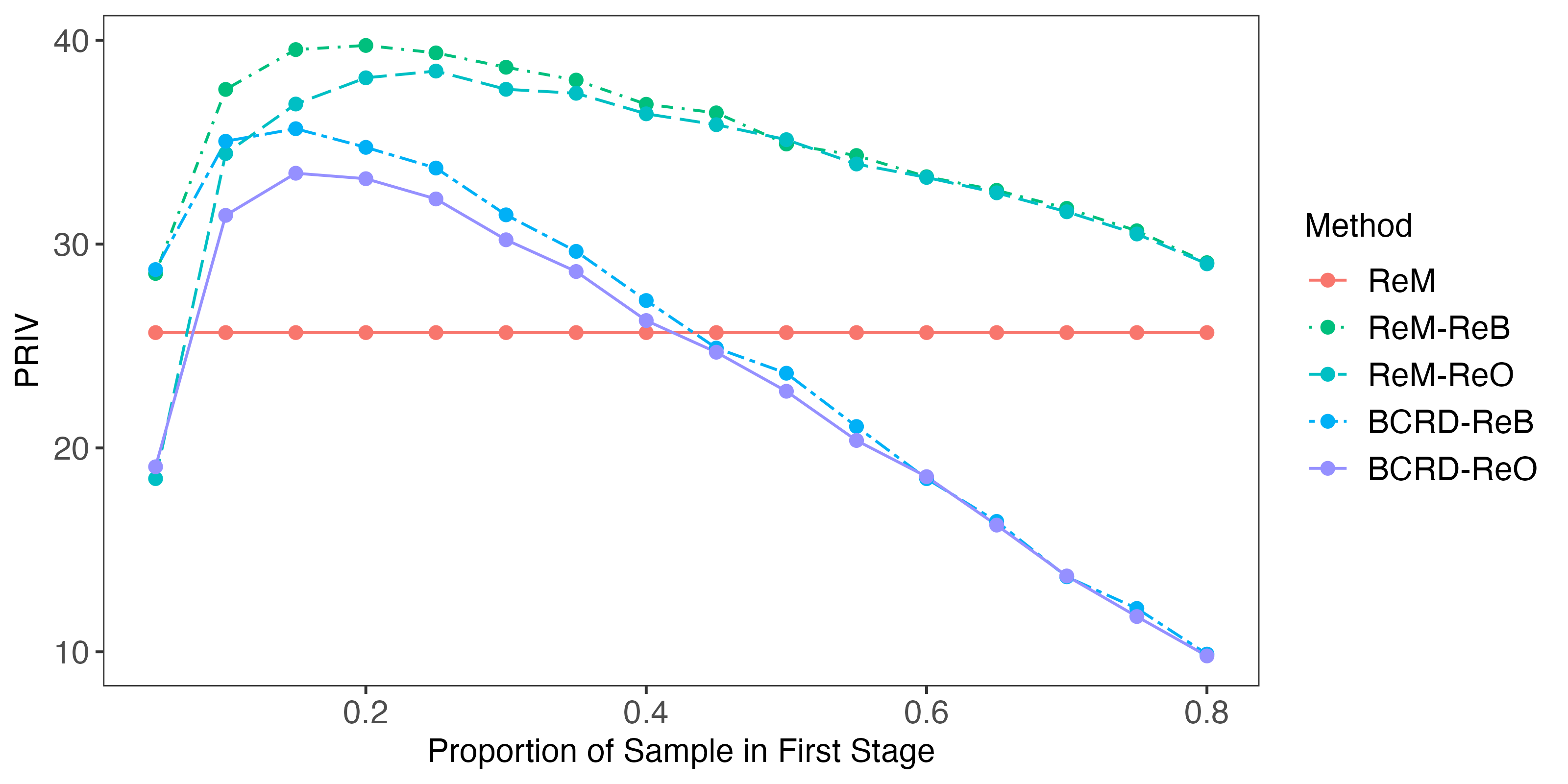}
\caption{Average PRIV of ReM, ReM-ReB, ReM-ReO, BCRD-ReB and BCRD-ReO with different proportion of sample in first stage $r$ when total sample size $N=1200$ and covariate dimensionality $p=20$ under non-linear model.}
    \label{fig:r_plot3}
\end{figure}

The effect of the proportion of sample in first stage on the performance of two-stage ReB procedures can also be studied by conducting experiment on ReM, BCRD-ReO, BCRD-ReB, ReM-ReO and ReM-ReB with $r$ ranging from 0.05 to 0.8 when $N=1200$, $p=20$, $\bar{R}^2=0.5$ and $\rho=0$. Similar to the experiments in linear setting, we set the acceptance rate of ReM be $\alpha=0.05$ and calculate the acceptance probabilities of other re-randomization procedures numerically. The running time ratio of re-randomization procedures over ReM is shown in Figure \ref{fig:nonlinear_r_time}.
Average PRIVs of these methods under different specifications of $r$ are plotted in Figure \ref{fig:r_plot3} and it is easy to see that ReM-ReB still exhibits the best robustness to the proportion of sample in first stage and similar conclusions as in linear settings can be drawn.

\subsection{Running time ratio in real-data analysis}\label{sec:realdata_time}

\begin{figure}[t]
    \centering
\includegraphics[scale=0.55]{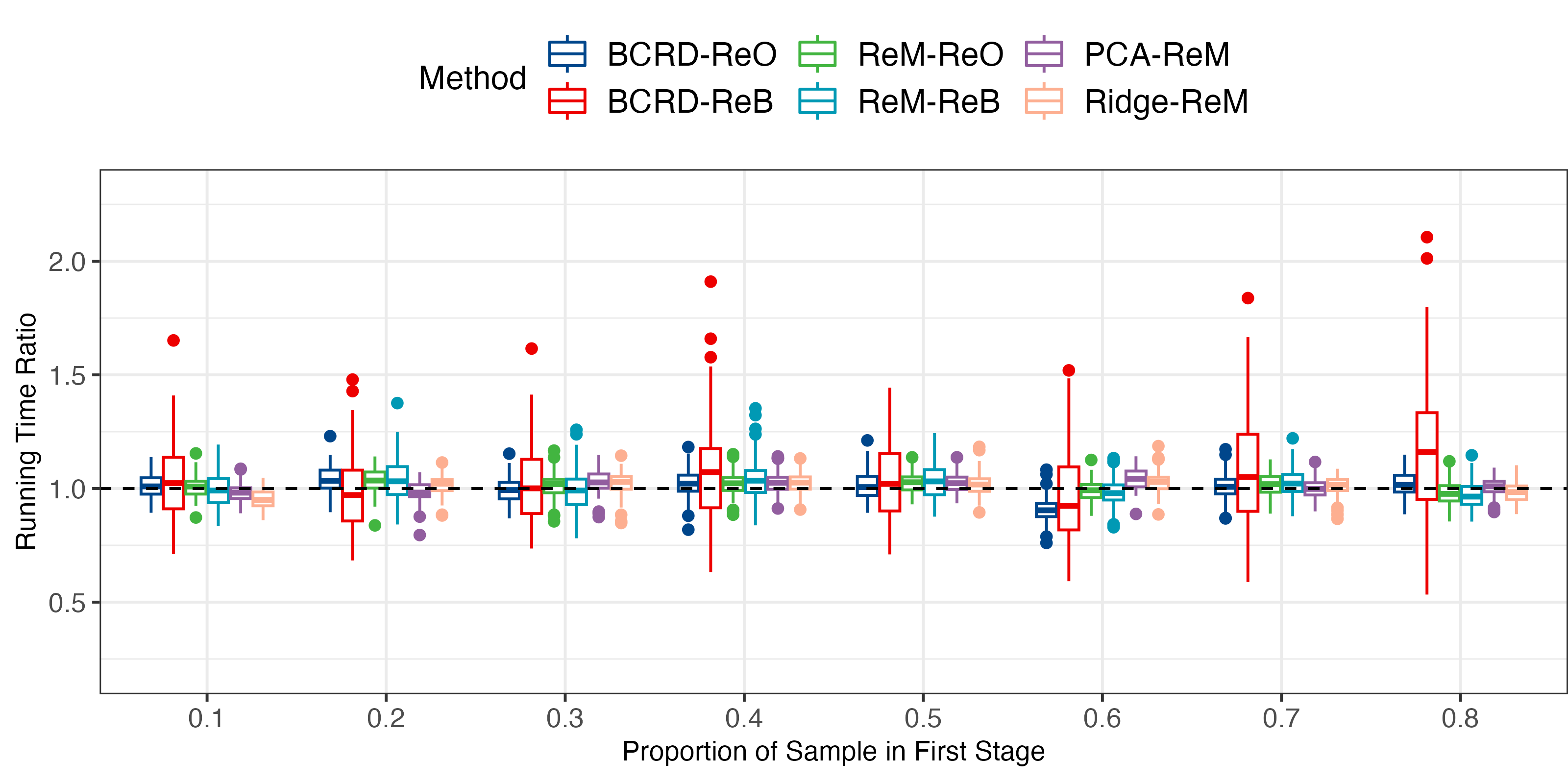}
    \caption{Running time ratio of re-randomization procedures over ReM under different proportion of sample in first stage $r$ in real-data analysis.}
    \label{fig:real_time}
\end{figure}

Figure \ref{fig:real_time} shows the running time ratio of re-randomization procedures against ReM under different proportion of sample in first stage $r$ in real application, i.e., Section \ref{sec:RealDataAnalysis}.

\end{document}